\documentclass[%
reprint,
amsmath,amssymb,
aps,
]{revtex4-2}
\usepackage{xcolor}
\usepackage{graphicx}
\usepackage{dcolumn}
\graphicspath{{./graphics/}}
\usepackage{enumerate}
\usepackage{subcaption}
\usepackage{bm}

\graphicspath{{./figures_v5/}}

\begin{document}

\title{Collective behavior of squirmers in thin films} 

\author{Bohan Wu-Zhang, Dmitry A. Fedosov, and Gerhard Gompper}
\affiliation{Theoretical Physics of Living Matter, Institute of Biological Information Processing and Institute for Advanced Simulation,
Forschungszentrum J\"ulich, 52425 J\"ulich, Germany \\
Email: b.zhang@fz-juelich.de, d.fedosov@fz-juelich.de, g.gompper@fz-juelich.de}

\date{\today}

\begin{abstract}
Bacteria in biofilms form complex structures and can collectively migrate
within mobile aggregates, which is referred to as swarming. This behavior is influenced by 
a combination of various factors, including morphological characteristics and propulsive forces of swimmers, their volume fraction 
within a confined environment, and hydrodynamic and steric interactions between them. In our study, we employ the squirmer model 
for microswimmers and the dissipative particle dynamics method for fluid modeling to investigate the collective 
motion of swimmers in thin films. The film thickness permits a free orientation of non-spherical squirmers, but 
constraints them to form a two-layered structure at maximum. Structural and dynamic properties of squirmer suspensions confined 
within the slit are analyzed for different volume fractions of swimmers, motility types (e.g., pusher, neutral squirmer, puller), 
and the presence of a rotlet dipolar flow field, which mimics the counter-rotating flow generated by flagellated bacteria. 
Different states are characterized, including a gas-like phase, swarming, and motility-induced phase separation, 
as a function of increasing volume fraction. Our study highlights the importance of an anisotropic swimmer shape, hydrodynamic 
interactions between squirmers, and their interaction with the walls for the emergence of different collective behaviors. 
Interestingly, the formation of collective structures may not be symmetric with respect to the two walls.
Furthermore, the presence of a rotlet dipole significantly mitigates differences in the collective behavior between various 
swimmer types. These results contribute to a better understanding of the formation of bacterial biofilms and the emergence 
of collective states in confined active matter.
\end{abstract}


\maketitle


\section{Introduction}

Collective motion of microswimmers is a popular topic in the field of active matter due to its wide applicability 
in the context of both biological and artificial systems as well as the richness of observed behaviors and physical 
mechanisms \cite{Elgeti_PMS_2015,Bechinger_APC_2016,Gompper_MAM_2020}. A prominent example of the collective behavior 
of biological microswimmers is biofilms, which represent complex dynamic communities of microorganisms at surfaces
\cite{Hall_BBN_2004,Verstraeten_SBF_2008}. Biofilms are often associated with various infectious diseases, such as 
dental plaque formation on teeth \cite{Marsh_DPB_2006} and chronic wounds that resist healing \cite{Attinger_cabcw_2012},
driving the research to better understand surface colonization by bacteria and biofilm development 
\cite{Verstraeten_SBF_2008,Mazza_PBI_2016}. In the context of artificial microswimmers, there exist a variety of 
active systems, including collectives of diffusiophoretic and thermophoretic Janus particles 
\cite{Palacci_LCL_2013,Buttinoni_CPS_2013,Bauerle_FSS_2020}, active droplets \cite{Krueger_CBAE_2016,Thutupalli_FIPS_2018}, 
and Quincke rollers \cite{Bricard_PMC_2013}. Studies with artificial microswimmers primarily focus on understanding 
the emergence of collective motion and the governing physical mechanisms
\cite{Elgeti_PMS_2015,Bechinger_APC_2016,Gompper_MAM_2020}. 

One of the prominent examples of collective behavior is motility-induced phase separation (MIPS), which can occur
without any attractive or alignment interactions between self-propelled particles 
\cite{Fily_APS_2012,Bialke_MIPS_2013,Redner_MIPS_2013,Cates_MIPS_2015,Digregorio_ABD_2018}. MIPS are characterized by the co-existence 
of low and high density phases of active particles, where the latter is generally represented by nearly immobile large 
clusters of particles \cite{Palacci_LCL_2013,Buttinoni_CPS_2013,Liu_SDT_2019}. In the initial stage of biofilm formation, 
a collective motility of bacteria known as {\it swarming} is frequently observed \cite{Verstraeten_SBF_2008,Beer_PDBS_2020}.
In contrast to MIPS clusters, bacterial swarms are stable and highly mobile aggregates, which generally migrate along 
surfaces. The formation of swarms often requires some type of alignment interactions between microswimmers 
\cite{Peruani_CSPR_2006,Vicsek_CM_2012}. Another collective phenomenon observed for bacterial suspensions is called 
active turbulence \cite{Wioland_CBSSV_2013,Martinez_RBS_2020,Guo_SVB_2018}, which is similar to the traditional 
Kolmogorov-Kraichnan-type hydrodynamic turbulence \cite{Kraichnan_TDT_1980}, but occurs at very low Reynolds numbers. 
The state of active turbulence is 
observed for a sufficiently large density of microswimmers, which each generates a force-dipolar flow field around it 
\cite{Wensink_MST_2012,Stenhammar_CBM_2017,Bardfalvy_LB_AT_2019,Qi_EAT_2022}. Finally, the motion of microswimmers in 
confinement, such as in channels and pores, can result in a hydrodynamic instability and the formation of complex 
flow patterns \cite{Wioland_CBSSV_2013,Lushi_FLSOCS_2014,Theillard_ACM_2017}. 

Different properties of microswimmers and their environment affect the emerging collective behavior. These include 
shape anisotropy \cite{Suma_MIPS_2014,Baer_SPR_2020,Moran_PAT_2022}, the generated flow field around a swimmer 
\cite{Alarcon_SAO_2013,Theers_CMS_2018,Qi_EAT_2022}, hydrodynamic interactions 
\cite{Zoettl_HCM_2014,Theers_CMS_2018,Matas_Navarro_HPS_2014}, the presence of confinement 
\cite{Lushi_FLSOCS_2014,Kuhr_CDMS_2019}, and dimensionality \cite{Stenhammar_ABP_2014,Krueger_CBAE_2016}. Biological
microswimmers (e.g., bacteria) have complex geometries and propulsion mechanisms, such that their detailed modeling 
is computationally challenging, especially in systems with a large number of swimmers. One of the popular models of 
simplified swimmers is the squirmer model \cite{Lighthill_SMB_1952,Blake_SEA_1971}, which represents a microswimmer 
by a spherically- or spheroidally-shaped active particle with a prescribed slip velocity at its surface \cite{Theers_MSM_2016}. 
Despite its simplicity, the squirmer 
model is flexible enough to capture various swimming modes, including pushers (e.g., {\it E. coli}), neutral microswimmers
(e.g., {\it Paramecium}), and pullers (e.g., {\it Chlamydomonas}). In particular, the squirmer model mimics well 
both the far- and near-field flow of a variety of realistic microswimmers. 

The squirmer model has been used to study collective behavior of microswimmers in quasi two-dimensional (2D) systems 
(i.e., a  monolayer of swimmers) 
\cite{Theers_CMS_2018, Qi_EAT_2022, Kuhr_CDMS_2019, Kyoya_SMCM_2015, Alarcon_MCP_2017, Zoettl_HCM_2014, Yoshinaga_HIDS_2017, Blaschke_PSC_2016}
and three-dimensional (3D) settings with periodic boundary conditions (BCs)
\cite{Ishikawa_DSM_2007,Delmotte_LSS_2015,Ishikawa_DCS_2008,Evans_CSS_2011,Alarcon_SAO_2013}. 
In quasi-2D systems, the positions of spherical squirmers are restricted to a monolayer, while their orientation can still 
cover a large range of angles in 3D, depending on their aspect ratio \cite{Theers_CMS_2018,Kuhr_CDMS_2019,Blaschke_PSC_2016}. 
As a result, the in-plane 
velocity of squirmers is not constant, but has a wide distribution. Nevertheless, these systems show MIPS at large enough 
concentrations of squirmers,  which is qualitatively consistent with quasi-2D simulations where the squirmer orientation 
is restricted to a plane \cite{Kyoya_SMCM_2015,Alarcon_MCP_2017,Yoshinaga_HIDS_2017}. Hydrodynamic interactions are 
found to suppress MIPS, such that the phase separation for active Brownian particles takes place at lower volume 
fractions $\phi$ of particles \cite{Theers_CMS_2018,Matas_Navarro_HPS_2014}. Pullers show the onset of MIPS at lower 
$\phi$ values in comparison to pushers. For spheroidal squirmers in quasi-2D systems \cite{Theers_CMS_2018,Qi_EAT_2022}, 
the collective behavior is qualitatively similar to spherical squirmers, but a larger aspect ratio of the spheroidal body 
favors cluster formation, leading to shape-induced jamming and alignment. The alignment interaction due to the anisotropy 
of spheroidal shape can lead to swarming behavior at moderate $\phi$ values. For the 3D systems with periodic BCs 
\cite{Ishikawa_DSM_2007, Delmotte_LSS_2015, Ishikawa_DCS_2008, Evans_CSS_2011, Alarcon_SAO_2013}, the MIPS phase also 
develops at large volume fractions of squirmers. 

In our study, we investigate the collective behavior of spheroidal squirmers in thin films that are thick 
enough to allow a full freedom of squirmer orientation. However, the squirmers are still restricted in their 
vertical position, so that they can 
form only a two-layered structure. Thus, we address some aspects of biofilm formation beyond the monolayer 
structure, when the development of further bacterial layers takes place. In particular, we address the following questions: 
\begin{itemize}
    \item Do swimmers within a thin film primarily assume orientations parallel or perpendicular to the wall?
    \item Under which conditions can swimmers spontaneously leave the wall to form multi-layered structures?
    \item Does the collective behavior of squirmers in thin films differ qualitatively from that of quasi-2D systems?       
\end{itemize}
In our simulations, the volume fraction of squirmers, their swimming mode, and the strength $\lambda$ of rotlet dipole, which mimics 
a counter-rotating flow field of swimmers whose body and flagella bundle rotate in opposite directions, are varied.
In agreement with the quasi-2D systems, pullers display MIPS phase at lower $\phi$ values in comparison with pushers 
and neutral swimmers. At moderate volume fractions, all squirmers with rotlet dipole and pushers with $\lambda = 0$ 
show swarming behavior. Pullers prefer a nearly perpendicular orientation to the wall, which leads to the formation 
of flower-like structures at low $\phi$ and provides the nucleation for larger clusters as $\phi$ is increased. Pullers 
rarely switch between the two walls, while pushers do so frequently, indicating that a collection of pushers would
quickly develop a multi-layered structure. In all investigated cases, a two-layered structure is dominant with the 
orientation of squirmers parallel to the walls. Interestingly, the presence of rotlet dipole significantly reduces 
differences in the collective behavior of squirmers with different swimming modes. These results provide the first steps 
in bridging very confined quasi-2D systems with much less confined situations of the collective behavior of swimmers 
in 3D. 

The paper is organized as follows. Section \ref{sec:method} contains all necessary details about the employed methods
and models, including parameters used in simulations. In Section \ref{sec:struct}, structural properties of squirmer 
suspensions are analysed, including cluster size distribution, position and orientation of squirmers within the slit, 
and the radial distribution function. Dynamic properties are characterized in Section \ref{sec:dynamic}, where effective 
rotational diffusion, mean-squared displacement, and the average speed of squirmers are presented. The main results
are discussed in Section \ref{sec:discuss}, with short conclusions. 

\begin{figure}[htb!]
\centering
  \includegraphics[width=0.95\linewidth]{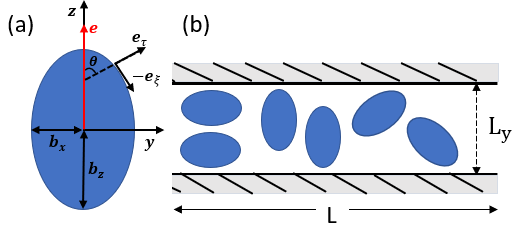}
  \caption{Schematic of the simulation setup. (a) Sketch of a spheroidal squirmer model. The orientation vector $\mathbf{e}$ is aligned with
    the squirmer's major axis $z$, and $b_x$ and $b_z=2 b_x$ denote minor and major radii of the spheroidal shape. $\mathbf{e}_{\tau}$
    and $\mathbf{e}_{\zeta}$ represent the local normal and tangential unit vectors, respectively. (b) Several squirmers within a slit of
    thickness $L_y$. Note that $L_y > 2 b_z$, so that the squirmers can freely rotate within the slit and form a two-layered structure. $L$ is
    the size of the simulation domain in the periodic dimensions $x$ and $z$.
    }
  \label{fgr: system_setup}
\end{figure}

\section{Methods and models}
\label{sec:method}

\subsection{Squirmer model} 

We consider a spheroidal squirmer model, whose surface is described by
\begin{equation} \label{eqn: head_eqn}
\left(\frac{x}{b_x}\right)^{2} + \left(\frac{y}{b_y}\right)^{2} + \left(\frac{z}{b_z}\right)^{2} = 1, 
\end{equation}
where $b_z$ and $b_x = b_y$ are the major and minor radii of the spheroidal squirmer [see Fig.~\ref{fgr: system_setup}(a)]. 
The orientational vector $\textbf{e}$ of the squirmer is aligned  with its major axis. The aspect ratio of the spheroidal shape 
is set to $b_z/b_x = 2$, which is similar to the aspect ratio of
$2 - 3$ for the body of \textit{E. coli} bacteria \cite{Darnton_ECI_2007}. The squirmer surface is discretized by $N_p$ particles
connected by springs into a triangulated network. To maintain the spheroidal shape, the squirmer consists of a
membrane-like surface with shear and curvature elasticity, and has constraints for its surface area and enclosed volume 
\cite{Fedosov_RBC_2010, Fedosov_MBF_2014}. Details of the membrane model are described in Appendix A.

Locomotion of a spheroidal squirmer is imposed through the prescribed surface slip velocity given by
\cite{Theers_CMS_2018,Zoettl_SSM_2018,Ishikawa_HIM_2006,Pagonabarraga_SRSS_2013}
\begin{equation} \label{eqn: usq}
  \textbf{u}_{sq} = -B_1(\textbf{e}_{\zeta} \cdot \textbf{e}_{z})(1 + \beta \zeta)\textbf{e}_{\zeta} +
  \frac{3\lambda z_s \bar{r}_s}{r_s^5}\textbf{e}_{\varphi},
\end{equation}
where $\textbf{e}_{z}$, $\textbf{e}_{\zeta}$, and $\textbf{e}_{\varphi}$ are unit vectors in Cartesian $(x, y, z)$ or spheroidal $(\zeta, \tau, \varphi)$
coordinates, which are related to each other as
\begin{align} 
x &= c \sqrt{\tau^2 - 1} \sqrt{1 - \zeta^2} \cos\varphi, \\
y &= c \sqrt{\tau^2 - 1}  \sqrt{1 - \zeta^2} \sin\varphi, \\
z &= c \tau \zeta. 
\end{align}
Here, $c= \sqrt{b_z^2 - b_x^2}$, and the spheroidal coordinates have the ranges $-1 \le \zeta \le 1$, $1 \le \tau < \infty$ and $0 \le \varphi < 2\pi$. 
The parameter $B_1$ determines self-propulsion speed of the squirmer $U_0 = B_1\tau_0(\tau_0 - (\tau_0^2 -1) \coth^{-1}\tau_0)$ with $\tau_0=b_z/c$
\cite{Theers_MSM_2016,Qi_RSS_2020}. The coefficient $\beta$ defines different swimming modes of the squirmer, including a pusher
($\beta < 0$), a neutral swimmer ($\beta = 0$), and a puller ($\beta > 0$). The second term in Eq.~(\ref{eqn: usq}) defines a rotlet dipole with
the strength $\lambda$, where $\textbf{r}_s = (x_s, y_s, z_s)$, $r_s = |\textbf{r}_s|$, $\bar{r}_s =  \sqrt{x^2_s + y^2_s}$
with the subscript $s$ denoting points at the squirmer surface. The rotlet dipole mimics a counter-rotating flow field, e.g. 
of \textit{E.~coli} whose body and flagella bundle rotate in opposite directions \cite{Hu_MMH_2015}.

\subsection{Boundary conditions}

The suspension of squirmers is confined between two walls in the $y$ direction, while periodic BCs are applied along 
the $x$ and $z$ directions. Fluid flow is modeled by the dissipative particle dynamics (DPD) method \cite{Hoogerbrugge_SMH_1992,Espanol_SMO_1995}, which is a particle-based
hydrodynamics simulation technique (see Appendix B for details).
Solid walls are represented by frozen DPD particles within a layer of thickness $r_c$ (the cutoff radius for DPD interactions)
with the same number density as that for the DPD fluid.
To prevent wall penetration by fluid particles, a reflective surface is placed at the fluid-solid interface, where bounce-back reflection
of fluid particles is enforced. No-slip BCs at the walls are imposed by the dissipative interaction between fluid and frozen-wall particles,
and through the bounce-back reflection at the interface.

To restrict the motion of squirmers between the two walls, particles within the squirmer discretization are subject to the repulsive
Lennard-Jones (LJ) potential
\begin{equation}
    U_{LJ}(r) = 4\epsilon_{LJ}\left[ \left(\frac{\sigma_{LJ}}{r}\right)^{12} - \left(\frac{\sigma_{LJ}}{r}\right)^6 \right]
\end{equation}
at the fluid-solid interface. Here, $\epsilon_{LJ}$ is the potential strength, $\sigma_{LJ}$ sets a characteristic repulsion length, and
$r$ is the distance to the wall. Note that only the repulsive part of the LJ potential is considered by setting the cutoff distance
to $2^{1/6}\sigma_{LJ}$. Excluded-volume interactions between different squirmers are also implemented through the LJ interactions,
where $r$ becomes a distance between two surface particles belonging to distinct squirmers. 

Squirmers are submerged within a DPD fluid, and also filled by DPD fluid particles due to their membrane-like representation. The membrane
surfaces of all suspended squirmers serve as a boundary separating DPD particles inside and outside of the membranes. This is achieved through
the reflection of fluid particles at the membrane surfaces from inside and outside. Note that the dissipative and random forces between
the internal and external fluid particles are deactivated, and only the conservative force is employed to maintain uniform fluid pressure
across the membranes. To enforce the slip velocity $\textbf{u}_{sq}$ at the squirmer surface, the dissipative interaction (see Appendix B) between the squirmer
particles and those of the surrounding fluid is altered as follows
\begin{equation}
  \textbf{F}^D(r_{ij}) = -\gamma W^D(r_{ij})(\hat{\textbf{r}}_{ij} \cdot \textbf{v}_{ij}^\ast) \hat{\textbf{r}}_{ij}, \quad  \quad 
  \textbf{v}_{ij}^\ast = \textbf{v}_i - \textbf{v}_j + \textbf{u}_{sq}^i,
\end{equation}
where $\hat{\textbf{r}}_{ij} = (\textbf{r}_i - \textbf{r}_j)/r_{ij}$, $r_{ij} = |\textbf{r}_i - \textbf{r}_j|$,  $\textbf{u}_{sq}^i$ 
is the slip velocity at the position of squirmer particle $i$, while $j$ corresponds to an outer-fluid DPD particle.
Furthermore, the friction coefficient $\gamma$ between fluid and squirmer particles is properly adjusted \cite{Fedosov_RBC_2010} to ensure
the imposition of $\textbf{u}_{sq}$ at the squirmer surface. 

\subsection{Simulation setup and parameters}

The simulation setup corresponds to a domain of dimensions $L \times L_y \times L$, where $L = 28 b_z$ and $L_y  = 2.5 b_z$
($b_z = 4$ in simulations), see Fig.~\ref{fgr: system_setup}(b). The number density of fluid particles is 
$n_f = 320/b_z^3$, with a particle mass $m=1$. The energy unit is $k_BT = 1$. Parameters selected for the DPD fluid 
(see Appendix B) yield a fluid dynamic viscosity of $\eta = 403.2\sqrt{mk_BT}/b_z^2$. 

Each squirmer in simulations consists of $N_p=1024$ particles. 
Excluded-volume interactions between different squirmers and between squirmers and the walls
are implemented through the LJ potential with parameters $\epsilon_{LJ} = k_BT$ and $\sigma_{LJ} = 0.125 b_z$. 
The computed three-dimensional rotational diffusion coefficient around the major axis of a spheroid 
without confinement is  
$D_R = 3.28 \times 10^{-4} \sqrt{k_BT/m} / b_z$, which is close to the theoretical prediction of $D_R = 3.52 \times 10^{-4}
\sqrt{k_BT/m} / b_z$ \cite{Theers_MSM_2016}. To present simulation results, the half length $b_z$ of the semi-major axis is 
chosen as a length scale, and the characteristic rotational time $\tau_R = 1/D_R$ as a time scale. In all simulations, 
the activity parameter $B_1$ is fixed at $56.8 b_zD_R$. We have verified that the swimming velocity of a single squirmer is $U_0= 45.5 b_zD_R$, in agreement with the theoretical prediction
\cite{Theers_MSM_2016}. Using these values, we can define a dimensionless P\'{e}clet number $Pe = U_0/(2b_z D_R) \approx 28.4$. Furthermore,
the Reynolds number $Re = 2b_z U_0 m n_f / \eta \approx 0.03$ is small enough to eliminate possible inertial effects.

Different volume fractions $\phi = 4 N_{sq} \pi b_x b_y b_z/ (3L^2 L_y) \in \{0.18, 0.35, 0.44, 0.56\}$ are considered,
corresponding to different numbers of squirmers $N_{sq} \in \{228, 456, 576, 720\}$. The three values of active stress $\beta \in \{-5, 0, 5\}$
represent simulation systems with pusher, neutral, and puller swimmers, respectively. Finally, we also consider two values of the non-dimensional rotlet
dipole strength $\tilde{\lambda} = \lambda/(b_z^4D_R) \in \{0, 133.5\}$. For $\tilde{\lambda}=133.5$, a single squirmer 
moves in a circular trajectory at a wall with a radius of approximately $2b_z$.
For comparison, the rotlet dipole strength 
of $\tilde{\lambda} \in \{0, 561.1\}$ has been used in Ref.~\cite{Qi_EAT_2022}, where squirmers were confined to a single layer.
All simulations are first run for a time of at least $1.5 \tau_R$ to reach a steady state, and afterwards various structural 
and dynamical characteristics of the system are measured during the time $1.5 \tau_R$.

\subsection{Mapping to experimental systems}

Squirmer parameters used in simulations can also be mapped to the properties of real swimmers. In the far-field approximation, the hydrodynamic field 
of a microswimmer is dominated by its force-dipole strength $\chi = f_d l_d/ (8\pi\eta)$ \cite{Solon_PPE_2015,Drescher_FDN_2011,Lauga_HSO_2009},
where $f_d$ and $l_d$ are the characteristic force and length of the dipole. The active stress $\beta$ of a spheroidal squirmer can be expressed as
\cite{Theers_MSM_2016}
\begin{equation} \label{eqn: beta}
   \beta = -\frac{\chi}{U_0(b_z^2 - b_x^2)}\frac{[3\tau_0 + (1-3\tau_0^2)\coth^{-1}\tau_0][\tau_0 - (\tau_0^2 -1)\coth^{-1}\tau_0]}
   {2/3 - \tau_0^2 + (\tau_0^2 -1)\coth^{-1}\tau_0}.
\end{equation}
Average properties of \textit{E. coli} bacteria swimming in water correspond to $U_0 = 29$ $\mu$m/s, $f_d = 0.42$ pN, $l_d = 1.9$ $\mu$m, 
$b_z = 1.5$ $\mu$m, $b_x = 0.5$ $\mu$m, and $\eta = 10^{-3}$ Pa$\cdot$s \cite{Drescher_FDN_2011,Hu_MMH_2015,Darnton_ECI_2007}, 
resulting in $\beta \approx -3$. Note that a direct comparison is possible only in the far-field limit, while the near-field 
flow of each swimmer depends on its geometric and propulsion details.

\section{Results}

\begin{figure*}[htb!]
  \centering
  \includegraphics[width=0.9\textwidth]{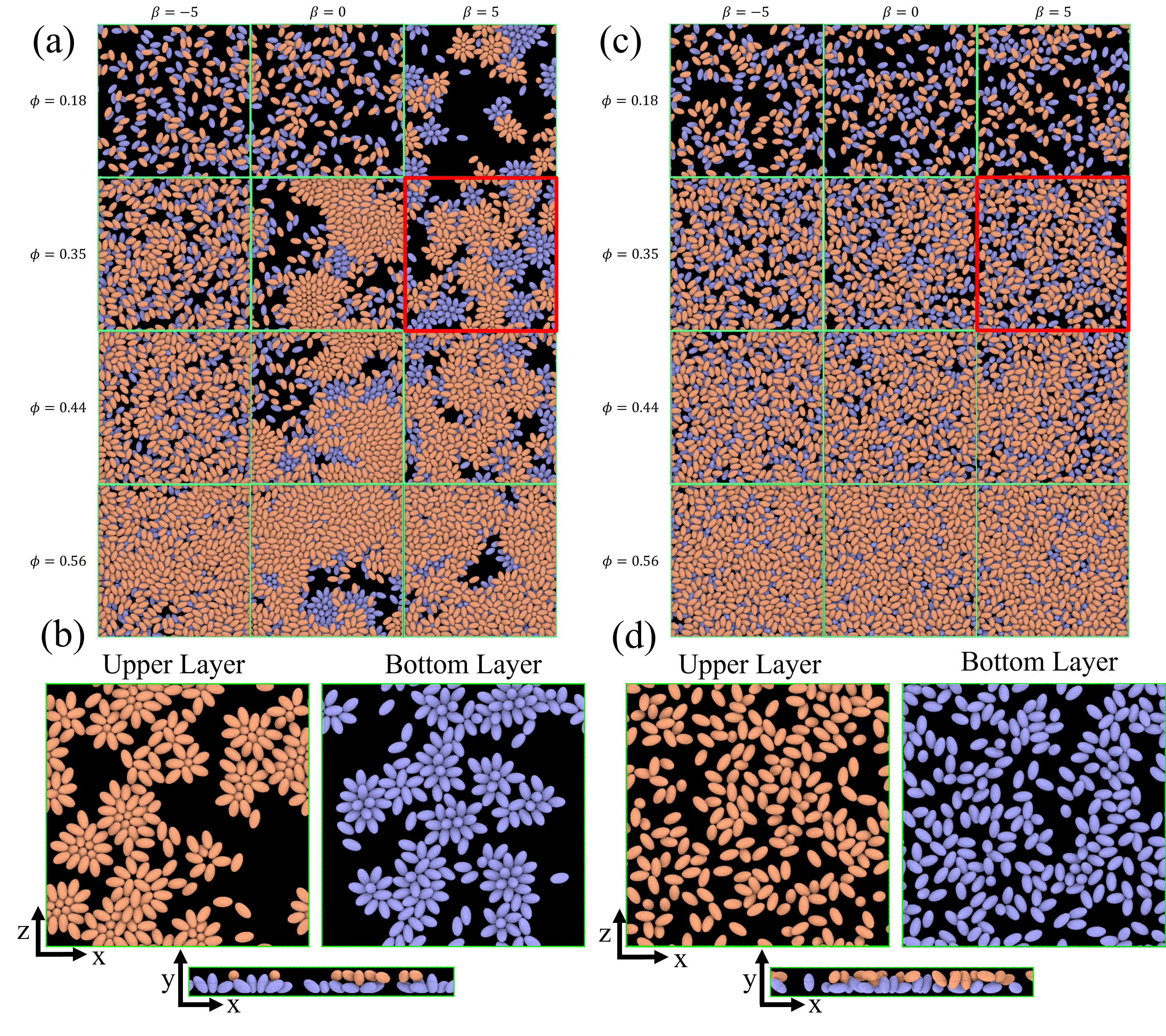}
  \caption{Snapshots of the emergent structures for different active stresses $\beta$, volume fractions $\phi$, and rotlet dipole strengths $\tilde{\lambda}$. (a) Simulated structures for $\tilde{\lambda} = 0$, $\phi = 0.18$ and $0.35$, and 
  $\beta \in \{0, \pm 5\}$. Squirmers in the upper (bottom) half of the slit are colored in orange (blue). See also corresponding 
  Movies S1 - S6. 
  (b) Squirmer structures within the upper and bottom layers of the simulated system for $\phi = 0.35$, $\beta = 5$, and $\tilde{\lambda} = 0$. See also Movie S6. 
  (c) Snapshots of the simulated systems for $\tilde{\lambda} = 133.5$, $\phi = 0.35$ and $\beta \in \{0, \pm 5\}$. 
  See also Movies S7 - S9. 
  (d) Squirmer structures within the upper and bottom layers of the slit for $\phi = 0.35$, $\beta = 5$, 
  and $\tilde{\lambda} = 133.5$. Squirmers belong to the upper (orange) or bottom (blue) layers, when 
  the $y$-coordinate of their center of mass is in the upper or lower half of the slit, respectively. See also Movie S9.}
  \label{fgr: snapshots}
\end{figure*}

In our investigation, we analyze the collective structural and dynamical properties of swimmer ensembles in thin fluid films 
as a function of volume fraction $\phi$ and squirmer characteristics of active stress $\beta$ and rotlet dipole strength $\tilde{\lambda}$. Snapshots of the emergent structures are displayed in Fig.~\ref{fgr: snapshots}. 
We characterize these phases through the analysis of cluster sizes and radial distribution functions of the squirmers. Furthermore,
the distribution of squirmers along the $y$-direction (perpendicular to the walls) and their orientation are considered to 
distinguish a two-layered arrangement with an average orientation within the $x-z$ plane from the stacked packing 
with the orientation perpendicular to the walls. Finally, dynamical properties, such as velocity distribution, and effective 
rotational and translational diffusivities, are computed to characterize squirmer motility within collective states.   

An important question for any two-layered structure is to which extent the 
two layers interact with and affect each other. The snapshots in Fig.~\ref{fgr: snapshots} nicely 
demonstrate that the correlation between the two layers strongly depends on volume fraction and
strength of the rotlet dipole. Without rotlet dipole, i.e. for $\tilde{\lambda}=0$, the layer correlation is 
very weak for small $\phi$, but becomes very significant for large $\phi$, where the clusters in the top and
bottom layers are essentially in registry. With strong rotlet dipole, i.e. for $\tilde{\lambda}=133.5$, 
the situation is very different, as there is hardly any visible correlation between the two layers.

Previous studies of structure formation of artificial and biological microswimmers in two- and 
three-dimensional systems \cite{Elgeti_PMS_2015, Qi_EAT_2022, Bechinger_APC_2016, Bialke_CDS_2012, Levis_CHDMC_2014} report the existence of various phases, including
\begin{itemize}
\item a gas of small clusters, where the distribution of swimmers is homogeneous and dynamic clusters are formed by a few swimmers;   
\item large clusters or motility-induced phase separation (MIPS), where the cluster size is often comparable with the size
  of the entire system; such large clusters are nearly immobile; 
\item swarming and flocking, which is characterized by the collective locomotion of dense swimmer clusters with swirling and
  streaming patterns.
\end{itemize}
A similar behavior is found in our system (see Fig.~\ref{fgr: snapshots}); however, the possibility of the formation of two
distinct layers at the walls gives rise to novel structures and dynamics.

\begin{figure*}[htb!]
  \centering
  \includegraphics[width=0.8\textwidth]{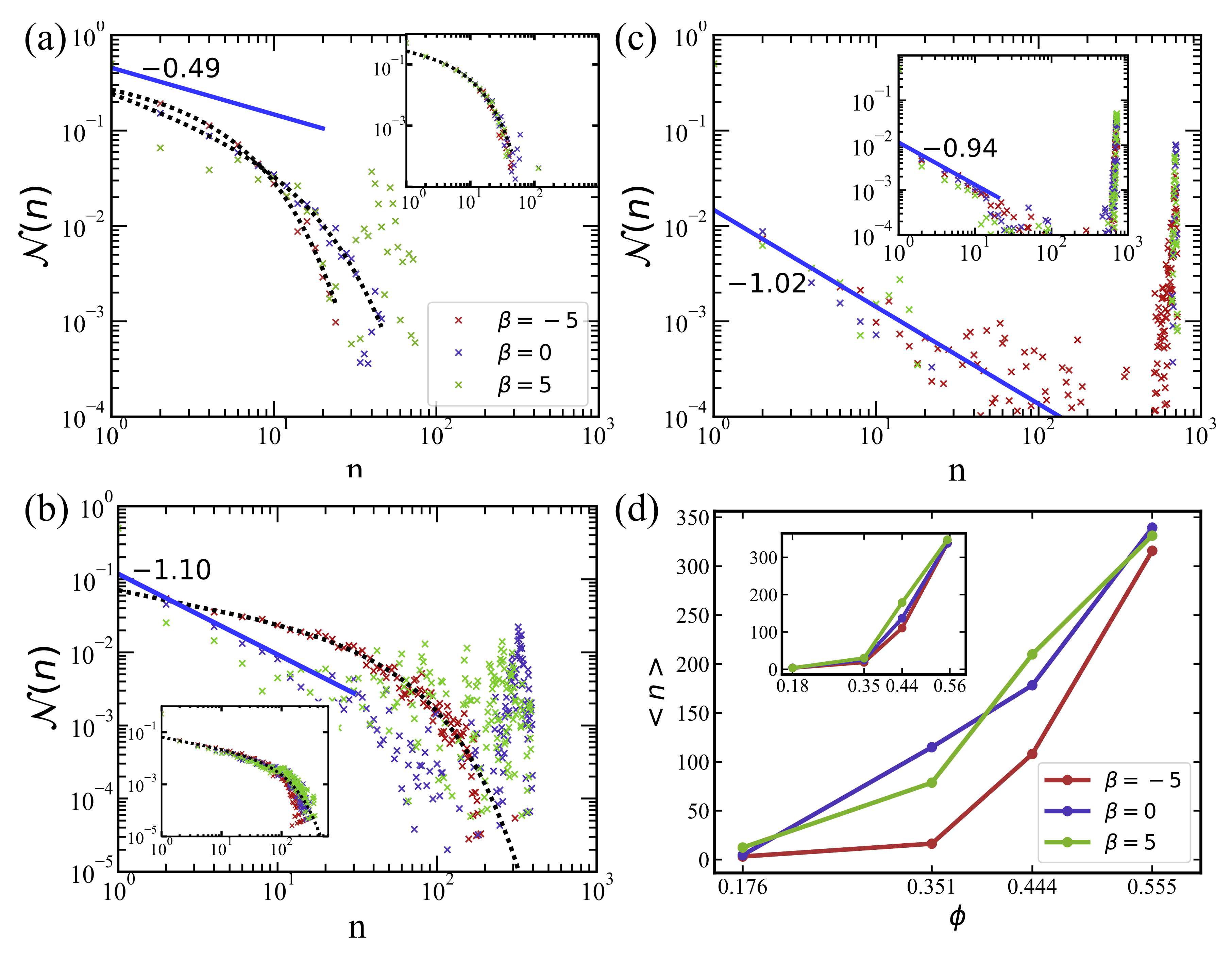}
  \caption{Cluster size distributions for different volume fractions of squirmers. (a) $\phi = 0.18$, 
    (b) $\phi = 0.35$, and 
    (c) $\phi = 0.56$. The curves with different colors represent various swimming modes: pusher with 
    $\beta = -5$ (red), neutral with $\beta = 0$ (blue), and puller with $\beta = 5$ (green). 
    The blue solid lines are the power-law fits, while the black-dotted lines
    are the fits using Eq.~(\ref{eqn: cl_fit}) with the parameters shown in Table~\ref{tab:table_cl}. 
    (d) The average cluster size for various conditions as a function of volume fraction. The 
    average cluster size is calculated using Eq.~(\ref{eqn: average_cl}).   
    All plots are for the simulations without rotlet dipole ($\tilde{\lambda} = 0$), while the 
    insets show data from the corresponding simulations with $\tilde{\lambda} = 133.5$.}
  \label{fgr: cl_plots}
\end{figure*}

\subsection{Structural properties}
\label{sec:struct}

\subsubsection{Cluster size distribution}
\label{sec: cl_size}

The cluster-size distribution function $\mathcal{N}(n)$ is calculated as
\begin{equation} \label{eqn: cl}
\mathcal{N}(n) = \frac{1}{N_{sq}}np(n),
\end{equation}
where $\mathcal{N}(n)$ represents the fraction of squirmers belonging to clusters of size $n$, and $p(n)$ denotes the number of
clusters of size $n$. The distribution is normalized, such that $\sum_{n} \mathcal{N}(n) = 1$. Different squirmers belong to the
same cluster when the nearest surface-to-surface distance $d_s$ between them satisfies $d_s/b_z < 0.25$. 
The average cluster size $\langle n \rangle$ is then \cite{Levis_CHDMC_2014}: 
\begin{equation} \label{eqn: average_cl}
<n> = \sum_{n} n \mathcal{N}(n).
\end{equation}
The cluster-size distribution is used to distinguish the different collective phases \cite{Levis_CHDMC_2014,Beer_PDBS_2020} 
mentioned above.  
For the gas of small clusters, $\mathcal{N}(n)$ exhibits an exponential decay, as shown in Fig.~\ref{fgr: cl_plots}(a)
for $\phi = 0.18$. For MIPS or large clusters, a bimodal distribution of $\mathcal{N}(n)$ is observed, with the second peak 
at large $n$ signaling the presence of large clusters, see Fig.~\ref{fgr: cl_plots}(b,c) for $\phi = 0.35$ and $\phi = 0.56$. 
Finally, swarming can be characterized by a power-law decay of
$\mathcal{N}(n)$ with an exponential cutoff and without the presence of a distinct peak \cite{Alarcon_MCP_2017,Qi_EAT_2022}. 
This implies a slow decline of cluster sizes over a broad range, compared to the fast decay for a gas of small clusters. 

\begin{table}[htb!]
	\centering
	\begin{tabular}{|c|c|c|c|c|c|c|}
		\hline
		$\phi$ &  $\beta$ & $\tilde{\lambda}$ & $A$ & $\alpha$ & $\gamma$ & phase \\
		\hline
		$0.18$      & $-5$ & $0$         & $0.33$ & $5.07$  & $0.19$  & gas \\
		$0.18$      & $0$  & $0$         & $0.26$ & $13.1$  & $0.57$  & gas \\
		$0.18$      & $5$  & $0$         & $0.09$ &  $-$      & $0.49$  & MIPS \\
		$0.18$      & $-5$ & $133.5$  & $0.32$ & $7.1$   & $0.4$    & gas \\
		$0.18$      & $0$  & $133.5$  & $0.31$ & $7.61$ & $0.43$  & gas \\
		$0.18$      & $5$  & $133.5$  & $0.34$ & $8.55$ & $0.54$  & gas \\
		\hline
		$0.35$      & $-5$ & $0$         & $0.07$ & $49.4$  & $0.4$    & swarming \\
		$0.35$      & $0$  & $0$         & $0.1$   &  $-$     & $1.1$    & MIPS \\
		$0.35$      & $5$  & $0$         & $0.05$ & $-$      & $0.99$  & MIPS \\
		$0.35$      & $-5$ & $133.5$  & $0.07$ & $45$    & $0.45$    & swarming \\
		$0.35$      & $0$  & $133.5$  & $0.07$ & $72.5$ & $0.58$   & swarming \\
		$0.35$      & $5$  & $133.5$  & $0.07$ & $90$    & $0.62$   & swarming \\
		\hline
        $0.44$      & $-5$ & $0$         & $0.04$ & $-$   & $0.89$    & MIPS \\
		$0.44$      & $0$  & $0$         & $0.02$  &  $-$  & $0.93$    & MIPS \\
		$0.44$      & $5$  & $0$         & $0.03$ & $-$    & $1.28$   & MIPS \\
		$0.44$      & $-5$ & $133.5$  & $0.03$ & $-$    & $0.62$    & MIPS \\
		$0.44$      & $0$  & $133.5$  & $0.03$ & $-$    & $0.69$   & MIPS \\
		$0.44$      & $5$  & $133.5$  & $0.03$ & $-$    & $0.83$   & MIPS \\
		\hline
		$0.56$      & $-5$ & $0$         & $0.01$ & $-$   & $1.02$    & MIPS \\
		$0.56$      & $0$  & $0$         & $0.03$  &  $-$  & $1.64$    & MIPS \\
		$0.56$      & $5$  & $0$         & $0.01$ & $-$    & $0.81$   & MIPS \\
		$0.56$      & $-5$ & $133.5$  & $0.01$ & $-$    & $0.9$    & MIPS \\
		$0.56$      & $0$  & $133.5$  & $0.01$ & $-$    & $0.94$   & MIPS \\
		$0.56$      & $5$  & $133.5$  & $0.01$ & $-$    & $1.04$   & MIPS \\
		\hline
	\end{tabular}
	\caption{Parameters of the cluster size distributions, obtained from fitting the simulation result 
        with Eq.~(\ref{eqn: cl_fit}), for various packing fractions $\phi$,
          active stresses $\beta$, and dimensionless rotlet dipole strengths $\tilde{\lambda} = \lambda/(b_z^4D_R)$.  A "$-$" sign in the $\alpha$-column indicates that
          the exponential term in Eq.~(\ref{eqn: cl_fit}) is omitted, so that the fitting function becomes $f(x) = Ax^{-\gamma}$.
          The last column provides the classification of different simulation cases into the defined phases.}
	\label{tab:table_cl}
\end{table}


The described characteristics of different phases can be extracted by fitting $\mathcal{N}(n)$ with the function \cite{Alarcon_MCP_2017}
\begin{equation} \label{eqn: cl_fit}
f(x) = Ax^{-\gamma}\exp^{-x/ \alpha},
\end{equation}
where $A$, $\gamma$, and $\alpha$ are the fitting parameters. Table~\ref{tab:table_cl} presents these parameters for different
simulated conditions. 

For $\phi = 0.18$, the simulated systems exhibit a homogeneous gas-like phase, except for the case of pullers
($\beta = 5$) without rotlet dipole ($\tilde{\lambda} = 0$), shown in Fig.~\ref{fgr: snapshots}(a). Here, the exponential 
term dominates, as can be seen well in Fig.~\ref{fgr: cl_plots}(a). At low packing fractions of squirmers, there is a 
limited chance for squirmer collisions, leading to a gas-like phase, compare also Fig.~\ref{fgr: snapshots}. 
For $\beta = 5$ at $\phi = 0.18$, $\mathcal{N}(n)$ resembles a bimodal distribution,  
indicating the existence of large clusters ($n \approx 60$), whose formation is primarily governed 
by the attractive hydrodynamic field around pullers.

Swarming-like behavior is observed in several cases at $\phi = 0.35$, characterized by the relatively large values of $\alpha$ and
moderate values of $\gamma \in [0.4, 0.7]$, indicating the dominance of the power-law term in Eq.~(\ref{eqn: cl_fit}). 
Note that for the cases of neutral swimmers and pullers (without rotlet dipole), the value of $\gamma$ is close to unity,
and $\mathcal{N}(n)$ exhibits a bimodal distribution in Fig.~\ref{fgr: cl_plots}(b), suggesting the presence of a MIPS phase. 
For even larger volume fraction $\phi = 0.56$, Fig.~\ref{fgr: cl_plots}(c) shows a prominent peak in $\mathcal{N}(n)$ at large $n$ 
for all cases, which is the main characteristic for the MIPS phase. This is also confirmed in Table~\ref{tab:table_cl} 
through $\gamma$ values close to unity. At large packing fractions, steric interactions between squirmers dominate due to
crowding, so that the differences in $\mathcal{N}(n)$ for different swimming modes nearly disappear. 

For squirmers without rotlet dipole ($\tilde{\lambda} = 0$), the swimming mode affects their collective behavior when the 
volume fraction is small enough, i.e. $\phi \lesssim 0.4$. In this case, the local hydrodynamic flow field generated by the 
squirmers is relevant for cluster formation and dynamics, in agreement with previous studies \cite{Theers_CMS_2018,Qi_EAT_2022}. 
However, when the rotlet dipole is activated ($\tilde{\lambda} = 133.5$), differences in $\mathcal{N}(n)$ for various active
stresses $\beta$ nearly disappear, as displayed in the insets of Fig.~\ref{fgr: cl_plots}. With rotlet dipole, we obtain the 
gas phase at $\phi = 0.18$, swarming at $\phi = 0.35$, and MIPS at $\phi = 0.56$, independently of active stress -- see also 
Fig.~\ref{fgr: snapshots}(c).
Note that the systems with $\phi = 0.56$ are very crowded, and MIPS is more difficult to recognize for 
$\tilde{\lambda} = 133.5$ in Fig.~\ref{fgr: snapshots}(c) than for $\tilde{\lambda} = 0$ 
in Fig.~\ref{fgr: snapshots}(a).  Additional simulations (not shown) indicate that the suspension of
squirmers with $\tilde{\lambda} = 133.5$ transits from swarming to the MIPS at $\phi \approx 0.4$.

The average cluster size $\langle n \rangle$ as a function of $\phi$ is shown in 
Fig.~\ref{fgr: cl_plots}(d). For $\phi \gtrsim 0.18$, $\langle n \rangle$ increases rapidly with 
increasing volume fraction of squirmers for all $\beta$ and $\lambda$ values. The simulation systems 
with different active stresses result in a similar range of average cluster sizes for both rotlet 
dipole strengths.

\begin{figure}[htb!]
  \centering
  \includegraphics[width=0.9\linewidth]{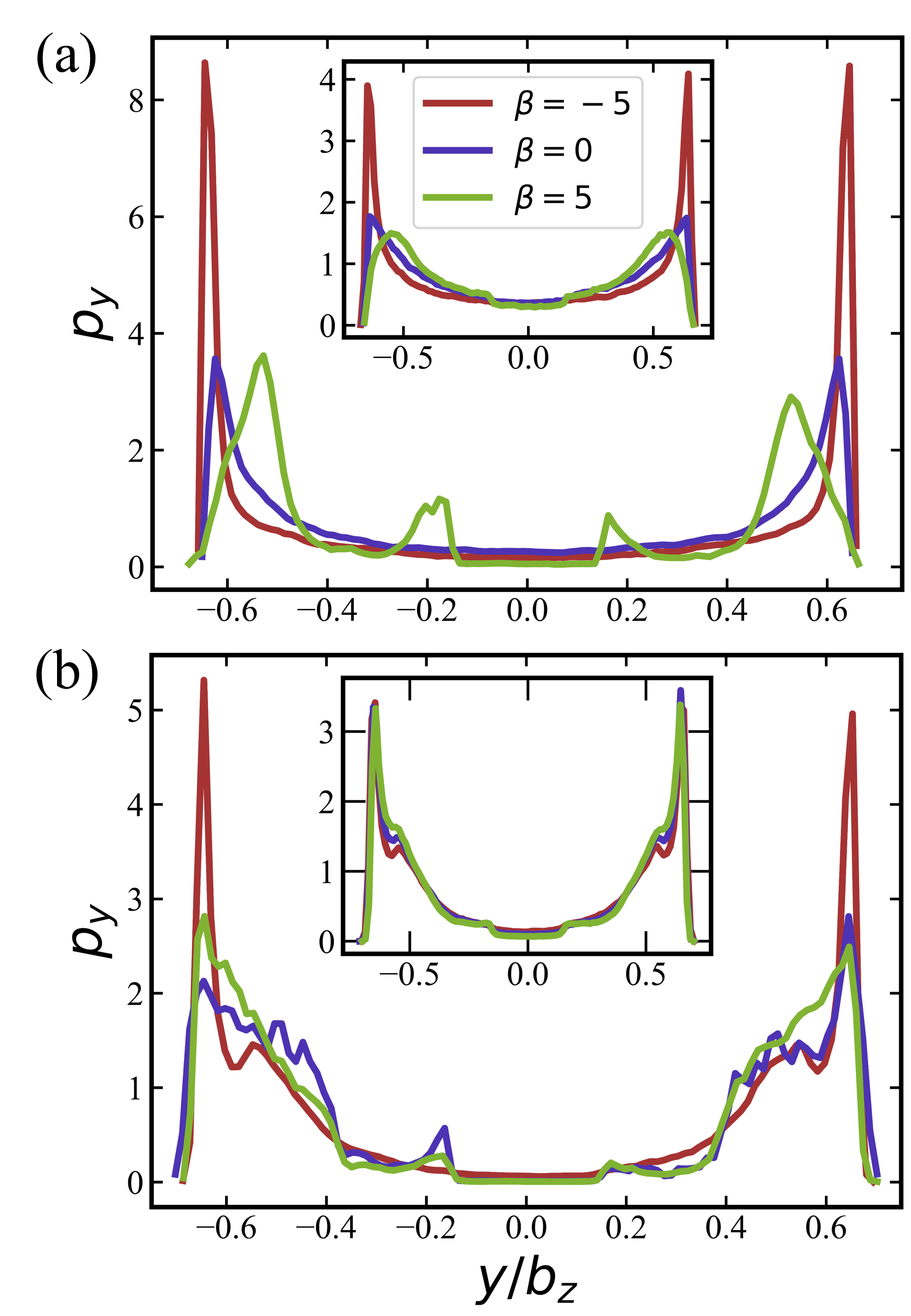}
  \caption{Distributions $p_y$ of the squirmers' centre-of-mass position along the   slit height for volume fractions (a) $\phi = 0.18$ and (b)
    $\phi=0.56$. The walls are located at $y/b_z = \pm 1.25$. Colors indicate the swimming 
    modes: pusher with $\beta = -5$ (red), neutral with $\beta = 0$ (blue),
    and puller with $\beta = 5$ (green). Main figures are for the simulations without rotlet dipole $\tilde{\lambda} = 0$, while the insets
    show the corresponding simulations with rotlet dipole $\tilde{\lambda} = 133.5$.
    }
  \label{fgr: y_dist}
\end{figure}

\subsubsection{Squirmer distribution between the walls} 
\label{sec: y_pos}

Figure ~\ref{fgr: y_dist} shows distributions $p_y$ of the squirmers' center-of-mass position along the $y$ direction between the two confining
walls. Notably, the squirmers exhibit a distinct affinity for the walls, especially in the case of pushers, 
as indicated by the pronounced peaks at $y/b_z \approx \pm 0.6$. This phenomenon is primarily attributed 
to the well-known wall-trapping effect, which is on the one hand due to motion, steric interactions, 
and slow reorientation \cite{Li_AMS_2009, Volpe_MSE_2011, Elgeti_WAS_2013}, on the other hand due to hydrodynamic attraction \cite{Berke_HAS_2008, Lauga_HSO_2009, Drescher_FDN_2011}.

\begin{figure}[htb!]
	\centering
	\includegraphics[width=1.0\linewidth]{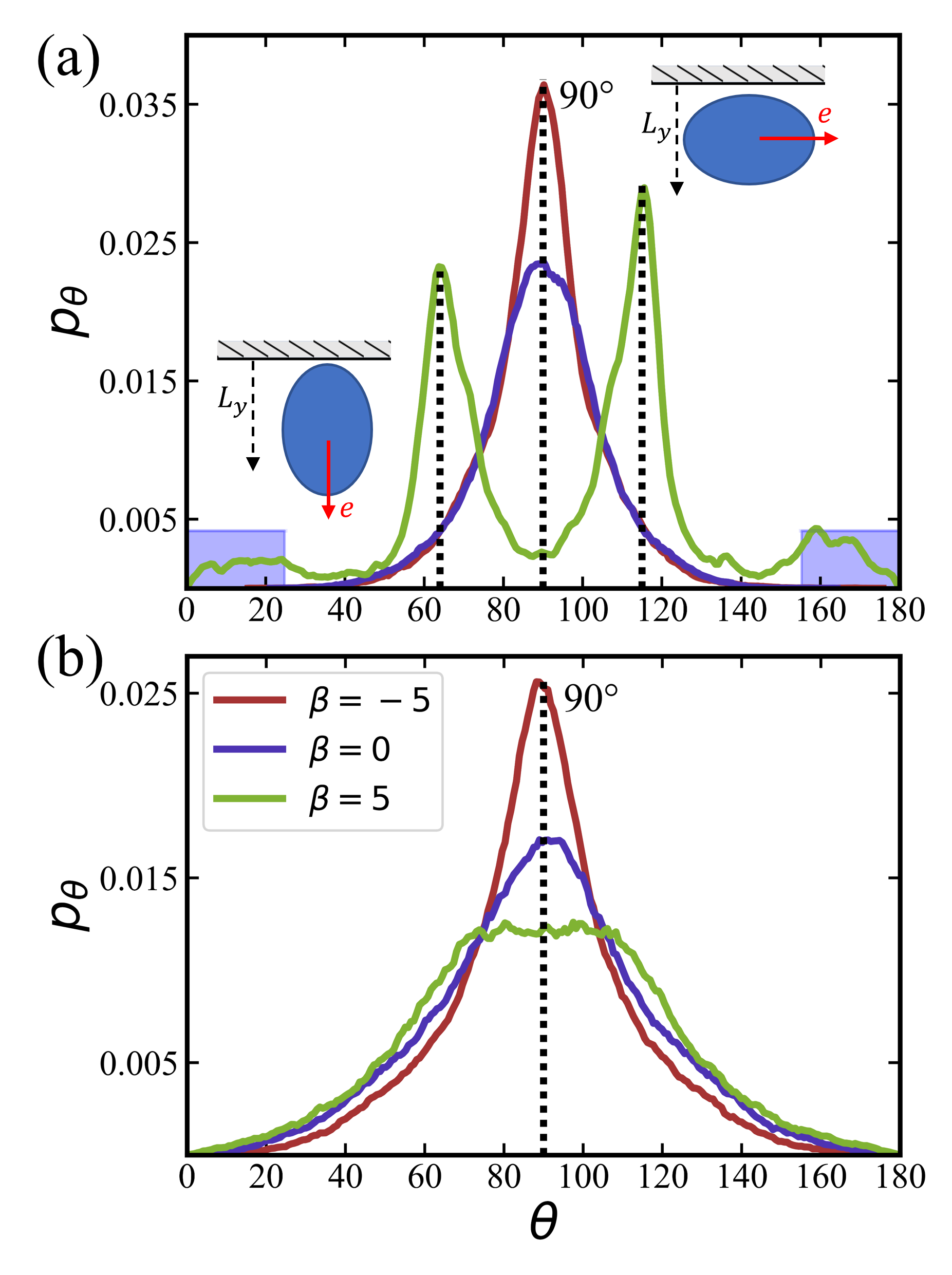}
	\caption{Orientational distribution $p_\theta$ of squirmers, where the angle $\theta$ is between the swimmer orientation
          vector and the $y$ axis (wall normal axis). $\theta = 0$ and $\theta = 180$ correspond to the orientation perpendicular to the walls, while
          $\theta = 90$ represent the orientation parallel to the walls, see the insets. $p_\theta$ is presented for $\phi=0.18$
          with the rotlet dipole strength (a) $\tilde{\lambda} = 0$ and (b) $\tilde{\lambda} = 133.5$. The three different $\beta$
          values correspond to pushers ($\beta = -5$), neutral swimmers ($\beta = 0$), and pullers ($\beta = 5$). The shaded blue
          areas in (a) mark the angle ranges $[0, 25]$ and $[155,180]$ used to compute the fraction of squirmers with a perpendicular 
	  orientation to the walls.
          }
	\label{fgr: odf_with_y}
\end{figure}

Pushers ($\beta=-5$) exhibit a distinctive two-layered structure with the orientation parallel to the 
walls, which is reminiscent of \textit{E. coli} behavior in \textit{in vitro} observations 
\cite{Frymier_3DMBS_1995, Junot_RTV_2022}. For pushers, the far-field flow 
re-orients the squirmers parallel to the wall \cite{Berke_HAS_2008, Lauga_HSO_2009, Drescher_FDN_2011}.
Furthermore, the elongated spheroidal body of the squirmers favors their orientation parallel to the walls
\cite{Vigeant_RIAEC_2002, Spagnolie_HSP_2012}.  For pullers ($\beta = 5$), 
Fig.~\ref{fgr: y_dist}(a)  show two small peaks at $y/b_z \approx \pm 0.2$ at small volume fraction 
$\phi = 0.18$, which indicates a nearly perpendicular orientation with respect to the
walls [see also corresponding conformations in Fig.~\ref{fgr: snapshots}(b)]. Note that the 
far-field flow generated by pullers favors such a perpendicular-to-the-wall orientation
\cite{Berke_HAS_2008, Lauga_HSO_2009, Drescher_FDN_2011}, while the spheroidal shape promotes 
parallel-to-the-wall orientation. Thus,
pullers in Fig.~\ref{fgr: y_dist}(a) show their major peaks in $p_y$ at $y/b_z \approx \pm 0.5$, 
such that their most probable orientation has about $40$ degree angle with respect to the walls 
[see also Fig.~\ref{fgr: odf_with_y}(a)]. $p_y$ for neutral swimmers at $\phi = 0.18$ is close to 
that for pushers.   

At the higher volume fraction of $\phi = 0.56$, differences in $p_y$ for the three swimming modes 
significantly diminish, and all cases essentially show a two-layered structure, see 
Fig.~\ref{fgr: y_dist}(b). Note that the two peaks in $p_y$ at $y/b_z \approx \pm 0.6$ become broader 
than those for $\phi = 0.18$, which can be attributed to an increasing importance of steric interactions 
at large volume fractions, such that close packing introduces a larger deviation to the two-layered 
structure. Insets in Fig.~\ref{fgr: y_dist} demonstrate that the activation
of rotlet dipole mitigates the influence of active stress on $p_y$, in agreement with the results for
the cluster-size distribution in Section \ref{sec: cl_size}.

\subsubsection{Squirmer orientation in the slit}
\label{sec:orient}

The orientational distribution function $p_\theta(\theta)$ of the angle of the squirmer 
orientation vector $\textbf{e}$ with the wall normal ($y$ axis) are presented in Fig.~\ref{fgr: odf_with_y}. 
A perpendicular orientation to the walls corresponds to $\theta = 0$ and $\theta = 180$, a 
parallel-to-the-wall orientation to $\theta = 90$. 
Figure~\ref{fgr: odf_with_y}(a)
shows that pushers and neutral swimmers with $\tilde{\lambda} = 0$ at $\phi=0.18$ are primarily 
oriented parallel to the wall, in agreement with the results in Fig.~\ref{fgr: y_dist}(a). 
In contrast, pullers show several peaks at $\theta \approx 20^o\, \& \,160^o$ (small peaks)
and $\theta \approx 60^o\, \& \,120^o$ (large peaks), which correspond to a nearly 
perpendicular-to-the-wall orientation and a tilted orientation
with about $40^o$ angle to the walls, respectively [compare also snapshots in Fig.~\ref{fgr: snapshots}(b)]. These results are consistent with the position distributions $p_y$ discussed
in Section \ref{sec: y_pos}. 

Figure ~\ref{fgr: odf_with_y}(b) shows $p_\theta$ for a non-zero rotlet dipole ($\tilde{\lambda} = 133.5$) 
at $\phi=0.18$, which again demsontrates a significant reducuction of differences between different 
swimming modes due to the rotlet dipole. In particular, for all $\beta$, a single peak in
$p_\theta$ centered around $90^o$ is observed, confirming the preferred squirmer orientation parallel 
to the walls, as illustrated in Fig.~\ref{fgr: snapshots}(d).
As the volume fraction of squirmers increases, the dominant role of steric interactions also leads to a diminished effect of
different swimming modes (even for $\tilde{\lambda} = 0$) with the formation of a two-layered structure. 

\begin{figure}[htb!]
	\centering
	\includegraphics[width=1.0\linewidth]{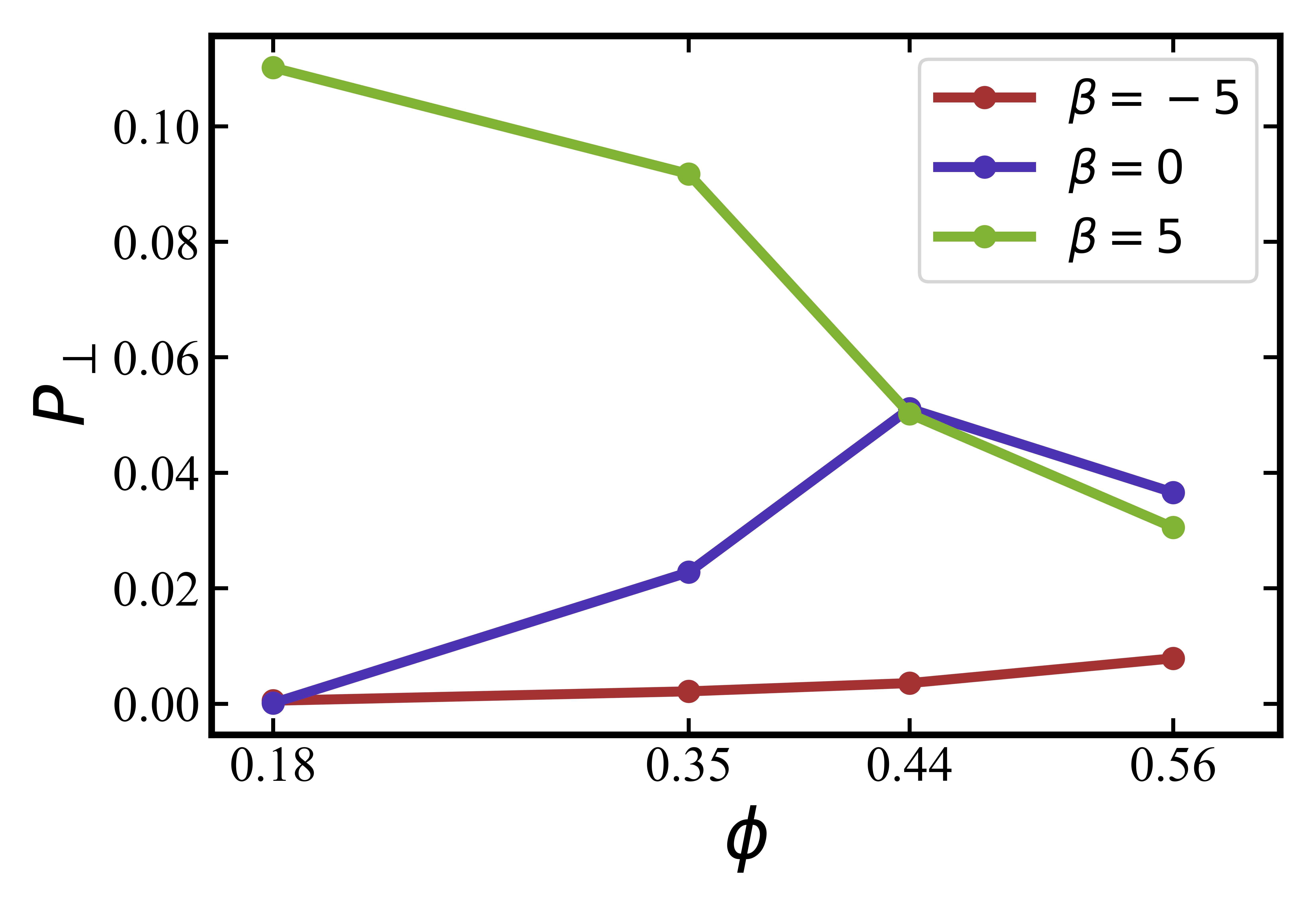}
	\caption{Probability $P_\perp$ of squirmers to be aligned with the $y$ axis (i.e., wall normal vector) for different volume fractions and 
		swimming modes at $\tilde{\lambda} = 0$. Data for pushers with $\beta = -5$ (red), neutral squirmers with $\beta = 0$ (blue), 
		and pullers with $\beta = 5$ (green). The probability is calculated by integrating the orientation distributions in 
		Fig.~\ref{fgr: odf_with_y}(a) over the ranges $[0,\,25]$ and $[155,\,180]$ degrees.
                }
	\label{fgr:perp_frac}
\end{figure}

The probability $P_\perp$ of squirmers to be oriented nearly perpendicular to the walls is displayed in Fig.~\ref{fgr:perp_frac} for different values 
of $\beta$ as a function of $\phi$. The probability is computed by integrating the orientation distributions over the ranges 
$[0,\,25^o]$ and $[155^o,\,180^o]$, as indicated by the blue shaded areas in Fig.~\ref{fgr: odf_with_y}(a).  At low $\phi$, 
the probability of perpendicular orientation to the walls is zero for pushers and neutral squirmers, while $P_\perp \approx 0.11$ 
for pullers. This simply confirms the tendency of  pushers and neutral swimmers to align with the walls, due to their hydrodynamic 
interactions with the walls. Interestingly, the majority of pullers does not have a perpendicular-to-the-wall orientation, despite of such
predictions for a single puller \cite{Berke_HAS_2008, Lauga_HSO_2009}. This is due to collective effects between pullers, 
which will be discussed in Section~\ref{sec:rdf} below. 

As the volume fraction of squirmers  increases, $P_\perp$ for pullers decreases because steric interactions between squirmers become 
dominant, forcing the formation of a two-layered structure, as discussed above. For neural squirmers, $P_\perp$ first increases with increasing $\phi$, but 
then closely follows $P_\perp$ for pullers when $\phi \gtrsim 0.4$. For pushers, there is only a slight increase in $P_\perp$ with 
increasing $\phi$. Nevertheless, we expect that for large volume fractions $\phi \gtrsim 0.6$, orientational differences between 
squirmers with various swimming modes essentially disappear.

\begin{figure}[htb!]
	\centering
	\includegraphics[width=0.9\linewidth]{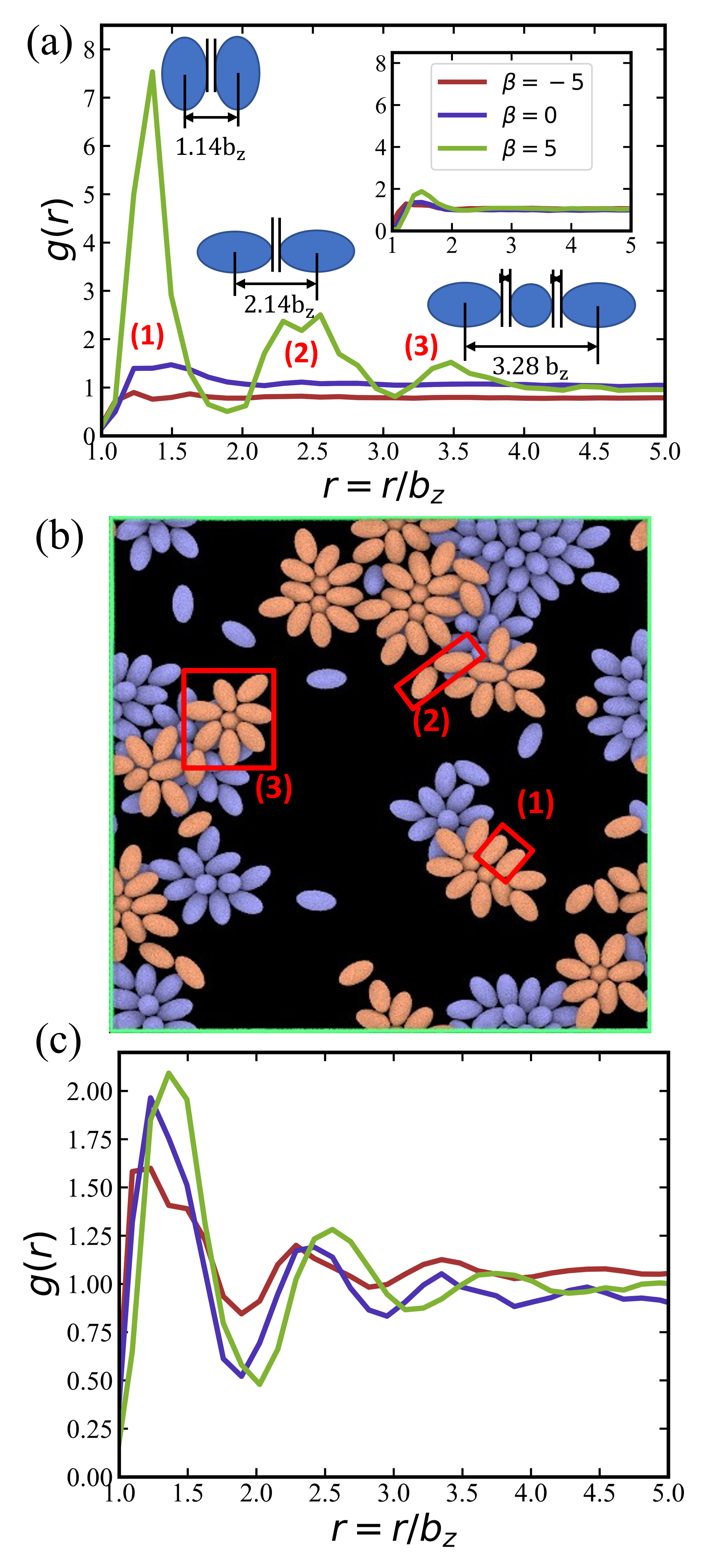}
	\caption{(a) 2D radial distribution function $g(r)$ of squirmers with
        $\phi = 0.18$ and $\tilde{\lambda} = 0$,  for pushers (red), neutral squirmers (blue), and pullers (green). The locations of peaks of $g(r)$ for pullers correlate well with local structural 
		arrangements illustrated next to the peaks. The inset presents the corresponding $g(r)$ functions for $\tilde{\lambda} = 133.5$
        and the same $\phi$. (b) Snapshot of the simulated system for $\phi = 0.18$, $\tilde{\lambda} = 0$, and $\beta = 5$, which illustrates the most
        frequent squirmer structures, see also Movie S3. (c) $g(r)$ of squirmers for $\phi = 0.56$ and $\tilde{\lambda} = 0$. The legend for the curves
        is the same as in (a). 
		}
	\label{fgr: rdf_plot}
\end{figure}

\subsubsection{2D radial distribution function parallel to the walls}
\label{sec:rdf}

To characterize the internal structure of the suspension of squirmers along the walls, we compute the 2D radial distribution function 
\begin{equation} \label{eqn: gr}
        g(r) = \frac{M}{N_{sq}^2}\left\langle\sum_{i}^{}\sum_{j\neq i}^{}\delta(\bm{r} - \bm{r}_{ij})\right\rangle ,
\end{equation}
where $\bm{r}_{ij}$ is the vector between two centers of mass of squirmer pairs, and $M = A/(2\pi r)$ is a normalization factor 
($A$ is the area of the considered 2D plane) such that $g(r) \to 1$ for $r\to \infty$. 
Note that $g(r)$ is calculated in 2D within the $x-z$ plane, where the measurements are performed 
within the two layers (upper and lower halves of the slit) separately. This is a reasonable 
simplification due to the prevalence of a two-layered structure as shown in Fig.~\ref{fgr: y_dist}.
  
Figure ~\ref{fgr: rdf_plot}(a) presents $g(r)$ for different swimming modes at low volume fraction 
$\phi = 0.18$ and without rotlet dipole, $\tilde{\lambda} = 0$. Only the suspension of pullers 
with $\beta = 5$ exhibits pronounced peaks in $g(r)$, which represent the most frequent structural 
elements illustrated in Fig.~\ref{fgr: rdf_plot}(b). The existence of structure for pullers, but not for pushers and neutral 
squirmers, is primarily related to the fact that pullers form large clusters already at $\phi = 0.18$, while the other squirmer 
types yield a gas-like phase, see Table \ref{tab:table_cl}. Interestingly, the x-y perspective plot in Fig.~\ref{fgr: snapshots}(b) 
shows that pullers frequently form flower-like arrangements with one or two squirmers oriented perpendicular to the walls and 
surrounded by several "petal" squirmers [also illustrated in Fig.~\ref{fgr: rdf_plot}(b)].  These structures form due 
to the attractive flow field generated by pullers, which slide along the walls until they collide and form the flower-like structures through the attractive hydrodynamic interactions. 
When the rotlet dipole is turned on at $\phi = 0.18$ [see the inset in Fig.~\ref{fgr: rdf_plot}(a)], the suspension of pullers becomes  
gas-like and thus looses its internal structure.   

Figure ~\ref{fgr: rdf_plot}(c) shows $g(r)$ at the high volume fraction $\phi = 0.56$, where several peaks are observed for all squirmer types. At this high 
volume fraction, all squirmer suspensions are in MIPS phase (see Table \ref{tab:table_cl}), so that large clusters are present with 
an internal fluid structure. Furthermore, due to the dominance of steric interactions, radial distribution functions are again very similar for different 
$\beta$, indicating that the internal structure is nearly independent of the swimming mode at large $\phi$. 

\begin{figure}[h]
	\centering
	\includegraphics[width=0.9\linewidth]{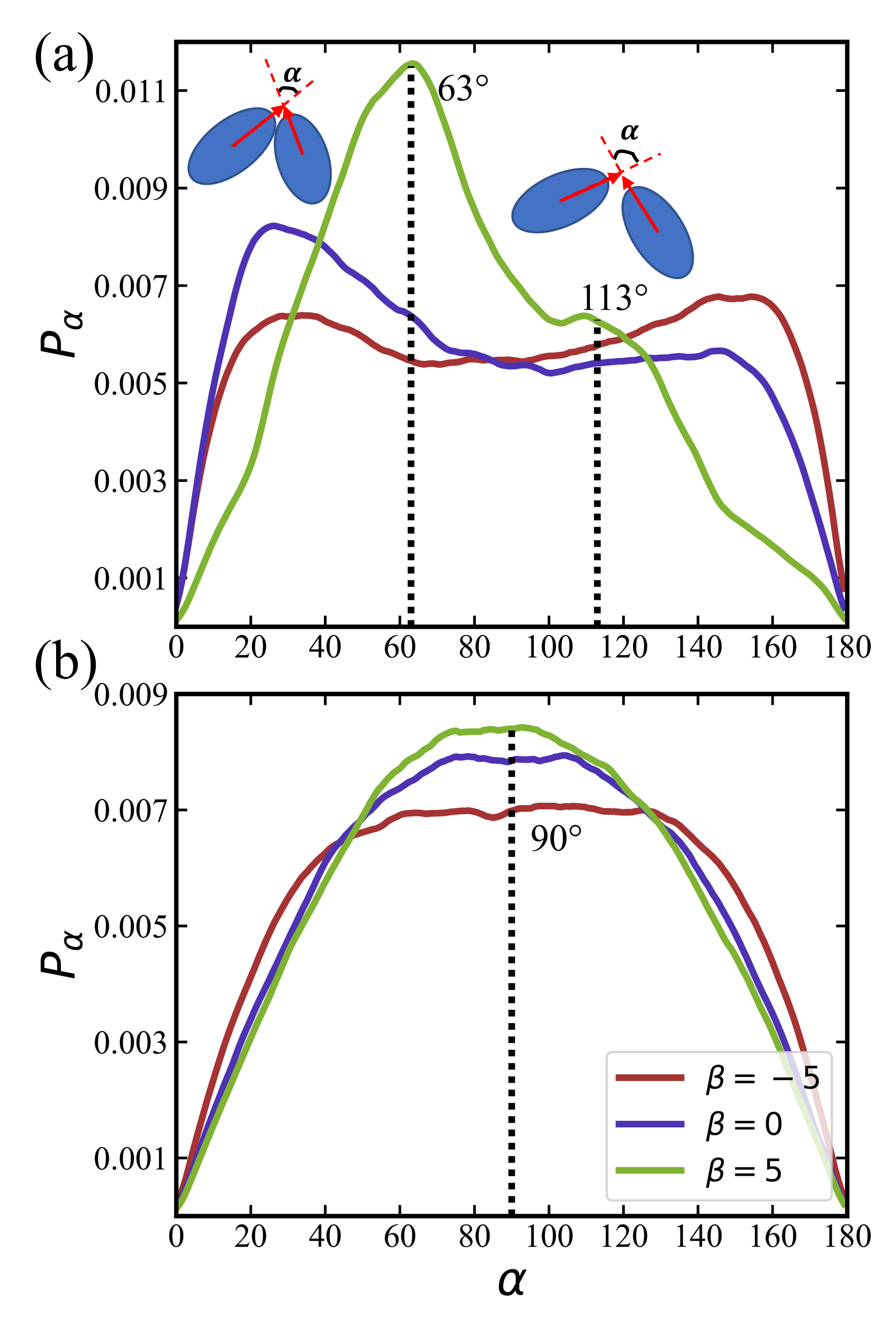}
	\caption{Distribution $p_\alpha$ of relative angle between two neighboring squirmers for different swimming modes at $\phi = 0.18$. 
		$\alpha$ is the angle between orientation vectors of a pair of squirmers, whose separation is smaller than $3b_z$.  
		(a) $p_\alpha$ for the case of no rotlet dipole ($\tilde{\lambda} = 0$), with the insets illustrating configurations of squirmers at the 
	  two peaks with $\alpha \approx 63^o$ and  $\alpha \approx 113^o$. (b) Relative angle distribution for $\tilde{\lambda} = 133.5$.
         }
	\label{fgr: relative_angle}
\end{figure}

\subsubsection{Angle between two neighboring squirmers}

Previous simulation studies \cite{Theers_MSM_2016, Kyoya_SMCM_2015} for a monolayer of squirmers in a thin fluid between two walls have analyzed 
the angles between pairs of squirmers for different swimming modes. Theers et al.~\cite{Theers_MSM_2016} report that pairs of 
pullers tend to adopt a fixed relative angle after their initial encounter, which is equal to approximately $45^o$. The stable angle 
formation does not occur for neutral squirmers and pushers. Similarly, Kyoya et al.~\cite{Kyoya_SMCM_2015} suggest that neutral 
squirmer pairs tend to align with each other, puller pairs assume a stable angle between them in the range of $0^o$ to $90^o$, and 
pusher pairs do not show any order and swim away from each other after a brief encounter. 

We calculate distributions $p_\alpha$ of the relative angle $\alpha$ between orientation vectors  
of two neighboring squirmers within a defined cutoff radius as 
\begin{equation} \label{eqn: fa}
        p_\alpha = \frac{1}{N_{sq}C}\left\langle\sum_{i}^{}\sum_{j\neq i}^{|\bm{r}_{j} - \bm{r}_{j}| \leq 3b_z}\delta(\alpha - \alpha_{ij})\right\rangle,
\end{equation}
where and $C$ is the normalization factor so that the integral of $p_\alpha$ is equal to unity. 
Figure~\ref{fgr: relative_angle}(a) shows  
the distribution $p_\alpha$ for various $\beta$ at $\phi = 0.18$. $p_\alpha$ for pullers displays 
two peaks located at $\alpha \approx 63^o$ and $\alpha \approx 113^o$, where the former value is not far from $45^o$ 
found for a monolayer of squirmers in Ref.~\cite{Theers_MSM_2016}. Notably, the peak at $\alpha \approx 63^o$ is quite prominent, since it represents the flower-like arrangements illustrated in
Fig.~\ref{fgr: rdf_plot}(b). However, the distribution for pullers spans a wide range of relative angles. Neutral squirmers and pushers yield broad and nearly flat distributions, indicating that no stable
structures are present. The existence of two small peaks at $\alpha \approx 25^o$ and $\alpha \approx 155^o$ suggests that pushers
and neutral squirmers tend to swim parallel to each other, which is consistent with previous 
predictions for a squirmer pair \cite{Ishikawa_HIM_2006, Goetze_FGRC_2011}.

The distribution of $p_\alpha$ for an active rotlet dipole with $\tilde{\lambda} = 133.5$ is shown in Fig.~\ref{fgr: relative_angle}(b). 
For all $\beta$ values, $p_\alpha$ is centered at $\alpha \approx 90^o$, indicating that hydrodynamic interactions favor the
perpendicular orientation between squirmer pairs. Note that for pushers, the central region in $p_\alpha$ is flatter than that for pullers and neutral squirmers. 

\begin{figure}[htb!]
	\centering
	\includegraphics[width=0.9\linewidth]{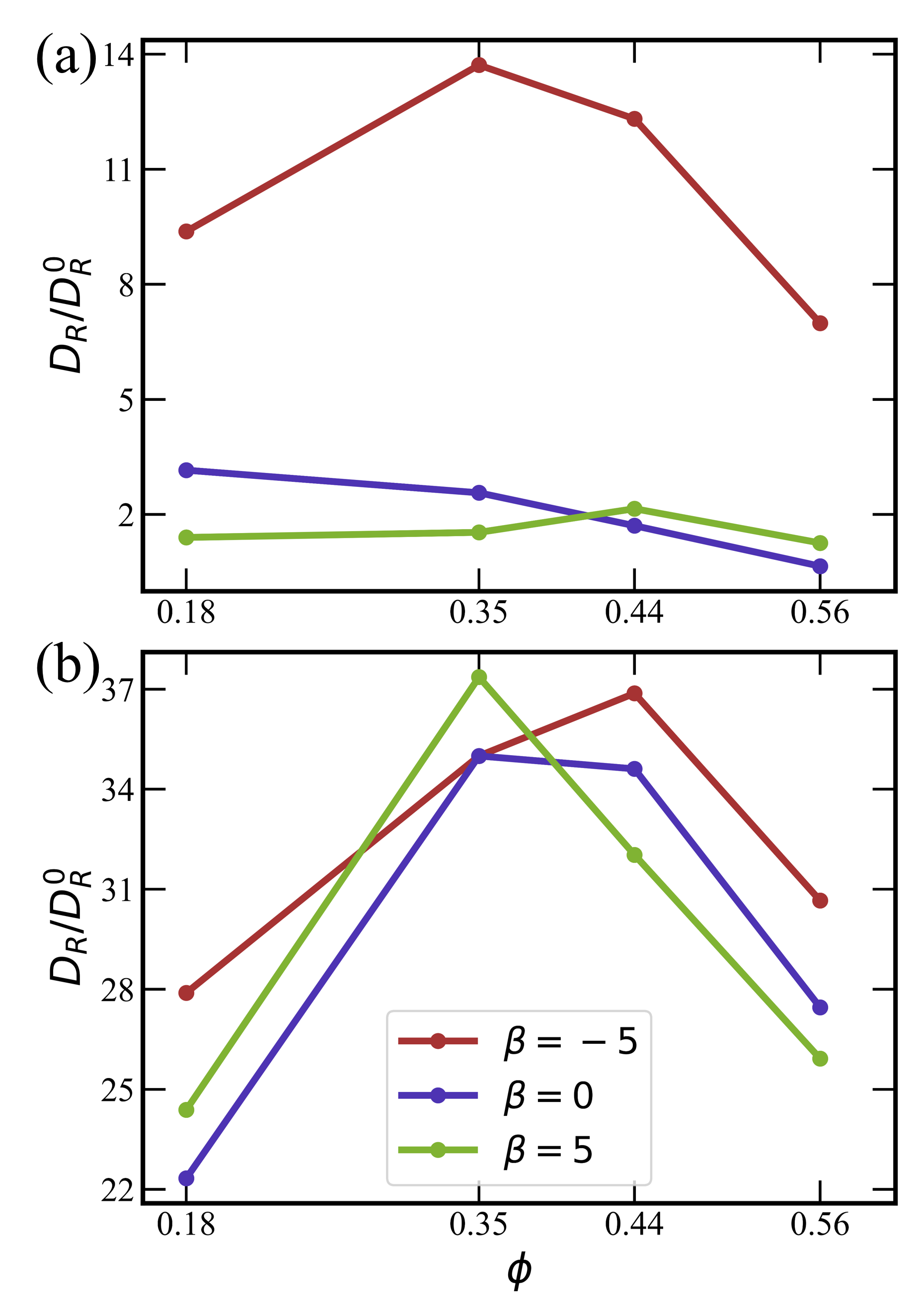}
	\caption{Effective rotational diffusion $D_R$ as a function of $\phi$ for (a) $\tilde{\lambda} = 0$ and (b)
          $\tilde{\lambda} = 133.5$. $D_R$ is obtained by fitting auto-correlation functions of squirmer orientation
          using Eq.~(\ref{eqn: exp_rd}), and normalized by the rotational diffusion $D_R^0$ of a spheroid. Data for pushers
          (red), for neutral squirmers (blue), and for pullers (green).
    } 
	\label{fgr: rotation}
\end{figure}

\subsection{Dynamical properties}
\label{sec:dynamic}

\subsubsection{Effective rotational diffusion coefficient}\label{sec:Dr}

To characterize the rotational properties of squirmers, we compute their effective rotational 
diffusion coefficient $D_R$,  obtained by fitting the auto-correlation function of squirmer orientation 
\begin{equation} \label{eqn: exp_rd}
 <\textbf{e}(t) \cdot\textbf{e}(t + \Delta t)> = e^{-D_{R} \Delta t} ,
\end{equation}
where $<\dots>$ denotes the average over all squirmers and all times in the simulated system. 
Here, the exponential decay applies for small $\Delta t \lesssim 1/D_R$. Figure~\ref{fgr: rotation} 
shows the dependence of $D_R$ on $\phi$, $\beta$, and $\lambda$. In most cases, $D_R/D_R^0 > 1$, 
indicating that the rotational dynamics of squirmers within a suspension is enhanced in comparison with 
the rotational diffusion $D_R^0$ of a spheroid in an unconfined system. This is due to hydrodynamic 
interactions between squirmers, and their collisions during motion. Only for the case of 
$\beta = 0$, $\phi = 0.56$, and $\tilde{\lambda} = 0$, rotational diffusion is reduced, $D_R < D_R^0$, 
which is likely due to the crowding of squirmers at large volume fractions [compare
Fig.~\ref{fgr: snapshots}(a)].  

Figure~\ref{fgr: rotation}(a) ($\tilde{\lambda} = 0$) shows that the enhancement of rotational 
diffusion coefficient for pushers is significantly larger than for pullers and neutral squirmers. 
This occurs due to repulsive hydrodynamic interactions between pushers \cite{Theers_MSM_2016,Kyoya_SMCM_2015},
which lead to the enhanced scattering between squirmers. The dependence of  $D_R$ on $\phi$ first exhibits 
an increase, followed by a maximum and subsequent decrease. With increasing $\phi$, the collision rate 
between squirmers increases, resulting in the enhancement of $D_R$. 
However, at large $\phi$, squirmer clustering and the transition to the MIPS phase take place 
[compare Fig.~\ref{fgr: snapshots}(a)], so that the effective rotational diffusion is 
significantly slowed down.        

Noteworthy is the difference in $D_R$ for $\tilde{\lambda} = 0$ and $\tilde{\lambda} = 133.5$ in 
Fig.~\ref{fgr: rotation}(a) and (b), respectively, where the 
enhancement for squirmers with rotlet dipole is at least a factor three larger than that without. 
Interestingly, squirmers with $\tilde{\lambda}=133.5$ switch between the two layers (i.e. migrate 
from one wall to the other) much more frequently than squirmers with $\tilde{\lambda}=0$,
which will be discussed in Section \ref{sec:migration}. The ability of interchange between the two 
layers increases significantly the effective 
rotational diffusion. The $D_R$ curves for $\tilde{\lambda} > 0$ are similar for different swimming modes $\beta$, in agreement with the structural characteristics for suspensions of squirmers with the rotlet dipole,
as discussed in Sec.~\ref{sec:struct} above. 

\begin{figure}[htb!]
	\centering
	\includegraphics[width=0.9\linewidth]{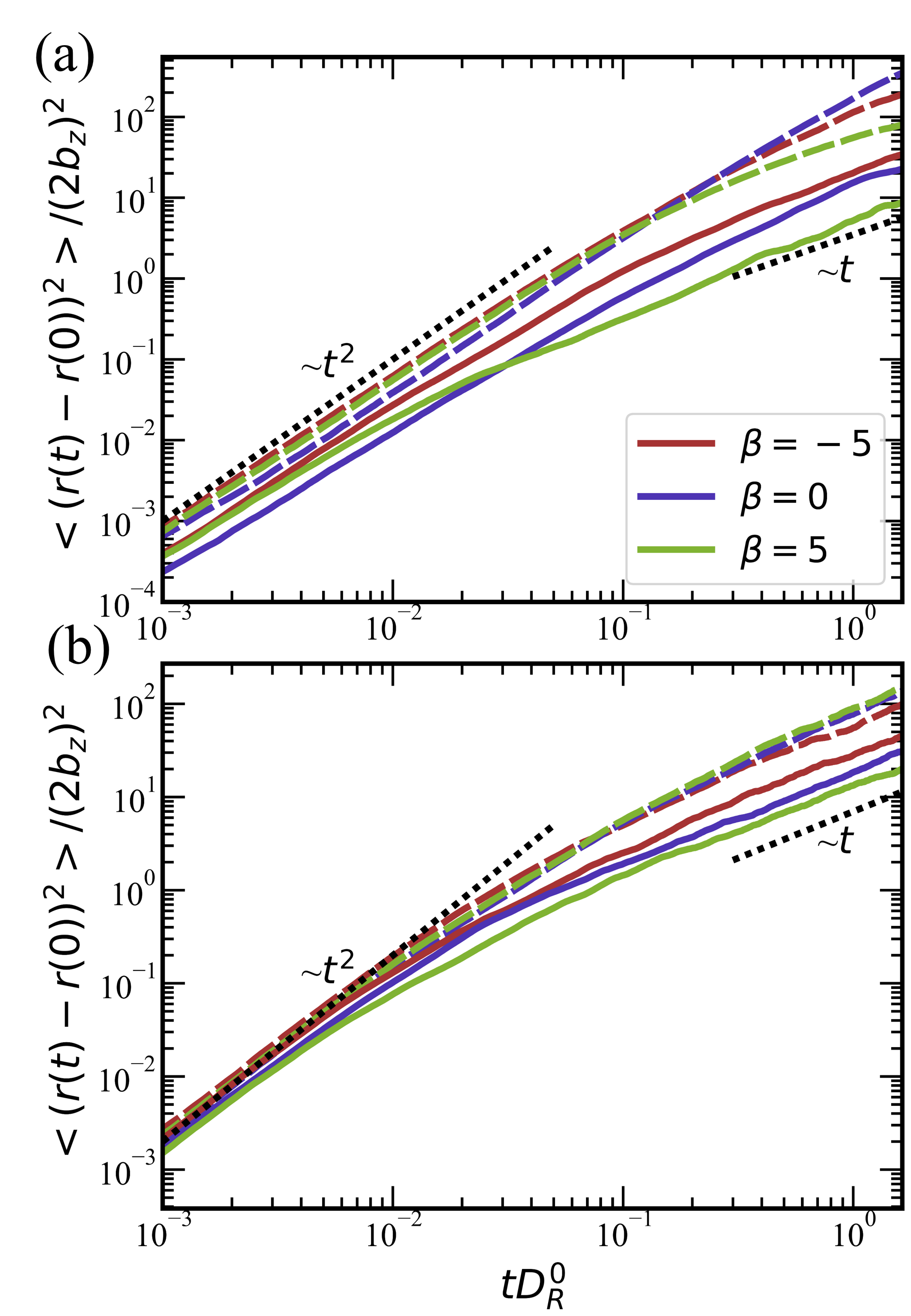}
	\caption{Mean-squared displacement of squirmers for $\phi = 0.18$ (dashed lines) and $\phi = 0.56$ (solid lines) (a) without the rotlet dipole $\tilde{\lambda} = 0$ 
		and (b) with the rotlet dipole $\tilde{\lambda} = 133.5$. Curves for pushers are in red, for neutral squirmers in blue, and for pullers in green. 
		Black dotted lines indicate power-laws $\sim t^2$ in the ballistic regime and $\sim t$ in the diffusive regime. 
       }
	\label{fgr: msd}
\end{figure}

\subsubsection{Mean-square displacement}

The mean-square displacement of squirmers for various conditions are shown in Fig.~\ref{fgr: msd}. 
The motion is ballistic at short times, with a quadratic time-dependence, and diffusive at long 
times, with $<(r(t) - r(0))>^2 \sim t$, as expected.
For $\phi = 0.18$ and $\tilde{\lambda} = 0$, pullers yield the lowest effective diffusivity, 
followed by neutral squirmers and pushers, because the suspension of pullers at these conditions is already 
in the MIPS phase (see Table \ref{tab:table_cl}). The effective diffusivity of squirmers at $\phi = 0.56$ 
is lower than at $\phi = 0.18$ due to crowding and the formation of large clusters.

The transition from the ballistic to the diffusive regime for $\tilde{\lambda} = 0$ in Fig.~\ref{fgr: msd}(a) 
occurs around 
$t D_R^0 \simeq 0.2-0.5$. Theoretically, this transition should take place around $tD_R \simeq 1$ 
\cite{Romanczuk_ABP_2012}, which is consistent with the values of $D_R/ D_R^0 \simeq 2$ 
in Fig.~\ref{fgr: rotation}(a). Note that the ballistic-to-diffusive 
transition for squirmers with rotlet dipole occurs at $t D_R^0 \gtrsim 0.1$, which is also consistent 
with the values of $D_R / D_R^0 \simeq 10$ in Fig.~\ref{fgr: rotation}(b). Furthermore, 
mean-square-displacement curves for $\tilde{\lambda} = 133.5$ in Fig.~\ref{fgr: msd}(b)
are similar to each other for various $\beta$ at a fixed volume fraction, confirming once more that the 
rotlet dipole nearly offsets the effects of different swimming modes. 

\begin{figure}[htb!]
	\centering
	\includegraphics[width=0.9\linewidth]{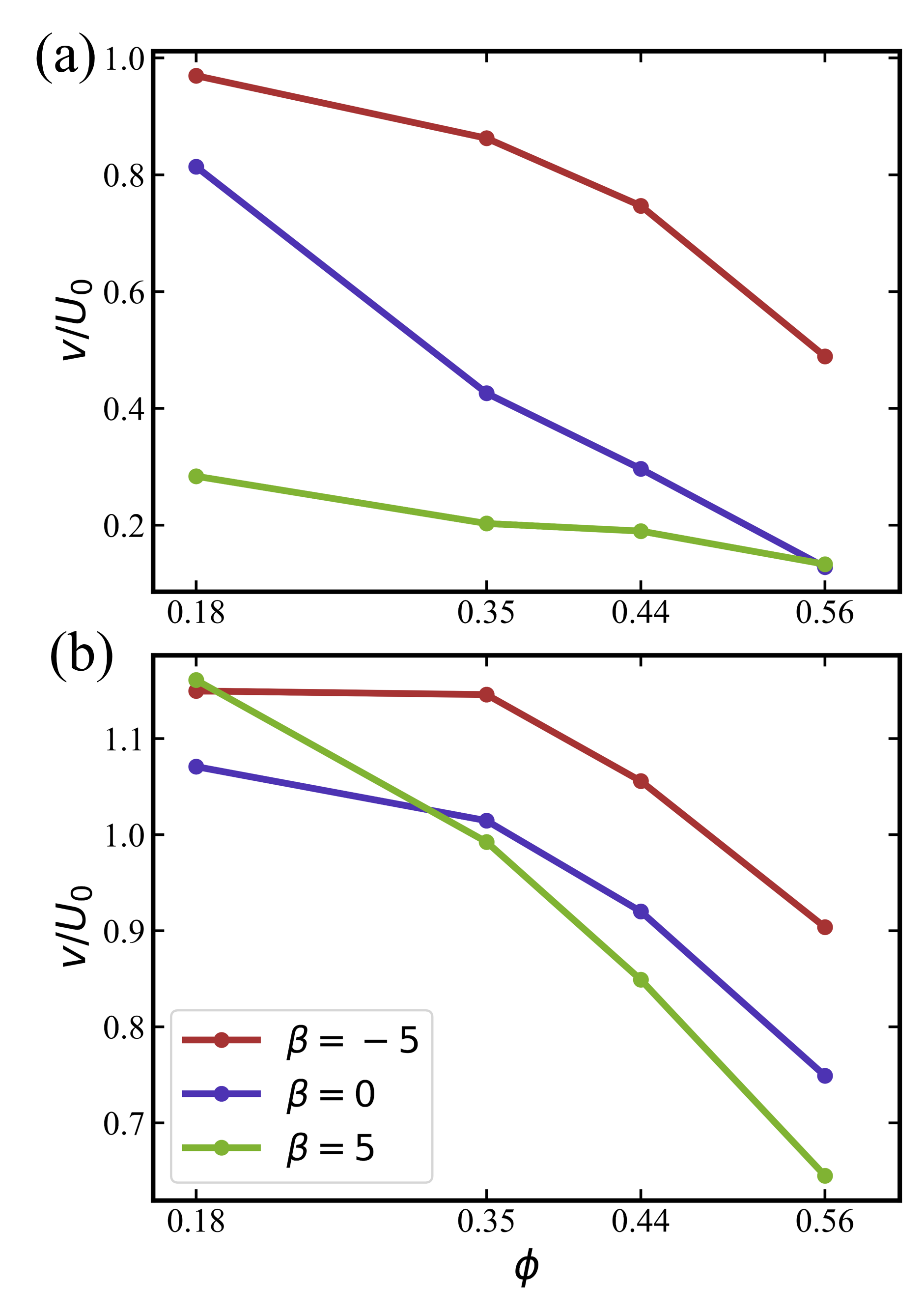}
	\caption{Average squirmer speed $\bar{v}$ as a function of $\phi$ for (a) $\tilde{\lambda} = 0$ and (b) $\tilde{\lambda} = 133.5$. 
		The speed is normalized by the swimming speed $U_0$ of a squirmer in an unconfined system. Data for pushers
		(red), for neutral squirmers (blue), and for pullers (green).} 
	\label{fgr: average_speed}
\end{figure}

\subsubsection{Average squirmer speed}

To further characterize the mobility of squirmers for different conditions, we also study the average 
squirmer speed $\bar{v}$. 
For a single squirmer $i$, the instantaneous swim speed is calculated as 
\begin{equation}
v_i(t) = |\mathbf{r}_i(t + \Delta t) - \mathbf{r}_i (t)| / \Delta t,
\end{equation}
with time interval $\Delta t = 0.022/D_R$ (significantly smaller than $1/D_R$), during which 
a free squirmer moves a distance $b_z$. $\bar{v}$ is then obtained by an ensemble and time average 
as a function of the volume fraction $\phi$.  

As shown in Fig.~\ref{fgr: average_speed}, pushers are the most mobile, displaying the largest average 
speed $\bar{v}$ among all squirmer types. 
For the case without rotlet dipole in Fig.~\ref{fgr: average_speed}(a), $\bar{v}$ decreases with 
increasing $\phi$ due to crowding at large volume fractions and the transition to MIPS or clustering 
regime. For all studied $\beta$ values, the average speed of squirmers is below $U_0$, which is the speed 
of a single squirmer in an unconfined system. 

The average speeds of squirmers with rotlet dipole in Fig.~\ref{fgr: average_speed}(b) are larger than 
those with $\tilde{\lambda} =0$. 
Furthermore, the decay in $\bar{v}$ for $\tilde{\lambda} = 133.5$ with increasing $\phi$ is slower than for 
$\tilde{\lambda} =0$. Interestingly, for $\phi \lesssim 0.4$, the average speed of squirmers with rotlet 
dipole is slightly larger than $U_0$, indicating motility enhancement due to interactions with 
the walls. An increased mobility near walls was  
reported previously for pushers without rotlet dipole \cite{Kyoya_SMCM_2015, Lintuvuori_HORW_2016, Spagnolie_HSP_2012}. The results in 
Fig.~\ref{fgr: average_speed} confirm that the presence of rotlet dipole enhances the mobility of 
squirmers and delays the transition to MIPS as a function of $\phi$.  

\begin{figure}[htb!]
	\centering
	\includegraphics[width=0.9\linewidth]{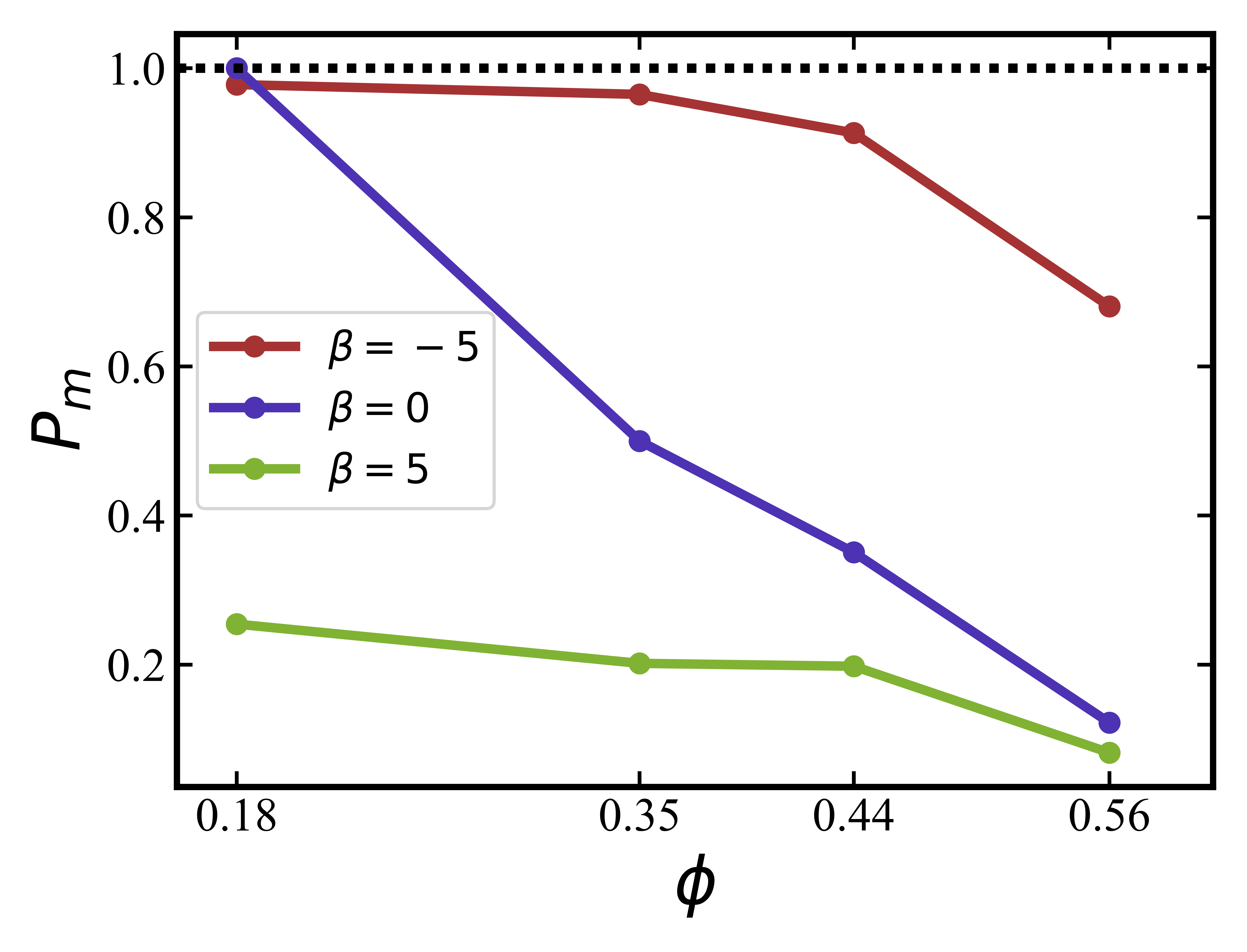}
	\caption{Fraction $P_m$ of squirmers that switch between the two layers at least once during a simulation 
      as a function of $\phi$ for $\tilde{\lambda} = 0$. Curves for pushers (red), for neutral squirmers (blue), and for pullers (green). The black dotted line marks $P_m = 1$, which is the fraction of squirmers 
      that migrate between the two layers for a non-zero rotlet dipole with $\tilde{\lambda} = 133.5$, independently of $\phi$ and $\beta$. 
      }
	\label{fgr: swimming}
\end{figure}

\subsubsection{Migration between the two layers}
\label{sec:migration}

Due to the dominance of two-layered squirmer structures for nearly all conditions (see Fig.~\ref{fgr: y_dist}), 
an interesting issue is the switching freqnuncy of individual squirmers between the layers. 
Figure~\ref{fgr: swimming} presents the fraction $P_m$ of squirmers that switch between the two layers 
at least once during the simulation time of $1.5\tau_R$ for different $\phi$ without rotlet dipole. 
For pushers, $P_m$ is close 
to unity at small $\phi$, and decreases to $P_m \approx 0.6$ at $\phi = 0.56$. Previous studies 
\cite{Ishimoto_SDB_2013, Theers_MSM_2016, Lintuvuori_HORW_2016, Zoettl_SSM_2018} have shown that the residence of pushers near a wall 
is unstable due to local hydrodynamic interactions which orient them slightly away from the wall. 
The drop in $P_m$ at large $\phi$ for pushers is due to significant crowding, which limits the migration 
between the two layers. For neutral squirmers, $P_m$ reduces from unity to $P_m \approx 0.1$ as the volume 
fraction is increased, indicating that their residence at the walls is more long-lived than for pushers.  

For pullers, notably a relatively low fraction migrates between the two layers.   
This is consistent with the preferred orientation toward a wall, compare Sec.~\ref{sec:orient}, ass well as
the formation of stable flower-like structures, as shown in Sec.~\ref{sec:rdf}. Thus, switching fo pullers 
between the two layers by pullers is facilitated by squirmer collisions and does not 
occur spontaneously. 

Finally, for squirmers with rotlet dipole $\tilde{\lambda} = 133.5$, $P_m = 1$ for all simulated 
$\beta$ and $\phi$ values. Thus, a rotlet dipole significantly destabilizes the residence of squirmers near 
the walls, and thereby enhances squirmer motility, as shown in Figs.~\ref{fgr: rotation} 
and \ref{fgr: average_speed}.

\section{Discussion and conclusions}
\label{sec:discuss}

In this study, we have performed mesoscale hydrodynamic simulations of a confined system of squirmers 
in a thin fluid film between two parallel walls, in order to better understand 
the intricate interplay between hydrodynamic and steric interactions and their impact on the collective behavior 
of squirmers. In contrast, previous studies that considered hydrodynamic interactions between swimmers have mainly focused on quasi-2D systems (i.e., a monolayer of squirmers) 
\cite{Theers_CMS_2018, Qi_EAT_2022, Kyoya_SMCM_2015, Alarcon_MCP_2017, Zoettl_HCM_2014, Yoshinaga_HIDS_2017, Blaschke_PSC_2016} 
or 3D systems with periodic boundary conditions 
\cite{Ishikawa_DSM_2007, Delmotte_LSS_2015, Ishikawa_DCS_2008, Evans_CSS_2011, Alarcon_SAO_2013}. 
The considered film thickness is large enough to allow the formation of two-layered structures. 
This situation is of interest when the formation of a biofilm, which starts from a monolayer of bacteria, 
proceeds toward the development of further layers. In particular, an interesting question is whether bacteria 
can spontaneously leave the surface layer. 

By systematically varying relevant parameters, such as the volume fraction of squirmers, their active stress, and their rotlet dipole strength, we show the emergence of different structures and phases (Fig.~\ref{fgr: snapshots}). As expected, at low volume fractions
of squirmers, the system is in a gas-like phase. As $\phi$ is increased, a swarming state is observed, 
with the formation of mobile clusters with a wide range of sizes. The elongated shape of squirmers is essential 
for the emergence of these swarming clusters, as it provides an alignment interaction, which has been 
observed previously for self-propelled rods, spheroids, and semi-flexible filaments \cite{Peruani_CSPR_2006,Theers_CMS_2018,Duman_CDF_2018}. Swarming clusters form at intermediate 
values of $\phi$ for pushers and squirmers with the rotlet dipole. The absence of the swarming state for 
pullers with $\tilde{\lambda} = 0$ suggests that contractile hydrodynamic interactions between swimmers 
suppress the formation of swarming clusters. 
 
At large enough volume fractions, a single large cluster of squirmers emerges, surrounded by a few mobile 
swimmers (see Fig.~\ref{fgr: snapshots}(a) at $\phi = 0.44$ or $0.56$), indicating formation of
MIPS. Pullers exhibit MIPS already at low volume fraction, $\phi \approx 0.18$, while neutral squirmers 
and pushers require larger volume fractions, in agreement with previous studies \cite{Theers_CMS_2018, Kyoya_SMCM_2015}. 
Our simulations also suggest that hydrodynamic interactions suppress motility-induced clustering and 
phase separation, in agreement with previous reports
\cite{Theers_CMS_2018, Matas_Navarro_HPS_2014} for particles with aspect ratios not far from unity. 
Note that the suspension of pushers at $\phi = 0.56$ does not show 
a transition to the state of active turbulence, because the strength of the induced force dipole is likely too 
weak \cite{Stenhammar_CBM_2017, Martinez_RBS_2020}.

An interesting aspect of our investigation is the interplay of simultaneous hydrodynamic and steric 
interaction both between squirmers and of squirmers with the confining walls, and its effect on the 
collective behavior. 
In particular, pullers favor a nearly perpendicular orientation with the walls, resulting in the formation 
of flower-like clusters [Fig.~\ref{fgr: snapshots}(b)] at low volume frations, where pullers first 
slide along the surface, and upon collisions form a structure with a single puller oriented nearly 
perpendicular to the wall surrounded by a few petal pullers. This leads to the nucleation of clusters for pullers already at low $\phi$, promoting MIPS at low $\phi$. 
Pushers and neutral squirmers favor an orientation parallel to the walls, which does not significantly 
hinder their mobility. Therefore, pushers can frequently switch between the layers of a two-layer 
structure, substantially delaying the formation of large clusters with increasing $\phi$. Thus, a colony 
of pusher-like swimmers (e.g., \textit{E. coli}) should be able to spontaneously switch from an initial 
monolayer structure to a multi-layered
configuration. Recent experiments \cite{Junot_RTV_2022} suggest that \textit{E. coli} bacteria also employ 
tumbling (i.e., active turning due to the rotation reversal of one of its flagella) to leave a surface, 
indicating that the aspect ratio of a swimmer is very important for its interaction with the wall. In our simulations, $b_z/b_x = 2$, which is somewhat smaller than that for \textit{E. coli}. 

For all studied cases, there is a clear preference for the two-layered structure.
Only pullers at $\phi \lesssim 0.25$ show partial preference for the 
perpendicular-to-the-wall orientation due to their hydrodynamic interactions with the walls. Pushers and neutral squirmers display mostly parallel-to-wall alignment.
 
Our comprehensive simulations reveal that the the effect of a rotlet dipole is to offset the effect of
active stress, characterized by $\beta$, to a large extent. Even at relatively low volume fractions, 
the presence of a rotlet dipole of dimensionless strength $\tilde{\lambda} \neq 0$ nearly eliminates 
differences in the behavior of pullers, neutral squirmers, and pushers. 
With the rotlet dipole, squirmers assume a parallel-to-the-wall orientation and 
therefore, remain mobile up to moderate volume fractions, frequently switching between the two layers. 
This enhancement of layer switching can be attributed to the rotational flow field of rotors, which 
implies an induced motion around their center of mass.
Furthermore, the rotlet dipole leads to a circular trajectories of squirmers near the walls \cite{Lauga_SIC_2006,Hu_PSP_2015}, which further 
contributes to the mobility of squirmers within the confinement. In our study, a single squirmer 
with $\tilde{\lambda} = 133.5$ circles at the wall with a radius of approximately $2b_z$. 
However, the circling motion near a wall is difficult to detect 
at higher volume fractions due to frequent squimer-squirmer scattering.
As a result, the effective rotational diffusivity of squirmers with rotlet dipole is approximately one 
order of magnitude larger than that for squirmers with $\tilde{\lambda}=0$. Thus, swimmers 
with rotlet dipole are expected to spontaneously initiate multi-layered structures within a biofilm.

Finally, we would like to discuss some limitations of our study. Despite the fact that the squirmer 
model can produce various swimming modes, the near-field flow of real microswimmers, like bacteria, 
is unavoidably species-specific, arising from body shape, the geometry and dynamics of propelling 
flagella, and the ability to navigate (e.g., \textit{E. coli} tumbling). Furthermore, 
steric interactions between microswimmers including their flagella are likely different from interactions 
between idealized spheroidal shapes. However, modeling of swimmers with explicit appendages 
\cite{Hu_PSP_2015} is computationally expensive, significantly limiting the number of simulated swimmers 
in a study. Therefore, the direct comparison of our simulation results with experiments requires 
some calibration, like of the aspect ratio of squirmers, and the strengths of active stress and 
rotlet dipole. Nevertheless, we expect that our simulations provide a good qualitative characterization 
of swimmer suspensions under confinement. Future work should also consider thicker films to bridge 
these results with those from three-dimensional periodic systems 
\cite{Ishikawa_DSM_2007,Delmotte_LSS_2015,Ishikawa_DCS_2008,Evans_CSS_2011,Alarcon_SAO_2013}.

\section*{Appendix A: Vesicle-like squirmer model}

Each squirmer surface is represented by $N_p = 1024$ particles connected by springs into a triangulated network.
Shear elasticity of the squirmer membrane is supplied by the spring potential \cite{Fedosov_RBC_2010}
\begin{equation} 
U_{bond} = \frac{k_BT l_m}{4p}\frac{3x^2-2x^3}{1-x} + \frac{k_p}{l}, 
\end{equation}
where $x=l/l_m \in (0,1)$, $l$ is the spring length, $l_m$ is the maximum spring extension, $p$ is the persistence length,  $k_p$ is
the force coefficient, and $k_BT$ is the energy unit defined by the temperature $T$ in the simulated system. The curvature elasticity
that provides bending resistance is implemented through the discretisation of the Helfrich bending energy \cite{Helfrich_EPB_1973} as
\begin{equation}
U_{bend} = \frac{\kappa}{2} \sum_i \sigma_i(H_i-H_0^i)^2,
\end{equation} 
where $\kappa$ is the bending rigidity, $\sigma_i$ is the area corresponding to vertex $i$ in the membrane triangulation, $H_i$ is the mean
curvature at vertex $i$, and $H_0^i$ is the spontaneous curvature at vertex $i$. The mean curvature is discretized as $H_i = {\bf n}_{i}
\cdot \sum_{j(i)} \sigma_{i j} {\bf r}_{i j}/(\sigma_{i}r_{i j})$, where ${\bf n}_{i}$ is the unit normal at the vertex $i$,
$\sigma_{i} =\sum_{j(i)} \sigma_{i j}r_{i j}/4$, $j(i)$ spans all vertices linked to vertex $i$, and $\sigma_{i j}=r_{ij}(\cot\theta_1+ \cot\theta_2)/2$
is the bond length in the dual lattice with $\theta_1$ and $\theta_2$ being the angles at the two vertices opposite to the edge $ij$ in the dihedral.
The spontaneous curvature $H_0^i$ is set locally after the triangulation of the spheroidal surface of the squirmer. 

The area and volume conservation constraints are represented by the potential \cite{Fedosov_RBC_2010}
\begin{equation}
    U_{a+v} = \frac{k_a(A - A_0^{tot})^2}{2A_0^{tot}} + \sum_{m \in 1...N_t} \frac{k_d(A_m - A_0^m)^2}{2A_0^m} + \frac{k_v(V - V_0^{tot})^2}{2V_0^{tot}}, 
\end{equation}
where $k_a$, $k_d$, and $k_v$ are the coefficients of global area, local area and volume conservation constraints, respectively. $A$ and $V$ are
the instantaneous area and volume of the enclosed membrane, $A_0^{tot}$ and $V_0^{tot}$ are the targeted global area and volume which are defined
by the spheroidal shape. $A_m$ is the area of the $m$-th triangle (or face) within the triangulation, while $A_m^0$ is the targeted value.
$N_t$ is the number of triangles within the triangulated surface. 

The membrane-like representation of squirmers has the shear modulus 
$\mu_0=1.44\times 10^5 k_BT/b_z^2$, the bending modulus $\kappa = 250 k_BT$, the area-constraint coefficients $k_d = 1.6\times 10^3 k_BT/b_z^2$ and
$k_a= 8\times 10^3 k_BT/b_z^2$, the volume-constraint coefficient $k_v = 3.2\times 10^4 k_BT/b_z^3$, the total area $A_0^{tot} = 5.36 b_z^2$, and
the total volume $V_0^{tot} = 1.05 b_z^3$. In all simulations, $k_BT = 1$ and $b_z = 4$. 

\section*{Appendix B: Modeling fluid flow}

To model fluid flow, we employ the dissipative particle dynamics (DPD) method \cite{Hoogerbrugge_SMH_1992,Espanol_SMO_1995}, which is a mesoscopic
hydrodynamics simulation technique. DPD is a particle-based Lagrangian method, where each particle represents a small fluid volume. DPD particles
$i$ and $j$ interact through three types (conservative, dissipative, and random) of pairwise forces given by
\begin{align}
  \textbf{F}^C(r_{ij}) & = a W^C(r_{ij})\hat{\textbf{r}}_{ij}, \\
  \textbf{F}^D(r_{ij}) & = -\gamma W^D(r_{ij})(\hat{\textbf{r}}_{ij} \cdot \textbf{v}_{ij}) \hat{\textbf{r}}_{ij}, \\
  \textbf{F}^R(r_{ij}) & = \sigma W^R(r_{ij})\xi_{ij} \hat{\textbf{r}}_{ij}/ \sqrt{\Delta t}, 
\end{align} 
where $a$, $\gamma$, and $\sigma$ are the force amplitudes, $\textbf{r}_{ij} = \textbf{r}_i - \textbf{r}_j$ is the relative position vector,
$r_{ij} = |\textbf{r}_{ij}|$, $\hat{\textbf{r}}_{ij} = \textbf{r}_{ij}/r_{ij}$, and $\textbf{v}_{ij} = \textbf{v}_i - \textbf{v}_j$ is the
velocity difference. $\xi_{ij} = \xi_{ji}$ is a symmetric Gaussian random variable with zero mean and unit variance, and $\Delta t$ is the
time step. All forces act within a cutoff radius $r_c$ and vanish beyond it. The conservative force controlls fluid compressibility, while
the dissipative and random forces form a thermostat, so that the DPD fluid has an isotropic temperature $T$. Thus, $\textbf{F}^D$ and
$\textbf{F}^R$ are related through the fluctuation-dissipation theorem \cite{Espanol_SMO_1995} as
\begin{equation}
  \sigma^2 = 2\gamma k_B T, \quad \quad W^D(r_{ij}) = [W^R(r_{ij})]^2.
\end{equation}
The weight functions are defined as 
\begin{equation}
W(r_{ij}) = W^R(r_{ij}) = 
    \begin{cases}
      (1 - r_{ij}/r_c)^s, & r_{ij} < r_c,\\
      0, & r_{ij} \geq r_c,
    \end{cases}
\end{equation}
with an exponent $s$. For the conservative force, $W^C(r_{ij}) = W(r_{ij})$ with $s = 1$. 

Time evolution of each DPD particle follows the Newton's second law
\begin{equation}
  \frac{d \textbf{r}_i}{dt} = \textbf{v}_i, \quad \quad m_i \frac{d \textbf{v}_i}{dt} = \sum_{j \neq i}
  \left(\textbf{F}^C(r_{ij}) + \textbf{F}^D(r_{ij}) + \textbf{F}^R(r_{ij}) \right),
\end{equation}
where $m_i$ is the mass of particle $i$. Time integration is performed using the velocity-Verlet algorithm. 

DPD parameters for the interactions between fluid particles and between fluid and wall particles are $a = 200 k_BT / b_z$, 
$\gamma = 80 \sqrt{mk_BT}/b_z$ ($m=1$ in all simulations), $s = 0.15$, and $r_c = b_z/4$. Furthermore, friction coupling of squirmers
to fluid flow is performed using DPD parameters $a=0$, $\gamma = 100 \sqrt{mk_BT}/b_z$, $s = 0.1$, and $r_c = 0.175 b_z$.
The time step for integration is $1.25 \times 10^{-3} b_z \sqrt{m/k_BT}$.

\section*{Author Contributions}

G.G. and D.A.F. designed the research project. B.W.-Z. performed the simulations and analysed the obtained data.
All authors participated in the discussions and writing of the manuscript.

\section*{Conflicts of interest}

There are no conflicts to declare.

\section*{Acknowledgements}

We thank Roland G. Winkler for numerous discussions related to the study. We acknowledge funding from the ETN-PHYMOT 
"Physics of microbial motility" within the European Union’s Horizon 2020 research and innovation programme under 
the Marie Skłodowska-Curie grant agreement No 955910. The authors also gratefully appreciate 
computing time on the supercomputer JURECA \cite{jureca} at Forschungszentrum J{\"u}lich under grant no. actsys.



\begin{thebibliography}{85}%
\makeatletter
\providecommand \@ifxundefined [1]{%
 \@ifx{#1\undefined}
}%
\providecommand \@ifnum [1]{%
 \ifnum #1\expandafter \@firstoftwo
 \else \expandafter \@secondoftwo
 \fi
}%
\providecommand \@ifx [1]{%
 \ifx #1\expandafter \@firstoftwo
 \else \expandafter \@secondoftwo
 \fi
}%
\providecommand \natexlab [1]{#1}%
\providecommand \enquote  [1]{``#1''}%
\providecommand \bibnamefont  [1]{#1}%
\providecommand \bibfnamefont [1]{#1}%
\providecommand \citenamefont [1]{#1}%
\providecommand \href@noop [0]{\@secondoftwo}%
\providecommand \href [0]{\begingroup \@sanitize@url \@href}%
\providecommand \@href[1]{\@@startlink{#1}\@@href}%
\providecommand \@@href[1]{\endgroup#1\@@endlink}%
\providecommand \@sanitize@url [0]{\catcode `\\12\catcode `\$12\catcode
  `\&12\catcode `\#12\catcode `\^12\catcode `\_12\catcode `\%12\relax}%
\providecommand \@@startlink[1]{}%
\providecommand \@@endlink[0]{}%
\providecommand \url  [0]{\begingroup\@sanitize@url \@url }%
\providecommand \@url [1]{\endgroup\@href {#1}{\urlprefix }}%
\providecommand \urlprefix  [0]{URL }%
\providecommand \Eprint [0]{\href }%
\providecommand \doibase [0]{https://doi.org/}%
\providecommand \selectlanguage [0]{\@gobble}%
\providecommand \bibinfo  [0]{\@secondoftwo}%
\providecommand \bibfield  [0]{\@secondoftwo}%
\providecommand \translation [1]{[#1]}%
\providecommand \BibitemOpen [0]{}%
\providecommand \bibitemStop [0]{}%
\providecommand \bibitemNoStop [0]{.\EOS\space}%
\providecommand \EOS [0]{\spacefactor3000\relax}%
\providecommand \BibitemShut  [1]{\csname bibitem#1\endcsname}%
\let\auto@bib@innerbib\@empty
\bibitem [{\citenamefont {Elgeti}\ \emph {et~al.}(2015)\citenamefont {Elgeti},
  \citenamefont {Winkler},\ and\ \citenamefont {Gompper}}]{Elgeti_PMS_2015}%
  \BibitemOpen
  \bibfield  {author} {\bibinfo {author} {\bibfnamefont {J.}~\bibnamefont
  {Elgeti}}, \bibinfo {author} {\bibfnamefont {R.~G.}\ \bibnamefont
  {Winkler}},\ and\ \bibinfo {author} {\bibfnamefont {G.}~\bibnamefont
  {Gompper}},\ }\bibfield  {title} {\bibinfo {title} {Physics of microswimmers
  - single particle motion and collective behavior: a review},\ }\href@noop {}
  {\bibfield  {journal} {\bibinfo  {journal} {Rep. Prog. Phys.}\ }\textbf
  {\bibinfo {volume} {78}},\ \bibinfo {pages} {056601} (\bibinfo {year}
  {2015})}\BibitemShut {NoStop}%
\bibitem [{\citenamefont {Bechinger}\ \emph {et~al.}(2016)\citenamefont
  {Bechinger}, \citenamefont {Di~Leonardo}, \citenamefont {L{\"o}wen},
  \citenamefont {Reichhardt}, \citenamefont {Volpe},\ and\ \citenamefont
  {Volpe}}]{Bechinger_APC_2016}%
  \BibitemOpen
  \bibfield  {author} {\bibinfo {author} {\bibfnamefont {C.}~\bibnamefont
  {Bechinger}}, \bibinfo {author} {\bibfnamefont {R.}~\bibnamefont
  {Di~Leonardo}}, \bibinfo {author} {\bibfnamefont {H.}~\bibnamefont
  {L{\"o}wen}}, \bibinfo {author} {\bibfnamefont {C.}~\bibnamefont
  {Reichhardt}}, \bibinfo {author} {\bibfnamefont {G.}~\bibnamefont {Volpe}},\
  and\ \bibinfo {author} {\bibfnamefont {G.}~\bibnamefont {Volpe}},\ }\bibfield
   {title} {\bibinfo {title} {Active particles in complex and crowded
  environments},\ }\href@noop {} {\bibfield  {journal} {\bibinfo  {journal}
  {Rev. Mod. Phys.}\ }\textbf {\bibinfo {volume} {88}},\ \bibinfo {pages}
  {045006} (\bibinfo {year} {2016})}\BibitemShut {NoStop}%
\bibitem [{\citenamefont {Gompper}\ \emph {et~al.}(2020)\citenamefont
  {Gompper}, \citenamefont {Winkler}, \citenamefont {Speck}, \citenamefont
  {Solon}, \citenamefont {Nardini}, \citenamefont {Peruani}, \citenamefont
  {L{\"o}wen}, \citenamefont {Golestanian}, \citenamefont {Kaupp},
  \citenamefont {Alvarez}, \citenamefont {Kiorboe}, \citenamefont {Lauga},
  \citenamefont {Poon}, \citenamefont {DeSimone}, \citenamefont
  {Mui{\~n}os-Landin}, \citenamefont {Fischer}, \citenamefont {S{\"o}ker},
  \citenamefont {Cichos}, \citenamefont {Kapral}, \citenamefont {Gaspard},
  \citenamefont {Ripoll}, \citenamefont {Sagues}, \citenamefont
  {Doostmohammadi}, \citenamefont {Yeomans}, \citenamefont {Aranson},
  \citenamefont {Bechinger}, \citenamefont {Stark}, \citenamefont {Hemelrijk},
  \citenamefont {Nedelec}, \citenamefont {Sarkar}, \citenamefont {Aryaksama},
  \citenamefont {Lacroix}, \citenamefont {Duclos}, \citenamefont {Yashunsky},
  \citenamefont {Silberzan}, \citenamefont {Arroyo},\ and\ \citenamefont
  {Kale}}]{Gompper_MAM_2020}%
  \BibitemOpen
  \bibfield  {author} {\bibinfo {author} {\bibfnamefont {G.}~\bibnamefont
  {Gompper}}, \bibinfo {author} {\bibfnamefont {R.~G.}\ \bibnamefont
  {Winkler}}, \bibinfo {author} {\bibfnamefont {T.}~\bibnamefont {Speck}},
  \bibinfo {author} {\bibfnamefont {A.}~\bibnamefont {Solon}}, \bibinfo
  {author} {\bibfnamefont {C.}~\bibnamefont {Nardini}}, \bibinfo {author}
  {\bibfnamefont {F.}~\bibnamefont {Peruani}}, \bibinfo {author} {\bibfnamefont
  {H.}~\bibnamefont {L{\"o}wen}}, \bibinfo {author} {\bibfnamefont
  {R.}~\bibnamefont {Golestanian}}, \bibinfo {author} {\bibfnamefont {U.~B.}\
  \bibnamefont {Kaupp}}, \bibinfo {author} {\bibfnamefont {L.}~\bibnamefont
  {Alvarez}}, \bibinfo {author} {\bibfnamefont {T.}~\bibnamefont {Kiorboe}},
  \bibinfo {author} {\bibfnamefont {E.}~\bibnamefont {Lauga}}, \bibinfo
  {author} {\bibfnamefont {W.~C.~K.}\ \bibnamefont {Poon}}, \bibinfo {author}
  {\bibfnamefont {A.}~\bibnamefont {DeSimone}}, \bibinfo {author}
  {\bibfnamefont {S.}~\bibnamefont {Mui{\~n}os-Landin}}, \bibinfo {author}
  {\bibfnamefont {A.}~\bibnamefont {Fischer}}, \bibinfo {author} {\bibfnamefont
  {N.~A.}\ \bibnamefont {S{\"o}ker}}, \bibinfo {author} {\bibfnamefont
  {F.}~\bibnamefont {Cichos}}, \bibinfo {author} {\bibfnamefont
  {R.}~\bibnamefont {Kapral}}, \bibinfo {author} {\bibfnamefont
  {P.}~\bibnamefont {Gaspard}}, \bibinfo {author} {\bibfnamefont
  {M.}~\bibnamefont {Ripoll}}, \bibinfo {author} {\bibfnamefont
  {F.}~\bibnamefont {Sagues}}, \bibinfo {author} {\bibfnamefont
  {A.}~\bibnamefont {Doostmohammadi}}, \bibinfo {author} {\bibfnamefont
  {J.~M.}\ \bibnamefont {Yeomans}}, \bibinfo {author} {\bibfnamefont {I.~S.}\
  \bibnamefont {Aranson}}, \bibinfo {author} {\bibfnamefont {C.}~\bibnamefont
  {Bechinger}}, \bibinfo {author} {\bibfnamefont {H.}~\bibnamefont {Stark}},
  \bibinfo {author} {\bibfnamefont {C.~K.}\ \bibnamefont {Hemelrijk}}, \bibinfo
  {author} {\bibfnamefont {F.~J.}\ \bibnamefont {Nedelec}}, \bibinfo {author}
  {\bibfnamefont {T.}~\bibnamefont {Sarkar}}, \bibinfo {author} {\bibfnamefont
  {T.}~\bibnamefont {Aryaksama}}, \bibinfo {author} {\bibfnamefont
  {M.}~\bibnamefont {Lacroix}}, \bibinfo {author} {\bibfnamefont
  {G.}~\bibnamefont {Duclos}}, \bibinfo {author} {\bibfnamefont
  {V.}~\bibnamefont {Yashunsky}}, \bibinfo {author} {\bibfnamefont
  {P.}~\bibnamefont {Silberzan}}, \bibinfo {author} {\bibfnamefont
  {M.}~\bibnamefont {Arroyo}},\ and\ \bibinfo {author} {\bibfnamefont
  {S.}~\bibnamefont {Kale}},\ }\bibfield  {title} {\bibinfo {title} {The 2020
  motile active matter roadmap},\ }\href@noop {} {\bibfield  {journal}
  {\bibinfo  {journal} {J. Phys.: Condens. Matter}\ }\textbf {\bibinfo {volume}
  {32}},\ \bibinfo {pages} {193001} (\bibinfo {year} {2020})}\BibitemShut
  {NoStop}%
\bibitem [{\citenamefont {Hall-Stoodley}\ \emph {et~al.}(2004)\citenamefont
  {Hall-Stoodley}, \citenamefont {Costerton},\ and\ \citenamefont
  {Stoodley}}]{Hall_BBN_2004}%
  \BibitemOpen
  \bibfield  {author} {\bibinfo {author} {\bibfnamefont {L.}~\bibnamefont
  {Hall-Stoodley}}, \bibinfo {author} {\bibfnamefont {J.~W.}\ \bibnamefont
  {Costerton}},\ and\ \bibinfo {author} {\bibfnamefont {P.}~\bibnamefont
  {Stoodley}},\ }\bibfield  {title} {\bibinfo {title} {Bacterial biofilms: from
  the natural environment to infectious diseases},\ }\href@noop {} {\bibfield
  {journal} {\bibinfo  {journal} {Nat. Rev. Microbiol.}\ }\textbf {\bibinfo
  {volume} {2}},\ \bibinfo {pages} {95} (\bibinfo {year} {2004})}\BibitemShut
  {NoStop}%
\bibitem [{\citenamefont {Verstraeten}\ \emph {et~al.}(2008)\citenamefont
  {Verstraeten}, \citenamefont {Braeken}, \citenamefont {Debkumari},
  \citenamefont {Fauvart}, \citenamefont {Fransaer}, \citenamefont {Vermant},\
  and\ \citenamefont {Michiels}}]{Verstraeten_SBF_2008}%
  \BibitemOpen
  \bibfield  {author} {\bibinfo {author} {\bibfnamefont {N.}~\bibnamefont
  {Verstraeten}}, \bibinfo {author} {\bibfnamefont {K.}~\bibnamefont
  {Braeken}}, \bibinfo {author} {\bibfnamefont {B.}~\bibnamefont {Debkumari}},
  \bibinfo {author} {\bibfnamefont {M.}~\bibnamefont {Fauvart}}, \bibinfo
  {author} {\bibfnamefont {J.}~\bibnamefont {Fransaer}}, \bibinfo {author}
  {\bibfnamefont {J.}~\bibnamefont {Vermant}},\ and\ \bibinfo {author}
  {\bibfnamefont {J.}~\bibnamefont {Michiels}},\ }\bibfield  {title} {\bibinfo
  {title} {Living on a surface: swarming and biofilm formation},\ }\href@noop
  {} {\bibfield  {journal} {\bibinfo  {journal} {Trends Microbiol.}\ }\textbf
  {\bibinfo {volume} {16}},\ \bibinfo {pages} {496} (\bibinfo {year}
  {2008})}\BibitemShut {NoStop}%
\bibitem [{\citenamefont {Marsh}(2006)}]{Marsh_DPB_2006}%
  \BibitemOpen
  \bibfield  {author} {\bibinfo {author} {\bibfnamefont {P.~D.}\ \bibnamefont
  {Marsh}},\ }\bibfield  {title} {\bibinfo {title} {Dental plaque as a biofilm
  and a microbial community--implications for health and disease},\ }\href@noop
  {} {\bibfield  {journal} {\bibinfo  {journal} {BMC Oral Health}\ }\textbf
  {\bibinfo {volume} {6}},\ \bibinfo {pages} {S14} (\bibinfo {year}
  {2006})}\BibitemShut {NoStop}%
\bibitem [{\citenamefont {Attinger}\ and\ \citenamefont
  {Wolcott}(2012)}]{Attinger_cabcw_2012}%
  \BibitemOpen
  \bibfield  {author} {\bibinfo {author} {\bibfnamefont {C.}~\bibnamefont
  {Attinger}}\ and\ \bibinfo {author} {\bibfnamefont {R.}~\bibnamefont
  {Wolcott}},\ }\bibfield  {title} {\bibinfo {title} {Clinically addressing
  biofilm in chronic wounds},\ }\href@noop {} {\bibfield  {journal} {\bibinfo
  {journal} {Adv. Wound Care}\ }\textbf {\bibinfo {volume} {1}},\ \bibinfo
  {pages} {127} (\bibinfo {year} {2012})}\BibitemShut {NoStop}%
\bibitem [{\citenamefont {Mazza}(2016)}]{Mazza_PBI_2016}%
  \BibitemOpen
  \bibfield  {author} {\bibinfo {author} {\bibfnamefont {M.~G.}\ \bibnamefont
  {Mazza}},\ }\bibfield  {title} {\bibinfo {title} {The physics of biofilms-an
  introduction},\ }\href@noop {} {\bibfield  {journal} {\bibinfo  {journal} {J.
  Phys. D: Appl. Phys.}\ }\textbf {\bibinfo {volume} {49}},\ \bibinfo {pages}
  {203001} (\bibinfo {year} {2016})}\BibitemShut {NoStop}%
\bibitem [{\citenamefont {Palacci}\ \emph {et~al.}(2013)\citenamefont
  {Palacci}, \citenamefont {Sacanna}, \citenamefont {Steinberg}, \citenamefont
  {Pine},\ and\ \citenamefont {Chaikin}}]{Palacci_LCL_2013}%
  \BibitemOpen
  \bibfield  {author} {\bibinfo {author} {\bibfnamefont {J.}~\bibnamefont
  {Palacci}}, \bibinfo {author} {\bibfnamefont {S.}~\bibnamefont {Sacanna}},
  \bibinfo {author} {\bibfnamefont {A.~P.}\ \bibnamefont {Steinberg}}, \bibinfo
  {author} {\bibfnamefont {D.~J.}\ \bibnamefont {Pine}},\ and\ \bibinfo
  {author} {\bibfnamefont {P.~M.}\ \bibnamefont {Chaikin}},\ }\bibfield
  {title} {\bibinfo {title} {Living crystals of light-activated colloidal
  surfers},\ }\href@noop {} {\bibfield  {journal} {\bibinfo  {journal}
  {Science}\ }\textbf {\bibinfo {volume} {339}},\ \bibinfo {pages} {936}
  (\bibinfo {year} {2013})}\BibitemShut {NoStop}%
\bibitem [{\citenamefont {Buttinoni}\ \emph {et~al.}(2013)\citenamefont
  {Buttinoni}, \citenamefont {Bialk{\'e}}, \citenamefont {K{\"u}mmel},
  \citenamefont {L{\"o}wen}, \citenamefont {Bechinger},\ and\ \citenamefont
  {Speck}}]{Buttinoni_CPS_2013}%
  \BibitemOpen
  \bibfield  {author} {\bibinfo {author} {\bibfnamefont {I.}~\bibnamefont
  {Buttinoni}}, \bibinfo {author} {\bibfnamefont {J.}~\bibnamefont
  {Bialk{\'e}}}, \bibinfo {author} {\bibfnamefont {F.}~\bibnamefont
  {K{\"u}mmel}}, \bibinfo {author} {\bibfnamefont {H.}~\bibnamefont
  {L{\"o}wen}}, \bibinfo {author} {\bibfnamefont {C.}~\bibnamefont
  {Bechinger}},\ and\ \bibinfo {author} {\bibfnamefont {T.}~\bibnamefont
  {Speck}},\ }\bibfield  {title} {\bibinfo {title} {Dynamical clustering and
  phase separation in suspensions of self-propelled colloidal particles},\
  }\href@noop {} {\bibfield  {journal} {\bibinfo  {journal} {Phys. Rev. Lett.}\
  }\textbf {\bibinfo {volume} {110}},\ \bibinfo {pages} {238301} (\bibinfo
  {year} {2013})}\BibitemShut {NoStop}%
\bibitem [{\citenamefont {B{\"a}uerle}\ \emph {et~al.}(2020)\citenamefont
  {B{\"a}uerle}, \citenamefont {L{\"o}ffler},\ and\ \citenamefont
  {Bechinger}}]{Bauerle_FSS_2020}%
  \BibitemOpen
  \bibfield  {author} {\bibinfo {author} {\bibfnamefont {T.}~\bibnamefont
  {B{\"a}uerle}}, \bibinfo {author} {\bibfnamefont {R.~C.}\ \bibnamefont
  {L{\"o}ffler}},\ and\ \bibinfo {author} {\bibfnamefont {C.}~\bibnamefont
  {Bechinger}},\ }\bibfield  {title} {\bibinfo {title} {Formation of stable and
  responsive collective states in suspensions of active colloids},\ }\href@noop
  {} {\bibfield  {journal} {\bibinfo  {journal} {Nat. Comm.}\ }\textbf
  {\bibinfo {volume} {11}},\ \bibinfo {pages} {2547} (\bibinfo {year}
  {2020})}\BibitemShut {NoStop}%
\bibitem [{\citenamefont {Kr{\"u}ger}\ \emph {et~al.}(2016)\citenamefont
  {Kr{\"u}ger}, \citenamefont {Bahr}, \citenamefont {Herminghaus},\ and\
  \citenamefont {Maass}}]{Krueger_CBAE_2016}%
  \BibitemOpen
  \bibfield  {author} {\bibinfo {author} {\bibfnamefont {C.}~\bibnamefont
  {Kr{\"u}ger}}, \bibinfo {author} {\bibfnamefont {C.}~\bibnamefont {Bahr}},
  \bibinfo {author} {\bibfnamefont {S.}~\bibnamefont {Herminghaus}},\ and\
  \bibinfo {author} {\bibfnamefont {C.~C.}\ \bibnamefont {Maass}},\ }\bibfield
  {title} {\bibinfo {title} {Dimensionality matters in the collective behaviour
  of active emulsions},\ }\href@noop {} {\bibfield  {journal} {\bibinfo
  {journal} {Eur. Phys. J. E}\ }\textbf {\bibinfo {volume} {39}},\ \bibinfo
  {pages} {64} (\bibinfo {year} {2016})}\BibitemShut {NoStop}%
\bibitem [{\citenamefont {Thutupalli}\ \emph {et~al.}(2018)\citenamefont
  {Thutupalli}, \citenamefont {Geyer}, \citenamefont {Singh}, \citenamefont
  {Adhikari},\ and\ \citenamefont {Stone}}]{Thutupalli_FIPS_2018}%
  \BibitemOpen
  \bibfield  {author} {\bibinfo {author} {\bibfnamefont {S.}~\bibnamefont
  {Thutupalli}}, \bibinfo {author} {\bibfnamefont {D.}~\bibnamefont {Geyer}},
  \bibinfo {author} {\bibfnamefont {R.}~\bibnamefont {Singh}}, \bibinfo
  {author} {\bibfnamefont {R.}~\bibnamefont {Adhikari}},\ and\ \bibinfo
  {author} {\bibfnamefont {H.~A.}\ \bibnamefont {Stone}},\ }\bibfield  {title}
  {\bibinfo {title} {Flow-induced phase separation of active particles is
  controlled by boundary conditions},\ }\href@noop {} {\bibfield  {journal}
  {\bibinfo  {journal} {Proc. Natl. Acad. Sci. USA}\ }\textbf {\bibinfo
  {volume} {115}},\ \bibinfo {pages} {5403} (\bibinfo {year}
  {2018})}\BibitemShut {NoStop}%
\bibitem [{\citenamefont {Bricard}\ \emph {et~al.}(2013)\citenamefont
  {Bricard}, \citenamefont {Caussin}, \citenamefont {Desreumaux}, \citenamefont
  {Dauchot},\ and\ \citenamefont {Bartolo}}]{Bricard_PMC_2013}%
  \BibitemOpen
  \bibfield  {author} {\bibinfo {author} {\bibfnamefont {A.}~\bibnamefont
  {Bricard}}, \bibinfo {author} {\bibfnamefont {J.-B.}\ \bibnamefont
  {Caussin}}, \bibinfo {author} {\bibfnamefont {N.}~\bibnamefont {Desreumaux}},
  \bibinfo {author} {\bibfnamefont {O.}~\bibnamefont {Dauchot}},\ and\ \bibinfo
  {author} {\bibfnamefont {D.}~\bibnamefont {Bartolo}},\ }\bibfield  {title}
  {\bibinfo {title} {Emergence of macroscopic directed motion in populations of
  motile colloids},\ }\href@noop {} {\bibfield  {journal} {\bibinfo  {journal}
  {Nature}\ }\textbf {\bibinfo {volume} {503}},\ \bibinfo {pages} {95}
  (\bibinfo {year} {2013})}\BibitemShut {NoStop}%
\bibitem [{\citenamefont {Fily}\ and\ \citenamefont
  {Marchetti}(2012)}]{Fily_APS_2012}%
  \BibitemOpen
  \bibfield  {author} {\bibinfo {author} {\bibfnamefont {Y.}~\bibnamefont
  {Fily}}\ and\ \bibinfo {author} {\bibfnamefont {M.~C.}\ \bibnamefont
  {Marchetti}},\ }\bibfield  {title} {\bibinfo {title} {Athermal phase
  separation of self-propelled particles with no alignment},\ }\href@noop {}
  {\bibfield  {journal} {\bibinfo  {journal} {Phys. Rev. Lett.}\ }\textbf
  {\bibinfo {volume} {108}},\ \bibinfo {pages} {235702} (\bibinfo {year}
  {2012})}\BibitemShut {NoStop}%
\bibitem [{\citenamefont {Bialk{\'e}}\ \emph {et~al.}(2013)\citenamefont
  {Bialk{\'e}}, \citenamefont {L{\"o}wen},\ and\ \citenamefont
  {Speck}}]{Bialke_MIPS_2013}%
  \BibitemOpen
  \bibfield  {author} {\bibinfo {author} {\bibfnamefont {J.}~\bibnamefont
  {Bialk{\'e}}}, \bibinfo {author} {\bibfnamefont {H.}~\bibnamefont
  {L{\"o}wen}},\ and\ \bibinfo {author} {\bibfnamefont {T.}~\bibnamefont
  {Speck}},\ }\bibfield  {title} {\bibinfo {title} {Microscopic theory for the
  phase separation of self-propelled repulsive disks},\ }\href@noop {}
  {\bibfield  {journal} {\bibinfo  {journal} {Europhys. Lett.}\ }\textbf
  {\bibinfo {volume} {103}},\ \bibinfo {pages} {30008} (\bibinfo {year}
  {2013})}\BibitemShut {NoStop}%
\bibitem [{\citenamefont {Redner}\ \emph {et~al.}(2013)\citenamefont {Redner},
  \citenamefont {Hagan},\ and\ \citenamefont {Baskaran}}]{Redner_MIPS_2013}%
  \BibitemOpen
  \bibfield  {author} {\bibinfo {author} {\bibfnamefont {G.~S.}\ \bibnamefont
  {Redner}}, \bibinfo {author} {\bibfnamefont {M.~F.}\ \bibnamefont {Hagan}},\
  and\ \bibinfo {author} {\bibfnamefont {A.}~\bibnamefont {Baskaran}},\
  }\bibfield  {title} {\bibinfo {title} {Structure and dynamics of a
  phase-separating active colloidal fluid},\ }\href@noop {} {\bibfield
  {journal} {\bibinfo  {journal} {Phys. Rev. Lett.}\ }\textbf {\bibinfo
  {volume} {110}},\ \bibinfo {pages} {055701} (\bibinfo {year}
  {2013})}\BibitemShut {NoStop}%
\bibitem [{\citenamefont {Cates}\ and\ \citenamefont
  {Tailleur}(2015)}]{Cates_MIPS_2015}%
  \BibitemOpen
  \bibfield  {author} {\bibinfo {author} {\bibfnamefont {M.~E.}\ \bibnamefont
  {Cates}}\ and\ \bibinfo {author} {\bibfnamefont {J.}~\bibnamefont
  {Tailleur}},\ }\bibfield  {title} {\bibinfo {title} {Motility-induced phase
  separation},\ }\href@noop {} {\bibfield  {journal} {\bibinfo  {journal}
  {Annu. Rev. Condens. Matter Phys.}\ }\textbf {\bibinfo {volume} {6}},\
  \bibinfo {pages} {219} (\bibinfo {year} {2015})}\BibitemShut {NoStop}%
\bibitem [{\citenamefont {Digregorio}\ \emph {et~al.}(2018)\citenamefont
  {Digregorio}, \citenamefont {Levis}, \citenamefont {Suma}, \citenamefont
  {Cugliandolo}, \citenamefont {Gonnella},\ and\ \citenamefont
  {Pagonabarraga}}]{Digregorio_ABD_2018}%
  \BibitemOpen
  \bibfield  {author} {\bibinfo {author} {\bibfnamefont {P.}~\bibnamefont
  {Digregorio}}, \bibinfo {author} {\bibfnamefont {D.}~\bibnamefont {Levis}},
  \bibinfo {author} {\bibfnamefont {A.}~\bibnamefont {Suma}}, \bibinfo {author}
  {\bibfnamefont {L.~F.}\ \bibnamefont {Cugliandolo}}, \bibinfo {author}
  {\bibfnamefont {G.}~\bibnamefont {Gonnella}},\ and\ \bibinfo {author}
  {\bibfnamefont {I.}~\bibnamefont {Pagonabarraga}},\ }\bibfield  {title}
  {\bibinfo {title} {Full phase diagram of active {B}rownian disks: from
  melting to motility-induced phase separation},\ }\href@noop {} {\bibfield
  {journal} {\bibinfo  {journal} {Phys. Rev. Lett.}\ }\textbf {\bibinfo
  {volume} {121}},\ \bibinfo {pages} {098003} (\bibinfo {year}
  {2018})}\BibitemShut {NoStop}%
\bibitem [{\citenamefont {Liu}\ \emph {et~al.}(2019)\citenamefont {Liu},
  \citenamefont {Patch}, \citenamefont {Bahar}, \citenamefont {Yllanes},
  \citenamefont {Welch}, \citenamefont {Marchetti}, \citenamefont
  {Thutupalli},\ and\ \citenamefont {Shaevitz}}]{Liu_SDT_2019}%
  \BibitemOpen
  \bibfield  {author} {\bibinfo {author} {\bibfnamefont {G.}~\bibnamefont
  {Liu}}, \bibinfo {author} {\bibfnamefont {A.}~\bibnamefont {Patch}}, \bibinfo
  {author} {\bibfnamefont {F.}~\bibnamefont {Bahar}}, \bibinfo {author}
  {\bibfnamefont {D.}~\bibnamefont {Yllanes}}, \bibinfo {author} {\bibfnamefont
  {R.~D.}\ \bibnamefont {Welch}}, \bibinfo {author} {\bibfnamefont {M.~C.}\
  \bibnamefont {Marchetti}}, \bibinfo {author} {\bibfnamefont {S.}~\bibnamefont
  {Thutupalli}},\ and\ \bibinfo {author} {\bibfnamefont {J.~W.}\ \bibnamefont
  {Shaevitz}},\ }\bibfield  {title} {\bibinfo {title} {Self-driven phase
  transitions drive {M}yxococcus xanthus fruiting body formation},\ }\href@noop
  {} {\bibfield  {journal} {\bibinfo  {journal} {Phys. Rev. Lett.}\ }\textbf
  {\bibinfo {volume} {122}},\ \bibinfo {pages} {248102} (\bibinfo {year}
  {2019})}\BibitemShut {NoStop}%
\bibitem [{\citenamefont {Be'er}\ \emph {et~al.}(2020)\citenamefont {Be'er},
  \citenamefont {Ilkanaiv}, \citenamefont {Gross}, \citenamefont {Kearns},
  \citenamefont {Heidenreich}, \citenamefont {B{\"a}r},\ and\ \citenamefont
  {Ariel}}]{Beer_PDBS_2020}%
  \BibitemOpen
  \bibfield  {author} {\bibinfo {author} {\bibfnamefont {A.}~\bibnamefont
  {Be'er}}, \bibinfo {author} {\bibfnamefont {B.}~\bibnamefont {Ilkanaiv}},
  \bibinfo {author} {\bibfnamefont {R.}~\bibnamefont {Gross}}, \bibinfo
  {author} {\bibfnamefont {D.~B.}\ \bibnamefont {Kearns}}, \bibinfo {author}
  {\bibfnamefont {S.}~\bibnamefont {Heidenreich}}, \bibinfo {author}
  {\bibfnamefont {M.}~\bibnamefont {B{\"a}r}},\ and\ \bibinfo {author}
  {\bibfnamefont {G.}~\bibnamefont {Ariel}},\ }\bibfield  {title} {\bibinfo
  {title} {A phase diagram for bacterial swarming},\ }\href@noop {} {\bibfield
  {journal} {\bibinfo  {journal} {Commun. Phys.}\ }\textbf {\bibinfo {volume}
  {3}},\ \bibinfo {pages} {66} (\bibinfo {year} {2020})}\BibitemShut {NoStop}%
\bibitem [{\citenamefont {Peruani}\ \emph {et~al.}(2006)\citenamefont
  {Peruani}, \citenamefont {Deutsch},\ and\ \citenamefont
  {B{\"a}r}}]{Peruani_CSPR_2006}%
  \BibitemOpen
  \bibfield  {author} {\bibinfo {author} {\bibfnamefont {F.}~\bibnamefont
  {Peruani}}, \bibinfo {author} {\bibfnamefont {A.}~\bibnamefont {Deutsch}},\
  and\ \bibinfo {author} {\bibfnamefont {M.}~\bibnamefont {B{\"a}r}},\
  }\bibfield  {title} {\bibinfo {title} {Nonequilibrium clustering of
  self-propelled rods},\ }\href@noop {} {\bibfield  {journal} {\bibinfo
  {journal} {Phys. Rev. E}\ }\textbf {\bibinfo {volume} {74}},\ \bibinfo
  {pages} {030904} (\bibinfo {year} {2006})}\BibitemShut {NoStop}%
\bibitem [{\citenamefont {Vicsek}\ and\ \citenamefont
  {Zafeiris}(2012)}]{Vicsek_CM_2012}%
  \BibitemOpen
  \bibfield  {author} {\bibinfo {author} {\bibfnamefont {T.}~\bibnamefont
  {Vicsek}}\ and\ \bibinfo {author} {\bibfnamefont {A.}~\bibnamefont
  {Zafeiris}},\ }\bibfield  {title} {\bibinfo {title} {Collective motion},\
  }\href@noop {} {\bibfield  {journal} {\bibinfo  {journal} {Phys. Rep.}\
  }\textbf {\bibinfo {volume} {517}},\ \bibinfo {pages} {71} (\bibinfo {year}
  {2012})}\BibitemShut {NoStop}%
\bibitem [{\citenamefont {Wioland}\ \emph {et~al.}(2013)\citenamefont
  {Wioland}, \citenamefont {Woodhouse}, \citenamefont {Dunkel}, \citenamefont
  {Kessler},\ and\ \citenamefont {Goldstein}}]{Wioland_CBSSV_2013}%
  \BibitemOpen
  \bibfield  {author} {\bibinfo {author} {\bibfnamefont {H.}~\bibnamefont
  {Wioland}}, \bibinfo {author} {\bibfnamefont {F.~G.}\ \bibnamefont
  {Woodhouse}}, \bibinfo {author} {\bibfnamefont {J.}~\bibnamefont {Dunkel}},
  \bibinfo {author} {\bibfnamefont {J.~O.}\ \bibnamefont {Kessler}},\ and\
  \bibinfo {author} {\bibfnamefont {R.~E.}\ \bibnamefont {Goldstein}},\
  }\bibfield  {title} {\bibinfo {title} {Confinement stabilizes a bacterial
  suspension into a spiral vortex},\ }\href@noop {} {\bibfield  {journal}
  {\bibinfo  {journal} {Phys. Rev. Lett.}\ }\textbf {\bibinfo {volume} {110}},\
  \bibinfo {pages} {268102} (\bibinfo {year} {2013})}\BibitemShut {NoStop}%
\bibitem [{\citenamefont {Martinez}\ \emph {et~al.}(2020)\citenamefont
  {Martinez}, \citenamefont {Cl{\'e}ment}, \citenamefont {Arlt}, \citenamefont
  {Douarche}, \citenamefont {Dawson}, \citenamefont {Schwarz-Linek},
  \citenamefont {Creppy}, \citenamefont {Skult{\'e}ty}, \citenamefont
  {Morozov}, \citenamefont {Auradou},\ and\ \citenamefont
  {Poon}}]{Martinez_RBS_2020}%
  \BibitemOpen
  \bibfield  {author} {\bibinfo {author} {\bibfnamefont {V.~A.}\ \bibnamefont
  {Martinez}}, \bibinfo {author} {\bibfnamefont {E.}~\bibnamefont
  {Cl{\'e}ment}}, \bibinfo {author} {\bibfnamefont {J.}~\bibnamefont {Arlt}},
  \bibinfo {author} {\bibfnamefont {C.}~\bibnamefont {Douarche}}, \bibinfo
  {author} {\bibfnamefont {A.}~\bibnamefont {Dawson}}, \bibinfo {author}
  {\bibfnamefont {J.}~\bibnamefont {Schwarz-Linek}}, \bibinfo {author}
  {\bibfnamefont {A.~K.}\ \bibnamefont {Creppy}}, \bibinfo {author}
  {\bibfnamefont {V.}~\bibnamefont {Skult{\'e}ty}}, \bibinfo {author}
  {\bibfnamefont {A.~N.}\ \bibnamefont {Morozov}}, \bibinfo {author}
  {\bibfnamefont {H.}~\bibnamefont {Auradou}},\ and\ \bibinfo {author}
  {\bibfnamefont {W.~C.~K.}\ \bibnamefont {Poon}},\ }\bibfield  {title}
  {\bibinfo {title} {A combined rheometry and imaging study of viscosity
  reduction in bacterial suspensions},\ }\href@noop {} {\bibfield  {journal}
  {\bibinfo  {journal} {Proc. Natl. Acad. Sci. USA}\ }\textbf {\bibinfo
  {volume} {117}},\ \bibinfo {pages} {2326} (\bibinfo {year}
  {2020})}\BibitemShut {NoStop}%
\bibitem [{\citenamefont {Guo}\ \emph {et~al.}(2018)\citenamefont {Guo},
  \citenamefont {Samanta}, \citenamefont {Peng}, \citenamefont {Xu},\ and\
  \citenamefont {Cheng}}]{Guo_SVB_2018}%
  \BibitemOpen
  \bibfield  {author} {\bibinfo {author} {\bibfnamefont {S.}~\bibnamefont
  {Guo}}, \bibinfo {author} {\bibfnamefont {D.}~\bibnamefont {Samanta}},
  \bibinfo {author} {\bibfnamefont {Y.}~\bibnamefont {Peng}}, \bibinfo {author}
  {\bibfnamefont {X.}~\bibnamefont {Xu}},\ and\ \bibinfo {author}
  {\bibfnamefont {X.}~\bibnamefont {Cheng}},\ }\bibfield  {title} {\bibinfo
  {title} {Symmetric shear banding and swarming vortices in bacterial
  superfluids},\ }\href@noop {} {\bibfield  {journal} {\bibinfo  {journal}
  {Proc. Natl. Acad. Sci. USA}\ }\textbf {\bibinfo {volume} {115}},\ \bibinfo
  {pages} {7212} (\bibinfo {year} {2018})}\BibitemShut {NoStop}%
\bibitem [{\citenamefont {Kraichnan}\ and\ \citenamefont
  {Montgomery}(1980)}]{Kraichnan_TDT_1980}%
  \BibitemOpen
  \bibfield  {author} {\bibinfo {author} {\bibfnamefont {R.~H.}\ \bibnamefont
  {Kraichnan}}\ and\ \bibinfo {author} {\bibfnamefont {D.}~\bibnamefont
  {Montgomery}},\ }\bibfield  {title} {\bibinfo {title} {Two-dimensional
  turbulence},\ }\href@noop {} {\bibfield  {journal} {\bibinfo  {journal} {Rep.
  Prog. Phys.}\ }\textbf {\bibinfo {volume} {43}},\ \bibinfo {pages} {547}
  (\bibinfo {year} {1980})}\BibitemShut {NoStop}%
\bibitem [{\citenamefont {Wensink}\ \emph {et~al.}(2012)\citenamefont
  {Wensink}, \citenamefont {Dunkel}, \citenamefont {Heidenreich}, \citenamefont
  {Drescher}, \citenamefont {Goldstein}, \citenamefont {L{\"o}wen},\ and\
  \citenamefont {Yeomans}}]{Wensink_MST_2012}%
  \BibitemOpen
  \bibfield  {author} {\bibinfo {author} {\bibfnamefont {H.~H.}\ \bibnamefont
  {Wensink}}, \bibinfo {author} {\bibfnamefont {J.}~\bibnamefont {Dunkel}},
  \bibinfo {author} {\bibfnamefont {S.}~\bibnamefont {Heidenreich}}, \bibinfo
  {author} {\bibfnamefont {K.}~\bibnamefont {Drescher}}, \bibinfo {author}
  {\bibfnamefont {R.~E.}\ \bibnamefont {Goldstein}}, \bibinfo {author}
  {\bibfnamefont {H.}~\bibnamefont {L{\"o}wen}},\ and\ \bibinfo {author}
  {\bibfnamefont {J.~M.}\ \bibnamefont {Yeomans}},\ }\bibfield  {title}
  {\bibinfo {title} {Meso-scale turbulence in living fluids},\ }\href@noop {}
  {\bibfield  {journal} {\bibinfo  {journal} {Proc. Natl. Acad. Sci. USA}\
  }\textbf {\bibinfo {volume} {109}},\ \bibinfo {pages} {14308} (\bibinfo
  {year} {2012})}\BibitemShut {NoStop}%
\bibitem [{\citenamefont {Stenhammar}\ \emph {et~al.}(2017)\citenamefont
  {Stenhammar}, \citenamefont {Nardini}, \citenamefont {Nash}, \citenamefont
  {Marenduzzo},\ and\ \citenamefont {Morozov}}]{Stenhammar_CBM_2017}%
  \BibitemOpen
  \bibfield  {author} {\bibinfo {author} {\bibfnamefont {J.}~\bibnamefont
  {Stenhammar}}, \bibinfo {author} {\bibfnamefont {C.}~\bibnamefont {Nardini}},
  \bibinfo {author} {\bibfnamefont {R.~W.}\ \bibnamefont {Nash}}, \bibinfo
  {author} {\bibfnamefont {D.}~\bibnamefont {Marenduzzo}},\ and\ \bibinfo
  {author} {\bibfnamefont {A.}~\bibnamefont {Morozov}},\ }\bibfield  {title}
  {\bibinfo {title} {Role of correlations in the collective behavior of
  microswimmer suspensions},\ }\href@noop {} {\bibfield  {journal} {\bibinfo
  {journal} {Phys. Rev. Lett}\ }\textbf {\bibinfo {volume} {119}},\ \bibinfo
  {pages} {028005} (\bibinfo {year} {2017})}\BibitemShut {NoStop}%
\bibitem [{\citenamefont {B{\'a}rdfalvy}\ \emph {et~al.}(2019)\citenamefont
  {B{\'a}rdfalvy}, \citenamefont {Nordanger}, \citenamefont {Nardini},
  \citenamefont {Morozov},\ and\ \citenamefont
  {Stenhammar}}]{Bardfalvy_LB_AT_2019}%
  \BibitemOpen
  \bibfield  {author} {\bibinfo {author} {\bibfnamefont {D.}~\bibnamefont
  {B{\'a}rdfalvy}}, \bibinfo {author} {\bibfnamefont {H.}~\bibnamefont
  {Nordanger}}, \bibinfo {author} {\bibfnamefont {C.}~\bibnamefont {Nardini}},
  \bibinfo {author} {\bibfnamefont {A.}~\bibnamefont {Morozov}},\ and\ \bibinfo
  {author} {\bibfnamefont {J.}~\bibnamefont {Stenhammar}},\ }\bibfield  {title}
  {\bibinfo {title} {Particle-resolved lattice {B}oltzmann simulations of
  3-dimensional active turbulence},\ }\href@noop {} {\bibfield  {journal}
  {\bibinfo  {journal} {Soft Matter}\ }\textbf {\bibinfo {volume} {15}},\
  \bibinfo {pages} {7747} (\bibinfo {year} {2019})}\BibitemShut {NoStop}%
\bibitem [{\citenamefont {Qi}\ \emph {et~al.}(2022)\citenamefont {Qi},
  \citenamefont {Westphal}, \citenamefont {Gompper},\ and\ \citenamefont
  {Winkler}}]{Qi_EAT_2022}%
  \BibitemOpen
  \bibfield  {author} {\bibinfo {author} {\bibfnamefont {K.}~\bibnamefont
  {Qi}}, \bibinfo {author} {\bibfnamefont {E.}~\bibnamefont {Westphal}},
  \bibinfo {author} {\bibfnamefont {G.}~\bibnamefont {Gompper}},\ and\ \bibinfo
  {author} {\bibfnamefont {R.~G.}\ \bibnamefont {Winkler}},\ }\bibfield
  {title} {\bibinfo {title} {Emergence of active turbulence in microswimmer
  suspensions due to active hydrodynamic stress and volume exclusion},\
  }\href@noop {} {\bibfield  {journal} {\bibinfo  {journal} {Commun. Phys.}\
  }\textbf {\bibinfo {volume} {5}},\ \bibinfo {pages} {49} (\bibinfo {year}
  {2022})}\BibitemShut {NoStop}%
\bibitem [{\citenamefont {Lushi}\ \emph {et~al.}(2014)\citenamefont {Lushi},
  \citenamefont {Wioland},\ and\ \citenamefont
  {Goldstein}}]{Lushi_FLSOCS_2014}%
  \BibitemOpen
  \bibfield  {author} {\bibinfo {author} {\bibfnamefont {E.}~\bibnamefont
  {Lushi}}, \bibinfo {author} {\bibfnamefont {H.}~\bibnamefont {Wioland}},\
  and\ \bibinfo {author} {\bibfnamefont {R.~E.}\ \bibnamefont {Goldstein}},\
  }\bibfield  {title} {\bibinfo {title} {Fluid flows created by swimming
  bacteria drive self-organization in confined suspensions},\ }\href@noop {}
  {\bibfield  {journal} {\bibinfo  {journal} {Proc. Natl. Acad. Sci. USA}\
  }\textbf {\bibinfo {volume} {111}},\ \bibinfo {pages} {9733} (\bibinfo {year}
  {2014})}\BibitemShut {NoStop}%
\bibitem [{\citenamefont {Theillard}\ \emph {et~al.}(2017)\citenamefont
  {Theillard}, \citenamefont {Alonso-Matillaa},\ and\ \citenamefont
  {Saintillan}}]{Theillard_ACM_2017}%
  \BibitemOpen
  \bibfield  {author} {\bibinfo {author} {\bibfnamefont {M.}~\bibnamefont
  {Theillard}}, \bibinfo {author} {\bibfnamefont {R.}~\bibnamefont
  {Alonso-Matillaa}},\ and\ \bibinfo {author} {\bibfnamefont {D.}~\bibnamefont
  {Saintillan}},\ }\bibfield  {title} {\bibinfo {title} {Geometric control of
  active collective motion},\ }\href@noop {} {\bibfield  {journal} {\bibinfo
  {journal} {Soft Matter}\ }\textbf {\bibinfo {volume} {13}},\ \bibinfo {pages}
  {363} (\bibinfo {year} {2017})}\BibitemShut {NoStop}%
\bibitem [{\citenamefont {Suma}\ \emph {et~al.}(2014)\citenamefont {Suma},
  \citenamefont {Gonnella}, \citenamefont {Marenduzzo},\ and\ \citenamefont
  {Orlandini}}]{Suma_MIPS_2014}%
  \BibitemOpen
  \bibfield  {author} {\bibinfo {author} {\bibfnamefont {A.}~\bibnamefont
  {Suma}}, \bibinfo {author} {\bibfnamefont {G.}~\bibnamefont {Gonnella}},
  \bibinfo {author} {\bibfnamefont {D.}~\bibnamefont {Marenduzzo}},\ and\
  \bibinfo {author} {\bibfnamefont {E.}~\bibnamefont {Orlandini}},\ }\bibfield
  {title} {\bibinfo {title} {Motility-induced phase separation in an active
  dumbbell fluid},\ }\href@noop {} {\bibfield  {journal} {\bibinfo  {journal}
  {Europhys. Lett.}\ }\textbf {\bibinfo {volume} {108}},\ \bibinfo {pages}
  {56004} (\bibinfo {year} {2014})}\BibitemShut {NoStop}%
\bibitem [{\citenamefont {B{\"a}r}\ \emph {et~al.}(2020)\citenamefont
  {B{\"a}r}, \citenamefont {Gro\ss{}mann}, \citenamefont {Heidenreich},\ and\
  \citenamefont {Peruani}}]{Baer_SPR_2020}%
  \BibitemOpen
  \bibfield  {author} {\bibinfo {author} {\bibfnamefont {M.}~\bibnamefont
  {B{\"a}r}}, \bibinfo {author} {\bibfnamefont {R.}~\bibnamefont
  {Gro\ss{}mann}}, \bibinfo {author} {\bibfnamefont {S.}~\bibnamefont
  {Heidenreich}},\ and\ \bibinfo {author} {\bibfnamefont {F.}~\bibnamefont
  {Peruani}},\ }\bibfield  {title} {\bibinfo {title} {Self-propelled rods:
  insights and perspectives for active matter},\ }\href@noop {} {\bibfield
  {journal} {\bibinfo  {journal} {Annu. Rev. Condens. Matter Phys.}\ }\textbf
  {\bibinfo {volume} {11}},\ \bibinfo {pages} {441} (\bibinfo {year}
  {2020})}\BibitemShut {NoStop}%
\bibitem [{\citenamefont {Moran}\ \emph {et~al.}(2022)\citenamefont {Moran},
  \citenamefont {Bruss}, \citenamefont {Sch{\"o}nh{\"o}fer},\ and\
  \citenamefont {Glotzer}}]{Moran_PAT_2022}%
  \BibitemOpen
  \bibfield  {author} {\bibinfo {author} {\bibfnamefont {S.~E.}\ \bibnamefont
  {Moran}}, \bibinfo {author} {\bibfnamefont {I.~R.}\ \bibnamefont {Bruss}},
  \bibinfo {author} {\bibfnamefont {P.~W.~A.}\ \bibnamefont
  {Sch{\"o}nh{\"o}fer}},\ and\ \bibinfo {author} {\bibfnamefont {S.~C.}\
  \bibnamefont {Glotzer}},\ }\bibfield  {title} {\bibinfo {title} {Particle
  anisotropy tunes emergent behavior in active colloidal systems},\ }\href@noop
  {} {\bibfield  {journal} {\bibinfo  {journal} {Soft Matter}\ }\textbf
  {\bibinfo {volume} {18}},\ \bibinfo {pages} {1044} (\bibinfo {year}
  {2022})}\BibitemShut {NoStop}%
\bibitem [{\citenamefont {Alarc\'{o}n}\ and\ \citenamefont
  {Pagonabarraga}(2013)}]{Alarcon_SAO_2013}%
  \BibitemOpen
  \bibfield  {author} {\bibinfo {author} {\bibfnamefont {F.}~\bibnamefont
  {Alarc\'{o}n}}\ and\ \bibinfo {author} {\bibfnamefont {I.}~\bibnamefont
  {Pagonabarraga}},\ }\bibfield  {title} {\bibinfo {title} {Spontaneous
  aggregation and global polar ordering in squirmer suspensions},\ }\href@noop
  {} {\bibfield  {journal} {\bibinfo  {journal} {J. Mol. Liquids}\ }\textbf
  {\bibinfo {volume} {185}},\ \bibinfo {pages} {56} (\bibinfo {year}
  {2013})}\BibitemShut {NoStop}%
\bibitem [{\citenamefont {Theers}\ \emph {et~al.}(2018)\citenamefont {Theers},
  \citenamefont {Westphal}, \citenamefont {Qi}, \citenamefont {Winkler},\ and\
  \citenamefont {Gompper}}]{Theers_CMS_2018}%
  \BibitemOpen
  \bibfield  {author} {\bibinfo {author} {\bibfnamefont {M.}~\bibnamefont
  {Theers}}, \bibinfo {author} {\bibfnamefont {E.}~\bibnamefont {Westphal}},
  \bibinfo {author} {\bibfnamefont {K.}~\bibnamefont {Qi}}, \bibinfo {author}
  {\bibfnamefont {R.~G.}\ \bibnamefont {Winkler}},\ and\ \bibinfo {author}
  {\bibfnamefont {G.}~\bibnamefont {Gompper}},\ }\bibfield  {title} {\bibinfo
  {title} {Clustering of microswimmers: interplay of shape and hydrodynamics},\
  }\href@noop {} {\bibfield  {journal} {\bibinfo  {journal} {Soft Matter}\
  }\textbf {\bibinfo {volume} {14}},\ \bibinfo {pages} {8590} (\bibinfo {year}
  {2018})}\BibitemShut {NoStop}%
\bibitem [{\citenamefont {Z\"{o}ttl}\ and\ \citenamefont
  {Stark}(2014)}]{Zoettl_HCM_2014}%
  \BibitemOpen
  \bibfield  {author} {\bibinfo {author} {\bibfnamefont {A.}~\bibnamefont
  {Z\"{o}ttl}}\ and\ \bibinfo {author} {\bibfnamefont {H.}~\bibnamefont
  {Stark}},\ }\bibfield  {title} {\bibinfo {title} {Hydrodynamics determines
  collective motion and phase behavior of active colloids in
  quasi-two-dimensional confinement},\ }\href@noop {} {\bibfield  {journal}
  {\bibinfo  {journal} {Phys. Rev. Lett.}\ }\textbf {\bibinfo {volume} {112}},\
  \bibinfo {pages} {118101} (\bibinfo {year} {2014})}\BibitemShut {NoStop}%
\bibitem [{\citenamefont {Matas-Navarro}\ \emph {et~al.}(2014)\citenamefont
  {Matas-Navarro}, \citenamefont {Golestanian}, \citenamefont {Liverpool},\
  and\ \citenamefont {Fielding}}]{Matas_Navarro_HPS_2014}%
  \BibitemOpen
  \bibfield  {author} {\bibinfo {author} {\bibfnamefont {R.}~\bibnamefont
  {Matas-Navarro}}, \bibinfo {author} {\bibfnamefont {R.}~\bibnamefont
  {Golestanian}}, \bibinfo {author} {\bibfnamefont {T.~B.}\ \bibnamefont
  {Liverpool}},\ and\ \bibinfo {author} {\bibfnamefont {S.~M.}\ \bibnamefont
  {Fielding}},\ }\bibfield  {title} {\bibinfo {title} {Hydrodynamic suppression
  of phase separation in active suspensions},\ }\href@noop {} {\bibfield
  {journal} {\bibinfo  {journal} {Phys. Rev. E}\ }\textbf {\bibinfo {volume}
  {90}},\ \bibinfo {pages} {032304} (\bibinfo {year} {2014})}\BibitemShut
  {NoStop}%
\bibitem [{\citenamefont {Kuhr}\ \emph {et~al.}(2019)\citenamefont {Kuhr},
  \citenamefont {R{\"u}hle},\ and\ \citenamefont {Stark}}]{Kuhr_CDMS_2019}%
  \BibitemOpen
  \bibfield  {author} {\bibinfo {author} {\bibfnamefont {J.-T.}\ \bibnamefont
  {Kuhr}}, \bibinfo {author} {\bibfnamefont {F.}~\bibnamefont {R{\"u}hle}},\
  and\ \bibinfo {author} {\bibfnamefont {H.}~\bibnamefont {Stark}},\ }\bibfield
   {title} {\bibinfo {title} {Collective dynamics in a monolayer of squirmers
  confined to a boundary by gravity},\ }\href@noop {} {\bibfield  {journal}
  {\bibinfo  {journal} {Soft Matter}\ }\textbf {\bibinfo {volume} {15}},\
  \bibinfo {pages} {5685} (\bibinfo {year} {2019})}\BibitemShut {NoStop}%
\bibitem [{\citenamefont {Stenhammar}\ \emph {et~al.}(2014)\citenamefont
  {Stenhammar}, \citenamefont {Marenduzzo}, \citenamefont {Allen},\ and\
  \citenamefont {Cates}}]{Stenhammar_ABP_2014}%
  \BibitemOpen
  \bibfield  {author} {\bibinfo {author} {\bibfnamefont {J.}~\bibnamefont
  {Stenhammar}}, \bibinfo {author} {\bibfnamefont {D.}~\bibnamefont
  {Marenduzzo}}, \bibinfo {author} {\bibfnamefont {R.~J.}\ \bibnamefont
  {Allen}},\ and\ \bibinfo {author} {\bibfnamefont {M.~E.}\ \bibnamefont
  {Cates}},\ }\bibfield  {title} {\bibinfo {title} {Phase behaviour of active
  {Brownian} particles: the role of dimensionality},\ }\href@noop {} {\bibfield
   {journal} {\bibinfo  {journal} {Soft Matter}\ }\textbf {\bibinfo {volume}
  {10}},\ \bibinfo {pages} {1489} (\bibinfo {year} {2014})}\BibitemShut
  {NoStop}%
\bibitem [{\citenamefont {Lighthill}(1952)}]{Lighthill_SMB_1952}%
  \BibitemOpen
  \bibfield  {author} {\bibinfo {author} {\bibfnamefont {M.~J.}\ \bibnamefont
  {Lighthill}},\ }\bibfield  {title} {\bibinfo {title} {On the squirming motion
  of nearly spherical deformable bodies through liquids at very small
  {R}eynolds numbers},\ }\href@noop {} {\bibfield  {journal} {\bibinfo
  {journal} {Commun. Pure Appl. Math.}\ }\textbf {\bibinfo {volume} {5}},\
  \bibinfo {pages} {109} (\bibinfo {year} {1952})}\BibitemShut {NoStop}%
\bibitem [{\citenamefont {Blake}(1971)}]{Blake_SEA_1971}%
  \BibitemOpen
  \bibfield  {author} {\bibinfo {author} {\bibfnamefont {J.~R.}\ \bibnamefont
  {Blake}},\ }\bibfield  {title} {\bibinfo {title} {A spherical envelope
  approach to ciliary propulsion},\ }\href@noop {} {\bibfield  {journal}
  {\bibinfo  {journal} {J. Fluid Mech.}\ }\textbf {\bibinfo {volume} {46}},\
  \bibinfo {pages} {199} (\bibinfo {year} {1971})}\BibitemShut {NoStop}%
\bibitem [{\citenamefont {Theers}\ \emph {et~al.}(2016)\citenamefont {Theers},
  \citenamefont {Westphal}, \citenamefont {Gompper},\ and\ \citenamefont
  {Winkler}}]{Theers_MSM_2016}%
  \BibitemOpen
  \bibfield  {author} {\bibinfo {author} {\bibfnamefont {M.}~\bibnamefont
  {Theers}}, \bibinfo {author} {\bibfnamefont {E.}~\bibnamefont {Westphal}},
  \bibinfo {author} {\bibfnamefont {G.}~\bibnamefont {Gompper}},\ and\ \bibinfo
  {author} {\bibfnamefont {R.~G.}\ \bibnamefont {Winkler}},\ }\bibfield
  {title} {\bibinfo {title} {Modeling a spheroidal microswimmer and cooperative
  swimming in a narrow slit},\ }\href@noop {} {\bibfield  {journal} {\bibinfo
  {journal} {Soft Matter}\ }\textbf {\bibinfo {volume} {12}},\ \bibinfo {pages}
  {7372} (\bibinfo {year} {2016})}\BibitemShut {NoStop}%
\bibitem [{\citenamefont {Kyoya}\ \emph {et~al.}(2015)\citenamefont {Kyoya},
  \citenamefont {Matsunaga}, \citenamefont {Imai}, \citenamefont {Omori},\ and\
  \citenamefont {Ishikawa}}]{Kyoya_SMCM_2015}%
  \BibitemOpen
  \bibfield  {author} {\bibinfo {author} {\bibfnamefont {K.}~\bibnamefont
  {Kyoya}}, \bibinfo {author} {\bibfnamefont {D.}~\bibnamefont {Matsunaga}},
  \bibinfo {author} {\bibfnamefont {Y.}~\bibnamefont {Imai}}, \bibinfo {author}
  {\bibfnamefont {T.}~\bibnamefont {Omori}},\ and\ \bibinfo {author}
  {\bibfnamefont {T.}~\bibnamefont {Ishikawa}},\ }\bibfield  {title} {\bibinfo
  {title} {Shape matters: near-field fluid mechanics dominate the collective
  motions of ellipsoidal squirmers},\ }\href@noop {} {\bibfield  {journal}
  {\bibinfo  {journal} {Phys. Rev. E}\ }\textbf {\bibinfo {volume} {92}},\
  \bibinfo {pages} {063027} (\bibinfo {year} {2015})}\BibitemShut {NoStop}%
\bibitem [{\citenamefont {Alarc\'{o}n}\ \emph {et~al.}(2017)\citenamefont
  {Alarc\'{o}n}, \citenamefont {Valeriani},\ and\ \citenamefont
  {Pagonabarraga}}]{Alarcon_MCP_2017}%
  \BibitemOpen
  \bibfield  {author} {\bibinfo {author} {\bibfnamefont {F.}~\bibnamefont
  {Alarc\'{o}n}}, \bibinfo {author} {\bibfnamefont {C.}~\bibnamefont
  {Valeriani}},\ and\ \bibinfo {author} {\bibfnamefont {I.}~\bibnamefont
  {Pagonabarraga}},\ }\bibfield  {title} {\bibinfo {title} {Morphology of
  clusters of attractive dry and wet self-propelled spherical particle
  suspensions},\ }\href@noop {} {\bibfield  {journal} {\bibinfo  {journal}
  {Soft Matter}\ }\textbf {\bibinfo {volume} {13}},\ \bibinfo {pages} {814}
  (\bibinfo {year} {2017})}\BibitemShut {NoStop}%
\bibitem [{\citenamefont {Yoshinaga}\ and\ \citenamefont
  {Liverpool}(2017)}]{Yoshinaga_HIDS_2017}%
  \BibitemOpen
  \bibfield  {author} {\bibinfo {author} {\bibfnamefont {N.}~\bibnamefont
  {Yoshinaga}}\ and\ \bibinfo {author} {\bibfnamefont {T.~B.}\ \bibnamefont
  {Liverpool}},\ }\bibfield  {title} {\bibinfo {title} {Hydrodynamic
  interactions in dense active suspensions: from polar order to dynamical
  clusters},\ }\href@noop {} {\bibfield  {journal} {\bibinfo  {journal} {Phys.
  Rev. E}\ }\textbf {\bibinfo {volume} {96}},\ \bibinfo {pages} {020603}
  (\bibinfo {year} {2017})}\BibitemShut {NoStop}%
\bibitem [{\citenamefont {Blaschke}\ \emph {et~al.}(2016)\citenamefont
  {Blaschke}, \citenamefont {Maurer}, \citenamefont {Menon}, \citenamefont
  {Z\"{o}ttl},\ and\ \citenamefont {Stark}}]{Blaschke_PSC_2016}%
  \BibitemOpen
  \bibfield  {author} {\bibinfo {author} {\bibfnamefont {J.}~\bibnamefont
  {Blaschke}}, \bibinfo {author} {\bibfnamefont {M.}~\bibnamefont {Maurer}},
  \bibinfo {author} {\bibfnamefont {K.}~\bibnamefont {Menon}}, \bibinfo
  {author} {\bibfnamefont {A.}~\bibnamefont {Z\"{o}ttl}},\ and\ \bibinfo
  {author} {\bibfnamefont {H.}~\bibnamefont {Stark}},\ }\bibfield  {title}
  {\bibinfo {title} {Phase separation and coexistence of hydrodynamically
  interacting microswimmers},\ }\href@noop {} {\bibfield  {journal} {\bibinfo
  {journal} {Soft Matter}\ }\textbf {\bibinfo {volume} {12}},\ \bibinfo {pages}
  {9821} (\bibinfo {year} {2016})}\BibitemShut {NoStop}%
\bibitem [{\citenamefont {Ishikawa}\ and\ \citenamefont
  {Pedley}(2007)}]{Ishikawa_DSM_2007}%
  \BibitemOpen
  \bibfield  {author} {\bibinfo {author} {\bibfnamefont {T.}~\bibnamefont
  {Ishikawa}}\ and\ \bibinfo {author} {\bibfnamefont {T.~J.}\ \bibnamefont
  {Pedley}},\ }\bibfield  {title} {\bibinfo {title} {Diffusion of swimming
  model micro-organisms in a semi-dilute suspension},\ }\href@noop {}
  {\bibfield  {journal} {\bibinfo  {journal} {J. Fluid Mech.}\ }\textbf
  {\bibinfo {volume} {588}},\ \bibinfo {pages} {437} (\bibinfo {year}
  {2007})}\BibitemShut {NoStop}%
\bibitem [{\citenamefont {Delmotte}\ \emph {et~al.}(2015)\citenamefont
  {Delmotte}, \citenamefont {Keaveny}, \citenamefont {Plourabou{\'e}},\ and\
  \citenamefont {Climent}}]{Delmotte_LSS_2015}%
  \BibitemOpen
  \bibfield  {author} {\bibinfo {author} {\bibfnamefont {B.}~\bibnamefont
  {Delmotte}}, \bibinfo {author} {\bibfnamefont {E.~E.}\ \bibnamefont
  {Keaveny}}, \bibinfo {author} {\bibfnamefont {F.}~\bibnamefont
  {Plourabou{\'e}}},\ and\ \bibinfo {author} {\bibfnamefont {E.}~\bibnamefont
  {Climent}},\ }\bibfield  {title} {\bibinfo {title} {Large-scale simulation of
  steady and time-dependent active suspensions with the force-coupling
  method},\ }\href@noop {} {\bibfield  {journal} {\bibinfo  {journal} {J. Comp.
  Phys.}\ }\textbf {\bibinfo {volume} {302}},\ \bibinfo {pages} {524} (\bibinfo
  {year} {2015})}\BibitemShut {NoStop}%
\bibitem [{\citenamefont {Ishikawa}\ \emph {et~al.}(2008)\citenamefont
  {Ishikawa}, \citenamefont {Locsei},\ and\ \citenamefont
  {Pedley}}]{Ishikawa_DCS_2008}%
  \BibitemOpen
  \bibfield  {author} {\bibinfo {author} {\bibfnamefont {T.}~\bibnamefont
  {Ishikawa}}, \bibinfo {author} {\bibfnamefont {J.~T.}\ \bibnamefont
  {Locsei}},\ and\ \bibinfo {author} {\bibfnamefont {T.~J.}\ \bibnamefont
  {Pedley}},\ }\bibfield  {title} {\bibinfo {title} {Development of coherent
  structures in concentrated suspensions of swimming model micro-organisms},\
  }\href@noop {} {\bibfield  {journal} {\bibinfo  {journal} {J. Fluid Mech.}\
  }\textbf {\bibinfo {volume} {615}},\ \bibinfo {pages} {401} (\bibinfo {year}
  {2008})}\BibitemShut {NoStop}%
\bibitem [{\citenamefont {Evans}\ \emph {et~al.}(2011)\citenamefont {Evans},
  \citenamefont {Ishikawa}, \citenamefont {Yamaguchi},\ and\ \citenamefont
  {Lauga}}]{Evans_CSS_2011}%
  \BibitemOpen
  \bibfield  {author} {\bibinfo {author} {\bibfnamefont {A.~A.}\ \bibnamefont
  {Evans}}, \bibinfo {author} {\bibfnamefont {T.}~\bibnamefont {Ishikawa}},
  \bibinfo {author} {\bibfnamefont {T.}~\bibnamefont {Yamaguchi}},\ and\
  \bibinfo {author} {\bibfnamefont {E.}~\bibnamefont {Lauga}},\ }\bibfield
  {title} {\bibinfo {title} {Orientational order in concentrated suspensions of
  spherical microswimmers},\ }\href@noop {} {\bibfield  {journal} {\bibinfo
  {journal} {Phys. Fluids}\ }\textbf {\bibinfo {volume} {23}},\ \bibinfo
  {pages} {111702} (\bibinfo {year} {2011})}\BibitemShut {NoStop}%
\bibitem [{\citenamefont {Darnton}\ \emph {et~al.}(2007)\citenamefont
  {Darnton}, \citenamefont {Turner}, \citenamefont {Rojevsky},\ and\
  \citenamefont {Berg}}]{Darnton_ECI_2007}%
  \BibitemOpen
  \bibfield  {author} {\bibinfo {author} {\bibfnamefont {N.~C.}\ \bibnamefont
  {Darnton}}, \bibinfo {author} {\bibfnamefont {L.}~\bibnamefont {Turner}},
  \bibinfo {author} {\bibfnamefont {S.}~\bibnamefont {Rojevsky}},\ and\
  \bibinfo {author} {\bibfnamefont {H.~C.}\ \bibnamefont {Berg}},\ }\bibfield
  {title} {\bibinfo {title} {On torque and tumbling in swimming {\it
  escherichia coli}},\ }\href@noop {} {\bibfield  {journal} {\bibinfo
  {journal} {J. Bacteriol.}\ }\textbf {\bibinfo {volume} {189}},\ \bibinfo
  {pages} {1756} (\bibinfo {year} {2007})}\BibitemShut {NoStop}%
\bibitem [{\citenamefont {Fedosov}\ \emph {et~al.}(2010)\citenamefont
  {Fedosov}, \citenamefont {Caswell},\ and\ \citenamefont
  {Karniadakis}}]{Fedosov_RBC_2010}%
  \BibitemOpen
  \bibfield  {author} {\bibinfo {author} {\bibfnamefont {D.~A.}\ \bibnamefont
  {Fedosov}}, \bibinfo {author} {\bibfnamefont {B.}~\bibnamefont {Caswell}},\
  and\ \bibinfo {author} {\bibfnamefont {G.~E.}\ \bibnamefont {Karniadakis}},\
  }\bibfield  {title} {\bibinfo {title} {A multiscale red blood cell model with
  accurate mechanics, rheology, and dynamics},\ }\href@noop {} {\bibfield
  {journal} {\bibinfo  {journal} {Biophys. J.}\ }\textbf {\bibinfo {volume}
  {98}},\ \bibinfo {pages} {2215} (\bibinfo {year} {2010})}\BibitemShut
  {NoStop}%
\bibitem [{\citenamefont {Fedosov}\ \emph {et~al.}(2014)\citenamefont
  {Fedosov}, \citenamefont {Noguchi},\ and\ \citenamefont
  {Gompper}}]{Fedosov_MBF_2014}%
  \BibitemOpen
  \bibfield  {author} {\bibinfo {author} {\bibfnamefont {D.~A.}\ \bibnamefont
  {Fedosov}}, \bibinfo {author} {\bibfnamefont {H.}~\bibnamefont {Noguchi}},\
  and\ \bibinfo {author} {\bibfnamefont {G.}~\bibnamefont {Gompper}},\
  }\bibfield  {title} {\bibinfo {title} {Multiscale modeling of blood flow:
  from single cells to blood rheology},\ }\href@noop {} {\bibfield  {journal}
  {\bibinfo  {journal} {Biomech. Model. Mechanobiol.}\ }\textbf {\bibinfo
  {volume} {13}},\ \bibinfo {pages} {239} (\bibinfo {year} {2014})}\BibitemShut
  {NoStop}%
\bibitem [{\citenamefont {Z\"{o}ttl}\ and\ \citenamefont
  {Stark}(2018)}]{Zoettl_SSM_2018}%
  \BibitemOpen
  \bibfield  {author} {\bibinfo {author} {\bibfnamefont {A.}~\bibnamefont
  {Z\"{o}ttl}}\ and\ \bibinfo {author} {\bibfnamefont {H.}~\bibnamefont
  {Stark}},\ }\bibfield  {title} {\bibinfo {title} {Simulating squirmers with
  multiparticle collision dynamics},\ }\href@noop {} {\bibfield  {journal}
  {\bibinfo  {journal} {Eur. Phys. J. E}\ }\textbf {\bibinfo {volume} {41}},\
  \bibinfo {pages} {61} (\bibinfo {year} {2018})}\BibitemShut {NoStop}%
\bibitem [{\citenamefont {Ishikawa}\ \emph {et~al.}(2006)\citenamefont
  {Ishikawa}, \citenamefont {Simmonds},\ and\ \citenamefont
  {Pedley}}]{Ishikawa_HIM_2006}%
  \BibitemOpen
  \bibfield  {author} {\bibinfo {author} {\bibfnamefont {T.}~\bibnamefont
  {Ishikawa}}, \bibinfo {author} {\bibfnamefont {M.~P.}\ \bibnamefont
  {Simmonds}},\ and\ \bibinfo {author} {\bibfnamefont {T.~J.}\ \bibnamefont
  {Pedley}},\ }\bibfield  {title} {\bibinfo {title} {Hydrodynamic interaction
  of two swimming model micro-organisms},\ }\href@noop {} {\bibfield  {journal}
  {\bibinfo  {journal} {J. Fluid Mech.}\ }\textbf {\bibinfo {volume} {568}},\
  \bibinfo {pages} {119} (\bibinfo {year} {2006})}\BibitemShut {NoStop}%
\bibitem [{\citenamefont {Pagonabarraga}\ and\ \citenamefont
  {Llopis}(2013)}]{Pagonabarraga_SRSS_2013}%
  \BibitemOpen
  \bibfield  {author} {\bibinfo {author} {\bibfnamefont {I.}~\bibnamefont
  {Pagonabarraga}}\ and\ \bibinfo {author} {\bibfnamefont {I.}~\bibnamefont
  {Llopis}},\ }\bibfield  {title} {\bibinfo {title} {The structure and rheology
  of sheared model swimmer suspensions},\ }\href@noop {} {\bibfield  {journal}
  {\bibinfo  {journal} {Soft Matter}\ }\textbf {\bibinfo {volume} {9}},\
  \bibinfo {pages} {7174} (\bibinfo {year} {2013})}\BibitemShut {NoStop}%
\bibitem [{\citenamefont {Qi}\ \emph {et~al.}(2020)\citenamefont {Qi},
  \citenamefont {Annepu}, \citenamefont {Gompper},\ and\ \citenamefont
  {Winkler}}]{Qi_RSS_2020}%
  \BibitemOpen
  \bibfield  {author} {\bibinfo {author} {\bibfnamefont {K.}~\bibnamefont
  {Qi}}, \bibinfo {author} {\bibfnamefont {H.}~\bibnamefont {Annepu}}, \bibinfo
  {author} {\bibfnamefont {G.}~\bibnamefont {Gompper}},\ and\ \bibinfo {author}
  {\bibfnamefont {R.~G.}\ \bibnamefont {Winkler}},\ }\bibfield  {title}
  {\bibinfo {title} {Rheotaxis of spheroidal squirmers in microchannel flow:
  interplay of shape, hydrodynamics, active stress, and thermal fluctuations},\
  }\href@noop {} {\bibfield  {journal} {\bibinfo  {journal} {Phys. Rev. Res.}\
  }\textbf {\bibinfo {volume} {2}},\ \bibinfo {pages} {033275} (\bibinfo {year}
  {2020})}\BibitemShut {NoStop}%
\bibitem [{\citenamefont {Hu}\ \emph {et~al.}(2015{\natexlab{a}})\citenamefont
  {Hu}, \citenamefont {Yang}, \citenamefont {Gompper},\ and\ \citenamefont
  {Winkler}}]{Hu_MMH_2015}%
  \BibitemOpen
  \bibfield  {author} {\bibinfo {author} {\bibfnamefont {J.}~\bibnamefont
  {Hu}}, \bibinfo {author} {\bibfnamefont {M.}~\bibnamefont {Yang}}, \bibinfo
  {author} {\bibfnamefont {G.}~\bibnamefont {Gompper}},\ and\ \bibinfo {author}
  {\bibfnamefont {R.~G.}\ \bibnamefont {Winkler}},\ }\bibfield  {title}
  {\bibinfo {title} {Modelling the mechanics and hydrodynamics of swimming {E.}
  coli},\ }\href@noop {} {\bibfield  {journal} {\bibinfo  {journal} {Soft
  Matter}\ }\textbf {\bibinfo {volume} {11}},\ \bibinfo {pages} {7867}
  (\bibinfo {year} {2015}{\natexlab{a}})}\BibitemShut {NoStop}%
\bibitem [{\citenamefont {Hoogerbrugge}\ and\ \citenamefont
  {Koelman}(1992)}]{Hoogerbrugge_SMH_1992}%
  \BibitemOpen
  \bibfield  {author} {\bibinfo {author} {\bibfnamefont {P.~J.}\ \bibnamefont
  {Hoogerbrugge}}\ and\ \bibinfo {author} {\bibfnamefont {J.~M. V.~A.}\
  \bibnamefont {Koelman}},\ }\bibfield  {title} {\bibinfo {title} {Simulating
  microscopic hydrodynamic phenomena with dissipative particle dynamics},\
  }\href@noop {} {\bibfield  {journal} {\bibinfo  {journal} {Europhys. Lett.}\
  }\textbf {\bibinfo {volume} {19}},\ \bibinfo {pages} {155} (\bibinfo {year}
  {1992})}\BibitemShut {NoStop}%
\bibitem [{\citenamefont {Espa\~{n}ol}\ and\ \citenamefont
  {Warren}(1995)}]{Espanol_SMO_1995}%
  \BibitemOpen
  \bibfield  {author} {\bibinfo {author} {\bibfnamefont {P.}~\bibnamefont
  {Espa\~{n}ol}}\ and\ \bibinfo {author} {\bibfnamefont {P.}~\bibnamefont
  {Warren}},\ }\bibfield  {title} {\bibinfo {title} {Statistical mechanics of
  dissipative particle dynamics},\ }\href@noop {} {\bibfield  {journal}
  {\bibinfo  {journal} {Europhys. Lett.}\ }\textbf {\bibinfo {volume} {30}},\
  \bibinfo {pages} {191} (\bibinfo {year} {1995})}\BibitemShut {NoStop}%
\bibitem [{\citenamefont {Solon}\ \emph {et~al.}(2015)\citenamefont {Solon},
  \citenamefont {Stenhammar}, \citenamefont {Wittkowski}, \citenamefont
  {Kardar}, \citenamefont {Kafri}, \citenamefont {Cates},\ and\ \citenamefont
  {Tailleur}}]{Solon_PPE_2015}%
  \BibitemOpen
  \bibfield  {author} {\bibinfo {author} {\bibfnamefont {A.~P.}\ \bibnamefont
  {Solon}}, \bibinfo {author} {\bibfnamefont {J.}~\bibnamefont {Stenhammar}},
  \bibinfo {author} {\bibfnamefont {R.}~\bibnamefont {Wittkowski}}, \bibinfo
  {author} {\bibfnamefont {M.}~\bibnamefont {Kardar}}, \bibinfo {author}
  {\bibfnamefont {Y.}~\bibnamefont {Kafri}}, \bibinfo {author} {\bibfnamefont
  {M.~E.}\ \bibnamefont {Cates}},\ and\ \bibinfo {author} {\bibfnamefont
  {J.}~\bibnamefont {Tailleur}},\ }\bibfield  {title} {\bibinfo {title}
  {Pressure and phase equilibria in interacting active {B}rownian spheres},\
  }\href@noop {} {\bibfield  {journal} {\bibinfo  {journal} {Phys. Rev. Lett.}\
  }\textbf {\bibinfo {volume} {114}},\ \bibinfo {pages} {198301} (\bibinfo
  {year} {2015})}\BibitemShut {NoStop}%
\bibitem [{\citenamefont {Drescher}\ \emph {et~al.}(2011)\citenamefont
  {Drescher}, \citenamefont {Dunkel}, \citenamefont {Cisneros}, \citenamefont
  {Ganguly},\ and\ \citenamefont {Goldstein}}]{Drescher_FDN_2011}%
  \BibitemOpen
  \bibfield  {author} {\bibinfo {author} {\bibfnamefont {K.}~\bibnamefont
  {Drescher}}, \bibinfo {author} {\bibfnamefont {J.}~\bibnamefont {Dunkel}},
  \bibinfo {author} {\bibfnamefont {L.~H.}\ \bibnamefont {Cisneros}}, \bibinfo
  {author} {\bibfnamefont {S.}~\bibnamefont {Ganguly}},\ and\ \bibinfo {author}
  {\bibfnamefont {R.~E.}\ \bibnamefont {Goldstein}},\ }\bibfield  {title}
  {\bibinfo {title} {Fluid dynamics and noise in bacterial cell-cell and
  cell-surface scattering},\ }\href@noop {} {\bibfield  {journal} {\bibinfo
  {journal} {Proc. Natl. Acad. Sci. USA}\ }\textbf {\bibinfo {volume} {108}},\
  \bibinfo {pages} {10940} (\bibinfo {year} {2011})}\BibitemShut {NoStop}%
\bibitem [{\citenamefont {Lauga}\ and\ \citenamefont
  {Powers}(2009)}]{Lauga_HSO_2009}%
  \BibitemOpen
  \bibfield  {author} {\bibinfo {author} {\bibfnamefont {E.}~\bibnamefont
  {Lauga}}\ and\ \bibinfo {author} {\bibfnamefont {T.~R.}\ \bibnamefont
  {Powers}},\ }\bibfield  {title} {\bibinfo {title} {The hydrodynamics of
  swimming microorganisms},\ }\href@noop {} {\bibfield  {journal} {\bibinfo
  {journal} {Rep. Prog. Phys.}\ }\textbf {\bibinfo {volume} {72}},\ \bibinfo
  {pages} {096601} (\bibinfo {year} {2009})}\BibitemShut {NoStop}%
\bibitem [{\citenamefont {Bialk{\'e}}\ \emph {et~al.}(2012)\citenamefont
  {Bialk{\'e}}, \citenamefont {Speck},\ and\ \citenamefont
  {L{\"o}wen}}]{Bialke_CDS_2012}%
  \BibitemOpen
  \bibfield  {author} {\bibinfo {author} {\bibfnamefont {J.}~\bibnamefont
  {Bialk{\'e}}}, \bibinfo {author} {\bibfnamefont {T.}~\bibnamefont {Speck}},\
  and\ \bibinfo {author} {\bibfnamefont {H.}~\bibnamefont {L{\"o}wen}},\
  }\bibfield  {title} {\bibinfo {title} {Crystallization in a dense suspension
  of self-propelled particles},\ }\href@noop {} {\bibfield  {journal} {\bibinfo
   {journal} {Phys. Rev. Lett.}\ }\textbf {\bibinfo {volume} {108}},\ \bibinfo
  {pages} {168301} (\bibinfo {year} {2012})}\BibitemShut {NoStop}%
\bibitem [{\citenamefont {Levis}\ and\ \citenamefont
  {Berthier}(2014)}]{Levis_CHDMC_2014}%
  \BibitemOpen
  \bibfield  {author} {\bibinfo {author} {\bibfnamefont {D.}~\bibnamefont
  {Levis}}\ and\ \bibinfo {author} {\bibfnamefont {L.}~\bibnamefont
  {Berthier}},\ }\bibfield  {title} {\bibinfo {title} {Clustering and
  heterogeneous dynamics in a kinetic {Monte Carlo} model of self-propelled
  hard disks},\ }\href@noop {} {\bibfield  {journal} {\bibinfo  {journal}
  {Phys. Rev. E}\ }\textbf {\bibinfo {volume} {89}},\ \bibinfo {pages} {062301}
  (\bibinfo {year} {2014})}\BibitemShut {NoStop}%
\bibitem [{\citenamefont {Li}\ and\ \citenamefont {Tang}(2009)}]{Li_AMS_2009}%
  \BibitemOpen
  \bibfield  {author} {\bibinfo {author} {\bibfnamefont {G.}~\bibnamefont
  {Li}}\ and\ \bibinfo {author} {\bibfnamefont {J.~X.}\ \bibnamefont {Tang}},\
  }\bibfield  {title} {\bibinfo {title} {Accumulation of microswimmers near a
  surface mediated by collision and rotational {B}rownian motion},\ }\href@noop
  {} {\bibfield  {journal} {\bibinfo  {journal} {Phys. Rev. Lett.}\ }\textbf
  {\bibinfo {volume} {103}},\ \bibinfo {pages} {078101} (\bibinfo {year}
  {2009})}\BibitemShut {NoStop}%
\bibitem [{\citenamefont {Volpe}\ \emph {et~al.}(2011)\citenamefont {Volpe},
  \citenamefont {Buttinoni}, \citenamefont {Vogt}, \citenamefont
  {K{\"u}mmerer},\ and\ \citenamefont {Bechinger}}]{Volpe_MSE_2011}%
  \BibitemOpen
  \bibfield  {author} {\bibinfo {author} {\bibfnamefont {G.}~\bibnamefont
  {Volpe}}, \bibinfo {author} {\bibfnamefont {I.}~\bibnamefont {Buttinoni}},
  \bibinfo {author} {\bibfnamefont {D.}~\bibnamefont {Vogt}}, \bibinfo {author}
  {\bibfnamefont {H.-J.}\ \bibnamefont {K{\"u}mmerer}},\ and\ \bibinfo {author}
  {\bibfnamefont {C.}~\bibnamefont {Bechinger}},\ }\bibfield  {title} {\bibinfo
  {title} {Microswimmers in patterned environments},\ }\href@noop {} {\bibfield
   {journal} {\bibinfo  {journal} {Soft Matter}\ }\textbf {\bibinfo {volume}
  {7}},\ \bibinfo {pages} {8810} (\bibinfo {year} {2011})}\BibitemShut
  {NoStop}%
\bibitem [{\citenamefont {Elgeti}\ and\ \citenamefont
  {Gompper}(2013)}]{Elgeti_WAS_2013}%
  \BibitemOpen
  \bibfield  {author} {\bibinfo {author} {\bibfnamefont {J.}~\bibnamefont
  {Elgeti}}\ and\ \bibinfo {author} {\bibfnamefont {G.}~\bibnamefont
  {Gompper}},\ }\bibfield  {title} {\bibinfo {title} {Wall accumulation of
  self-propelled spheres},\ }\href@noop {} {\bibfield  {journal} {\bibinfo
  {journal} {Europhys. Lett.}\ }\textbf {\bibinfo {volume} {101}},\ \bibinfo
  {pages} {48003} (\bibinfo {year} {2013})}\BibitemShut {NoStop}%
\bibitem [{\citenamefont {Berke}\ \emph {et~al.}(2008)\citenamefont {Berke},
  \citenamefont {Turner}, \citenamefont {Berg},\ and\ \citenamefont
  {Lauga}}]{Berke_HAS_2008}%
  \BibitemOpen
  \bibfield  {author} {\bibinfo {author} {\bibfnamefont {A.~P.}\ \bibnamefont
  {Berke}}, \bibinfo {author} {\bibfnamefont {L.}~\bibnamefont {Turner}},
  \bibinfo {author} {\bibfnamefont {H.~C.}\ \bibnamefont {Berg}},\ and\
  \bibinfo {author} {\bibfnamefont {E.}~\bibnamefont {Lauga}},\ }\bibfield
  {title} {\bibinfo {title} {Hydrodynamic attraction of swimming microorganisms
  by surfaces},\ }\href@noop {} {\bibfield  {journal} {\bibinfo  {journal}
  {Phys. Rev. Lett.}\ }\textbf {\bibinfo {volume} {101}},\ \bibinfo {pages}
  {038102} (\bibinfo {year} {2008})}\BibitemShut {NoStop}%
\bibitem [{\citenamefont {Frymier}\ \emph {et~al.}(1995)\citenamefont
  {Frymier}, \citenamefont {Ford}, \citenamefont {Berg},\ and\ \citenamefont
  {Cummings}}]{Frymier_3DMBS_1995}%
  \BibitemOpen
  \bibfield  {author} {\bibinfo {author} {\bibfnamefont {P.~D.}\ \bibnamefont
  {Frymier}}, \bibinfo {author} {\bibfnamefont {R.~M.}\ \bibnamefont {Ford}},
  \bibinfo {author} {\bibfnamefont {H.~C.}\ \bibnamefont {Berg}},\ and\
  \bibinfo {author} {\bibfnamefont {P.~T.}\ \bibnamefont {Cummings}},\
  }\bibfield  {title} {\bibinfo {title} {Three-dimensional tracking of motile
  bacteria near a solid planar surface},\ }\href@noop {} {\bibfield  {journal}
  {\bibinfo  {journal} {Proc. Natl. Acad. Sci. USA}\ }\textbf {\bibinfo
  {volume} {92}},\ \bibinfo {pages} {6195} (\bibinfo {year}
  {1995})}\BibitemShut {NoStop}%
\bibitem [{\citenamefont {Junot}\ \emph {et~al.}(2022)\citenamefont {Junot},
  \citenamefont {Darnige}, \citenamefont {Lindner}, \citenamefont {Martinez},
  \citenamefont {Arlt}, \citenamefont {Dawson}, \citenamefont {Poon},
  \citenamefont {Auradou},\ and\ \citenamefont {Cl{\'e}ment}}]{Junot_RTV_2022}%
  \BibitemOpen
  \bibfield  {author} {\bibinfo {author} {\bibfnamefont {G.}~\bibnamefont
  {Junot}}, \bibinfo {author} {\bibfnamefont {T.}~\bibnamefont {Darnige}},
  \bibinfo {author} {\bibfnamefont {A.}~\bibnamefont {Lindner}}, \bibinfo
  {author} {\bibfnamefont {V.~A.}\ \bibnamefont {Martinez}}, \bibinfo {author}
  {\bibfnamefont {J.}~\bibnamefont {Arlt}}, \bibinfo {author} {\bibfnamefont
  {A.}~\bibnamefont {Dawson}}, \bibinfo {author} {\bibfnamefont {W.~C.~K.}\
  \bibnamefont {Poon}}, \bibinfo {author} {\bibfnamefont {H.}~\bibnamefont
  {Auradou}},\ and\ \bibinfo {author} {\bibfnamefont {E.}~\bibnamefont
  {Cl{\'e}ment}},\ }\bibfield  {title} {\bibinfo {title} {Run-to-tumble
  variability controls the surface residence times of {E.} coli bacteria},\
  }\href@noop {} {\bibfield  {journal} {\bibinfo  {journal} {Phys. Rev. Lett.}\
  }\textbf {\bibinfo {volume} {128}},\ \bibinfo {pages} {248101} (\bibinfo
  {year} {2022})}\BibitemShut {NoStop}%
\bibitem [{\citenamefont {Vigeant}\ \emph {et~al.}(2002)\citenamefont
  {Vigeant}, \citenamefont {Ford}, \citenamefont {Wagner},\ and\ \citenamefont
  {Tamm}}]{Vigeant_RIAEC_2002}%
  \BibitemOpen
  \bibfield  {author} {\bibinfo {author} {\bibfnamefont {M.~A.-S.}\
  \bibnamefont {Vigeant}}, \bibinfo {author} {\bibfnamefont {R.~M.}\
  \bibnamefont {Ford}}, \bibinfo {author} {\bibfnamefont {M.}~\bibnamefont
  {Wagner}},\ and\ \bibinfo {author} {\bibfnamefont {L.~K.}\ \bibnamefont
  {Tamm}},\ }\bibfield  {title} {\bibinfo {title} {Reversible and irreversible
  adhesion of motile {\it escherichia coli} cells analyzed by total internal
  reflection aqueous fluorescence microscopy},\ }\href@noop {} {\bibfield
  {journal} {\bibinfo  {journal} {Appl. Environ. Microbiol.}\ }\textbf
  {\bibinfo {volume} {68}},\ \bibinfo {pages} {2794} (\bibinfo {year}
  {2002})}\BibitemShut {NoStop}%
\bibitem [{\citenamefont {Spagnolie}\ and\ \citenamefont
  {Lauga}(2012)}]{Spagnolie_HSP_2012}%
  \BibitemOpen
  \bibfield  {author} {\bibinfo {author} {\bibfnamefont {S.~E.}\ \bibnamefont
  {Spagnolie}}\ and\ \bibinfo {author} {\bibfnamefont {E.}~\bibnamefont
  {Lauga}},\ }\bibfield  {title} {\bibinfo {title} {Hydrodynamics of
  self-propulsion near a boundary: predictions and accuracy of far-field
  approximations},\ }\href@noop {} {\bibfield  {journal} {\bibinfo  {journal}
  {J. Fluid Mech.}\ }\textbf {\bibinfo {volume} {700}},\ \bibinfo {pages} {105}
  (\bibinfo {year} {2012})}\BibitemShut {NoStop}%
\bibitem [{\citenamefont {G{\"o}tze}\ and\ \citenamefont
  {Gompper}(2011)}]{Goetze_FGRC_2011}%
  \BibitemOpen
  \bibfield  {author} {\bibinfo {author} {\bibfnamefont {I.~O.}\ \bibnamefont
  {G{\"o}tze}}\ and\ \bibinfo {author} {\bibfnamefont {G.}~\bibnamefont
  {Gompper}},\ }\bibfield  {title} {\bibinfo {title} {Flow generation by
  rotating colloids in planar microchannels},\ }\href@noop {} {\bibfield
  {journal} {\bibinfo  {journal} {Europhys. Lett.}\ }\textbf {\bibinfo {volume}
  {92}},\ \bibinfo {pages} {64003} (\bibinfo {year} {2011})}\BibitemShut
  {NoStop}%
\bibitem [{\citenamefont {Romanczuk}\ \emph {et~al.}(2012)\citenamefont
  {Romanczuk}, \citenamefont {B{\"a}r}, \citenamefont {Ebeling}, \citenamefont
  {Lindner},\ and\ \citenamefont {Schimansky-Geier}}]{Romanczuk_ABP_2012}%
  \BibitemOpen
  \bibfield  {author} {\bibinfo {author} {\bibfnamefont {P.}~\bibnamefont
  {Romanczuk}}, \bibinfo {author} {\bibfnamefont {M.}~\bibnamefont {B{\"a}r}},
  \bibinfo {author} {\bibfnamefont {W.}~\bibnamefont {Ebeling}}, \bibinfo
  {author} {\bibfnamefont {B.}~\bibnamefont {Lindner}},\ and\ \bibinfo {author}
  {\bibfnamefont {L.}~\bibnamefont {Schimansky-Geier}},\ }\bibfield  {title}
  {\bibinfo {title} {Active brownian particles: From individual to collective
  stochastic dynamics},\ }\href@noop {} {\bibfield  {journal} {\bibinfo
  {journal} {Eur. Phys. J. Special Topics}\ }\textbf {\bibinfo {volume}
  {202}},\ \bibinfo {pages} {1} (\bibinfo {year} {2012})}\BibitemShut {NoStop}%
\bibitem [{\citenamefont {Lintuvuori}\ \emph {et~al.}(2016)\citenamefont
  {Lintuvuori}, \citenamefont {Brown}, \citenamefont {Stratford},\ and\
  \citenamefont {Marenduzzo}}]{Lintuvuori_HORW_2016}%
  \BibitemOpen
  \bibfield  {author} {\bibinfo {author} {\bibfnamefont {J.~S.}\ \bibnamefont
  {Lintuvuori}}, \bibinfo {author} {\bibfnamefont {A.~T.}\ \bibnamefont
  {Brown}}, \bibinfo {author} {\bibfnamefont {K.}~\bibnamefont {Stratford}},\
  and\ \bibinfo {author} {\bibfnamefont {D.}~\bibnamefont {Marenduzzo}},\
  }\bibfield  {title} {\bibinfo {title} {Hydrodynamic oscillations and variable
  swimming speed in squirmers close to repulsive walls},\ }\href@noop {}
  {\bibfield  {journal} {\bibinfo  {journal} {Soft Matter}\ }\textbf {\bibinfo
  {volume} {12}},\ \bibinfo {pages} {7959} (\bibinfo {year}
  {2016})}\BibitemShut {NoStop}%
\bibitem [{\citenamefont {Ishimoto}\ and\ \citenamefont
  {Gaffney}(2013)}]{Ishimoto_SDB_2013}%
  \BibitemOpen
  \bibfield  {author} {\bibinfo {author} {\bibfnamefont {K.}~\bibnamefont
  {Ishimoto}}\ and\ \bibinfo {author} {\bibfnamefont {E.~A.}\ \bibnamefont
  {Gaffney}},\ }\bibfield  {title} {\bibinfo {title} {Squirmer dynamics near a
  boundary},\ }\href@noop {} {\bibfield  {journal} {\bibinfo  {journal} {Phys.
  Rev. E}\ }\textbf {\bibinfo {volume} {88}},\ \bibinfo {pages} {062702}
  (\bibinfo {year} {2013})}\BibitemShut {NoStop}%
\bibitem [{\citenamefont {Duman}\ \emph {et~al.}(2018)\citenamefont {Duman},
  \citenamefont {Isele-Holder}, \citenamefont {Elgeti},\ and\ \citenamefont
  {Gompper}}]{Duman_CDF_2018}%
  \BibitemOpen
  \bibfield  {author} {\bibinfo {author} {\bibfnamefont {{\"O}.}~\bibnamefont
  {Duman}}, \bibinfo {author} {\bibfnamefont {R.~E.}\ \bibnamefont
  {Isele-Holder}}, \bibinfo {author} {\bibfnamefont {J.}~\bibnamefont
  {Elgeti}},\ and\ \bibinfo {author} {\bibfnamefont {G.}~\bibnamefont
  {Gompper}},\ }\bibfield  {title} {\bibinfo {title} {Collective dynamics of
  self-propelled semiflexible filaments},\ }\href@noop {} {\bibfield  {journal}
  {\bibinfo  {journal} {Soft Matter}\ }\textbf {\bibinfo {volume} {14}},\
  \bibinfo {pages} {4483} (\bibinfo {year} {2018})}\BibitemShut {NoStop}%
\bibitem [{\citenamefont {Lauga}\ \emph {et~al.}(2006)\citenamefont {Lauga},
  \citenamefont {DiLuzio}, \citenamefont {Whitesides},\ and\ \citenamefont
  {Stone}}]{Lauga_SIC_2006}%
  \BibitemOpen
  \bibfield  {author} {\bibinfo {author} {\bibfnamefont {E.}~\bibnamefont
  {Lauga}}, \bibinfo {author} {\bibfnamefont {W.~R.}\ \bibnamefont {DiLuzio}},
  \bibinfo {author} {\bibfnamefont {G.~M.}\ \bibnamefont {Whitesides}},\ and\
  \bibinfo {author} {\bibfnamefont {H.~A.}\ \bibnamefont {Stone}},\ }\bibfield
  {title} {\bibinfo {title} {Swimming in circles: motion of bacteria near solid
  boundaries},\ }\href@noop {} {\bibfield  {journal} {\bibinfo  {journal}
  {Biophys. J.}\ }\textbf {\bibinfo {volume} {90}},\ \bibinfo {pages} {400}
  (\bibinfo {year} {2006})}\BibitemShut {NoStop}%
\bibitem [{\citenamefont {Hu}\ \emph {et~al.}(2015{\natexlab{b}})\citenamefont
  {Hu}, \citenamefont {Wysocki}, \citenamefont {Winkler},\ and\ \citenamefont
  {Gompper}}]{Hu_PSP_2015}%
  \BibitemOpen
  \bibfield  {author} {\bibinfo {author} {\bibfnamefont {J.}~\bibnamefont
  {Hu}}, \bibinfo {author} {\bibfnamefont {A.}~\bibnamefont {Wysocki}},
  \bibinfo {author} {\bibfnamefont {R.~G.}\ \bibnamefont {Winkler}},\ and\
  \bibinfo {author} {\bibfnamefont {G.}~\bibnamefont {Gompper}},\ }\bibfield
  {title} {\bibinfo {title} {Physical sensing of surface properties by
  microswimmers -- directing bacterial motion via wall slip},\ }\href@noop {}
  {\bibfield  {journal} {\bibinfo  {journal} {Sci. Rep.}\ }\textbf {\bibinfo
  {volume} {5}},\ \bibinfo {pages} {9586} (\bibinfo {year}
  {2015}{\natexlab{b}})}\BibitemShut {NoStop}%
\bibitem [{\citenamefont {Helfrich}(1973)}]{Helfrich_EPB_1973}%
  \BibitemOpen
  \bibfield  {author} {\bibinfo {author} {\bibfnamefont {W.}~\bibnamefont
  {Helfrich}},\ }\bibfield  {title} {\bibinfo {title} {Elastic properties of
  lipid bilayers: theory and possible experiments},\ }\href@noop {} {\bibfield
  {journal} {\bibinfo  {journal} {Z. Naturforsch.}\ }\textbf {\bibinfo {volume}
  {28}},\ \bibinfo {pages} {693} (\bibinfo {year} {1973})}\BibitemShut
  {NoStop}%
\bibitem [{\citenamefont {{J{\"u}lich Supercomputing Centre}}(2021)}]{jureca}%
  \BibitemOpen
  \bibfield  {author} {\bibinfo {author} {\bibnamefont {{J{\"u}lich
  Supercomputing Centre}}},\ }\bibfield  {title} {\bibinfo {title} {{JURECA:
  Data Centric and Booster Modules implementing the Modular Supercomputing
  Architecture at J{\"u}lich Supercomputing Centre}},\ }\href@noop {}
  {\bibfield  {journal} {\bibinfo  {journal} {J. Large-Scale Res. Facil.}\
  }\textbf {\bibinfo {volume} {7}},\ \bibinfo {pages} {A182} (\bibinfo {year}
  {2021})}\BibitemShut {NoStop}%
\end{thebibliography}

%

\end{document}